\documentclass[10pt]{article}
\usepackage{graphicx}
\usepackage{amsmath}
\usepackage{amssymb}
\usepackage{caption2}
\begin{document}
\bibliographystyle{prsty}
\begin{center}
{\large {\bf \sc{  Analysis of  the $\bar{D}\Sigma_c$, $\bar{D}\Sigma_c^*$, $\bar{D}^{*}\Sigma_c$ and
$  \bar{D}^{*}\Sigma_c^*$  pentaquark molecular states  with QCD sum rules }}} \\[2mm]
Zhi-Gang Wang \footnote{E-mail: zgwang@aliyun.com.  }     \\
 Department of Physics, North China Electric Power University, Baoding 071003, P. R. China
\end{center}

\begin{abstract}
In this article, we  study  the $\bar{D}\Sigma_c$, $\bar{D}\Sigma_c^*$, $\bar{D}^{*}\Sigma_c$  and   $\bar{D}^{*}\Sigma_c^*$  pentaquark molecular states
with the QCD sum rules by carrying out the operator product expansion   up to   the vacuum condensates of dimension $13$ in a consistent way.
  The present calculations support assigning the $P_c(4312)$ to be the $\bar{D}\Sigma_c$ pentaquark molecular state with $J^P={\frac{1}{2}}^-$, assigning the $P_c(4380)$ to be the $\bar{D}\Sigma_c^*$ pentaquark molecular state with $J^P={\frac{3}{2}}^-$, assigning the $P_c(4440/4457)$ to be the $\bar{D}^{*}\Sigma_c$ pentaquark molecular state with $J^P={\frac{3}{2}}^-$ or the $\bar{D}^{*}\Sigma_c^*$ pentaquark molecular state with $J^P={\frac{5}{2}}^-$. Special attentions are payed to  the operator product expansion.
\end{abstract}

 PACS number: 12.39.Mk, 14.20.Lq, 12.38.Lg

Key words: Pentaquark molecular states, QCD sum rules

\section{Introduction}

In 2015,  the  LHCb collaboration studied the $\Lambda_b^0\to J/\psi K^- p$ decays and observed  two pentaquark candidates $P_c(4380)$ and $P_c(4450)$ in the $J/\psi p$  mass spectrum  with the significances of more than 9 standard deviations \cite{LHCb-4380}.
The  Breit-Wigner   masses and widths are  $M_{P_c(4380)}=4380\pm 8\pm 29\,\rm{MeV}$, $M_{P_c(4450)}=4449.8\pm 1.7\pm 2.5\,\rm{MeV}$, $\Gamma_{P_c(4380)}=205\pm 18\pm 86\,\rm{MeV}$, and  $\Gamma_{P_c(4450)}=39\pm 5\pm 19\,\rm{MeV}$, respectively.
The  preferred quantum numbers of the  $\left(P_c(4380), \,P_c(4450)\right)$ are   $J^P=\left({\frac{3}{2}}^-,\, {\frac{5}{2}}^+\right)$, respectively, while the quantum numbers  $J^P=\left({\frac{3}{2}}^+,\, {\frac{5}{2}}^-\right)$ and $\left({\frac{5}{2}}^+,\, {\frac{3}{2}}^-\right)$ are also acceptable solutions. More experimental data are still needed to determine the quantum numbers unambiguously.
In 2016, the LHCb collaboration  inspected the $\Lambda_b^0\to J/\psi K^- p$ decays for the presence of $J/\psi p$  or $J/\psi K^-$  contributions with minimal assumptions about $K^- p$  contributions  and obtained model-independent support for the evidences of the $P_c^+(4380/4500)$  \cite{LHCb-1604}.
Also in 2016, the LHCb collaboration obtained additional support for the existences of the two $P_c^+(4380/4450)$ in the $\Lambda_b^0\to J/\psi \pi^- p$ decays   \cite{LHCb-1606}.

 There have been  several  possible  assignments since the observations of the $P_c(4380)$ and $P_c(4450)$, such as the    pentaquark molecular  states \cite{RChen-PRL,HXChen-PRL,EOset-PRD,JHe-PLB,FKGuo,Penta-molecule,HXChen-EPJC,KAzizi} (or not the molecular pentaquark states \cite{Penta-molecule-No}), the diquark-triquark  type  pentaquark states \cite{di-tri-penta},  the diquark-diquark-antiquark type pentaquark states \cite{Maiani1507,di-di-anti-penta,Wang1508-EPJC,WangHuang1508}, re-scattering effects \cite{rescattering-penta}, etc. In Table 1, we present some typical assignments in the  scenario of pentaquark molecular  states, in this article, we will focus on this scenario, and examine the possible molecule assignments based on the QCD sum rules.

The QCD sum rules is a powerful theoretical tool in studying the ground state hadrons \cite{SVZ79,PRT85,ColangeloReview,NarisonBook}.  The diquark-diquark-antiquark type hidden-charm pentaquark states have been studied in details with the QCD sum rules by carrying out the operator product expansion up to the vacuum condensates of dimension $10$ in a consistent way  \cite{Wang1508-EPJC,WangHuang1508}. In Ref.\cite{HXChen-PRL}, Chen et al study the $\bar{D}^{*}\Sigma_c$          and $\bar{D}\Sigma_c^*-\bar{D}^*\Lambda_c$ pentaquark molecular states with the QCD sum rules by carrying out the operator product expansion up to the vacuum condensates of dimension $8$. In Ref.\cite{HXChen-EPJC}, Chen et al construct many interpolating currents to study the meson-baryon type pentaquark molecular states with the spin $J=\frac{1}{2}$, $\frac{3}{2}$ and $\frac{5}{2}$ extensively. In Ref.\cite{Azizi-PRD}, Azizi, Sarac and  Sundu study the $\bar{D}^{*}\Sigma_c$          and $\bar{D}\Sigma_c^*-\bar{D}^*\Lambda_c$ pentaquark molecular states with the QCD sum rules by carrying out the operator product expansion up to the vacuum condensates of dimension $6$. In Refs.\cite{HXChen-PRL,HXChen-EPJC,Azizi-PRD}, also in the QCD sum rules for the tetraquark states \cite{Nielsen-mc-1GeV}, the QCD spectral densities have two energy scales, $\mu=m_c$ for the $\overline{MS}$ mass  $m_{c}(m_c)$ and $\mu=1\,\rm{GeV}$ for other input parameters. In Refs.\cite{HXChen-PRL,HXChen-EPJC},
 $m_c(m_c)=1.23\,\rm{GeV}$, while in Ref.\cite{Azizi-PRD}, $m_c(m_c)=1.27\,\rm{GeV}$.

In Refs.\cite{Wang-tetraquark-QCDSR,Wang-molecule-QCDSR}, we  study the diquark-antidiquark type tetraquark states and meson-meson type molecular states with the QCD
sum rules by calculating the  vacuum condensates up to dimension-10  in
the operator product expansion in a systematic way, and explore the energy scale dependence of the hidden-charm (hidden-bottom) tetraquark states and molecular states in details for the first time, and suggest a  formula
\begin{eqnarray}
\mu&=&\sqrt{M^2_{X/Y/Z}-(2{\mathbb{M}}_Q)^2} \, ,
 \end{eqnarray}
 with the effective heavy quark  masses ${\mathbb{M}}_Q$ to determine  the optimal energy scales of the  QCD spectral densities, which works very well for the hidden-charm (hidden-bottom) tetraquark states and molecular states \cite{Wang-tetraquark-QCDSR,Wang-molecule-QCDSR,Wang-EPJC-update}, and hidden-charm pentaquark states \cite{Wang1508-EPJC,WangHuang1508}. In calculations, we take the $\overline{MS}$ masses $m_Q(m_Q)$ from the Particle Data Group \cite{PDG}. In the QCD sum rules for the multiquark states, it is difficult to satisfy the pole dominance or ground state dominance, the energy scale formula $\mu=\sqrt{M^2_{X/Y/Z}-(2{\mathbb{M}}_Q)^2}$ can enhance the pole contributions remarkably, and improve the convergent behaviors of the operator product expansion considerably.

    In this article, we extend our previous works \cite{Wang1508-EPJC,WangHuang1508,Wang-tetraquark-QCDSR,Wang-molecule-QCDSR,Wang-EPJC-update} to study the masses and pole residues of the  $\bar{D}\Sigma_c^*$,  $\bar{D}^{*}\Sigma_c$  and
  $\bar{D}^{*}\Sigma_c^*$  pentaquark molecular states    with the QCD sum rules by carrying out  the operator product expansion up to the vacuum condensates  of dimension $13$,  and revisit the assignments of the $P_c(4380)$ and $P_c(4450)$.
  In calculations,   we separate  the contributions of the negative parity and positive parity pentaquark  molecular  states  unambiguously,
and   study the hidden-charm pentaquark molecular states in three cases in details.

After the present work  was finished and submitted to https://arxiv.org/, and appeared as arXiv:1806.10384,
 the LHCb collaboration  observed a  narrow pentaquark candidate $P_c(4312)$ in the $ J/\psi  p$ mass spectrum with the statistical significance of  $7.3\sigma$, and confirmed the $P_c(4450)$ pentaquark structure, and observed that it consists  of two narrow overlapping peaks $P_c(4440)$ and $P_c(4457)$
  with  the statistical significance of  $5.4\sigma$ \cite{LHCb-Pc4312}.
   The measured  masses and widths are
\begin{flalign}
 &P_c(4312) : M = 4311.9\pm0.7^{+6.8}_{-0.6} \mbox{ MeV}\, , \, \Gamma = 9.8\pm2.7^{+ 3.7}_{- 4.5} \mbox{ MeV} \, , \nonumber \\
 & P_c(4440) : M = 4440.3\pm1.3^{+4.1}_{-4.7} \mbox{ MeV}\, , \, \Gamma = 20.6\pm4.9_{-10.1}^{+ 8.7} \mbox{ MeV} \, , \nonumber \\
 &P_c(4457) : M = 4457.3\pm0.6^{+4.1}_{-1.7} \mbox{ MeV} \, ,\, \Gamma = 6.4\pm2.0_{- 1.9}^{+ 5.7} \mbox{ MeV} \,   .
\end{flalign}
  The $P_c(4312)$ may be a $\bar{D}\Sigma_c$ pentaquark molecule candidate \cite{LHCb-Pc4312,Pc4312-mole}. We modify the assignments  according to the new experimental data and add the QCD sum rules for the $\bar{D}\Sigma_c$ pentaquark molecular state.

\begin{table}
\begin{center}
\begin{tabular}{|c|c|c|c|c|}\hline\hline
$P_c(4380)$                     &$P_c(4450)$                                 &References       \\ \hline

$\bar{D}^{*}\Sigma_c$          &$\bar{D}^{*}\Sigma_c^*$                      &\cite{RChen-PRL}   \\  \hline
$\bar{D}^{*}\Sigma_c$          &$\bar{D}\Sigma_c^*-\bar{D}^*\Lambda_c$       &\cite{HXChen-PRL}  \\  \hline
                               &$\bar{D}^{*}\Sigma_c-\bar{D}^{*}\Sigma_c^*$  &\cite{EOset-PRD}   \\ \hline
$\bar{D}\Sigma_c^{*}$          &$\bar{D}^{*}\Sigma_c$                        &\cite{JHe-PLB}   \\      \hline \hline
\end{tabular}
\end{center}
\caption{ Some typical pentaquark molecule assignments.}
\end{table}

 The article is arranged as follows:
  we derive the QCD sum rules for the masses and pole residues of the $\bar{D}\Sigma_c$, $\bar{D}\Sigma_c^*$,  $\bar{D}^{*}\Sigma_c$  and
  $\bar{D}^{*}\Sigma_c^*$  pentaquark molecular states  in Sect.2;  in Sect.3, we present the numerical results and discussions; and Sect.4 is reserved for our
conclusion.

\section{QCD sum rules for  the $\bar{D}\Sigma_c$, $\bar{D}\Sigma_c^*$, $\bar{D}^{*}\Sigma_c$ and
$  \bar{D}^{*}\Sigma_c^*$  pentaquark molecular states}

In the following, we write down  the two-point correlation functions $\Pi(p)$, $\Pi_{\mu\nu}(p)$ and $\Pi_{\mu\nu\alpha\beta}(p)$  in the QCD sum rules,
\begin{eqnarray}
\Pi(p)&=&i\int d^4x e^{ip \cdot x} \langle0|T\left\{J(x)\bar{J}(0)\right\}|0\rangle \, , \\
\Pi_{\mu\nu}(p)&=&i\int d^4x e^{ip \cdot x} \langle0|T\left\{J_{\mu}(x)\bar{J}_{\nu}(0)\right\}|0\rangle \, , \\
\Pi_{\mu\nu\alpha\beta}(p)&=&i\int d^4x e^{ip \cdot x} \langle0|T\left\{J_{\mu\nu}(x)\bar{J}_{\alpha\beta}(0)\right\}|0\rangle \, ,
\end{eqnarray}
where the currents $J(x)=J^{\bar{D}\Sigma_c}(x)$, $J_\mu(x)=J^{\bar{D}\Sigma_c^*}_{\mu}(x)$, $ J^{\bar{D}^*\Sigma_c}_{\mu}(x)$, $J_{\mu\nu}(x)=J^{\bar{D}^*\Sigma_c^*}_{\mu\nu}(x)$,
\begin{eqnarray}
 J^{\bar{D}\Sigma_c}(x)&=& \bar{c}(x)i\gamma_5 u(x)\, \varepsilon^{ijk}  u^T_i(x) C\gamma_\alpha d_j(x)\, \gamma^\alpha\gamma_5 c_{k}(x) \, ,\nonumber \\
 J^{\bar{D}\Sigma_c^*}_{\mu}(x)&=& \bar{c}(x)i\gamma_5 u(x)\, \varepsilon^{ijk}  u^T_i(x) C\gamma_\mu d_j(x)\, c_{k}(x) \, ,\nonumber \\
 J^{\bar{D}^*\Sigma_c}_{\mu}(x)&=& \bar{c}(x)\gamma_\mu u(x)\, \varepsilon^{ijk}  u^T_i(x) C\gamma_\alpha d_j(x)\, \gamma^\alpha\gamma_5 c_{k}(x) \, ,\nonumber \\
 J^{\bar{D}^*\Sigma_c^*}_{\mu\nu}(x)&=& \bar{c}(x)\gamma_\mu u(x)\, \varepsilon^{ijk}  u^T_i(x) C\gamma_\nu d_j(x)\,   c_{k}(x) +(\mu\leftrightarrow\nu)\, ,
\end{eqnarray}
the $i$, $j$, $k$ are color indices. In this article, we choose the color singlet-singlet type (or meson-baryon type) currents  $J^{\bar{D}\Sigma_c}(x)$,
 $J^{\bar{D}\Sigma_c^*}_{\mu}(x)$, $ J^{\bar{D}^*\Sigma_c}_{\mu}(x)$, $J^{\bar{D}^*\Sigma_c^*}_{\mu\nu}(x)$ to interpolate the $\bar{D}\Sigma_c$, $\bar{D}\Sigma_c^*$, $\bar{D}^{*}\Sigma_c$ and
$  \bar{D}^{*}\Sigma_c^*$  pentaquark molecular states with the spin-parity $J^P={\frac{1}{2}}^-$, ${\frac{3}{2}}^-$, ${\frac{3}{2}}^-$ and ${\frac{5}{2}}^-$, respectively.
A  five-quark state has many  Fock states,  we call it MB pentaquark molecular state if its dominant Fock state is of the meson(M)-baryon(B) type. For example,
the $\bar{D}\Sigma_c^*$ pentaquark molecular state maybe have  many other MB Fock components such as  $\bar{D}^{*}\Sigma_c$, $J/\psi p$, $\cdots$ beyond the dominant $\bar{D}\Sigma_c^*$ component. The current $J^{\bar{D}\Sigma_c^*}_{\mu}(x)$ couples dominantly to the $\bar{D}\Sigma_c^*$ pentaquark molecular state, although other currents with the same quantum numbers as the current $J^{\bar{D}\Sigma_c^*}_{\mu}(x)$ have non-vanishing couplings with
the $\bar{D}\Sigma_c^*$ pentaquark molecular state due to its small Fock components $\bar{D}^{*}\Sigma_c$, $J/\psi p$, $\cdots$, the couplings are expected to be weak enough to be neglected. We can obtain additional support by studying the two-body strong decays of the $\bar{D}\Sigma_c^*$ and $\bar{D}^{*}\Sigma_c$ pentaquark molecular states with the three-point QCD sum rules, this may be our next work.

On the other hand, we can perform Fierz re-arrangement  to the currents $J(x)$, $J_{\mu}(x)$ and $J_{\mu\nu}(x)$ both in the color and Dirac-spinor  spaces  to obtain the diquark-diquark-antiquark  type currents.
 The meson-baryon type current with special quantum numbers couples potentially (dominantly)  to a special pentaquark molecular
state, while the current can be re-arranged to a current as a special superposition of diquark-diquark-antiquark type currents, which  couple potentially
 to the pentaquark states respectively. The pentaquark molecular state can be taken as a special superposition of a series of the diquark-diquark-antiquark type  pentaquark states, and embodies
the net effects.

The currents $J(0)$, $J_\mu(0)$ and $J_{\mu\nu}(0)$ couple potentially to the ${\frac{1}{2}}^-$, ${\frac{1}{2}}^+$, ${\frac{3}{2}}^-$ and ${\frac{1}{2}}^-$, ${\frac{3}{2}}^+$, ${\frac{5}{2}}^-$  hidden-charm  pentaquark molecular  states $P_{\frac{1}{2}}^{-}$, $P_{\frac{1}{2}}^{+}$, $P_{\frac{3}{2}}^{-}$ and $P_{\frac{1}{2}}^{-}$, $P_{\frac{3}{2}}^{+}$, $P_{\frac{5}{2}}^{-}$, respectively,
\begin{eqnarray}
\langle 0| J (0)|P_{\frac{1}{2}}^{-}(p)\rangle &=&\lambda^{-}_{\frac{1}{2}} U^{-}(p,s) \, ,  \\
\langle 0| J_{\mu} (0)|P_{\frac{1}{2}}^{+}(p)\rangle &=&f^{+}_{\frac{1}{2}}p_\mu U^{+}(p,s) \, , \nonumber \\
\langle 0| J_{\mu} (0)|P_{\frac{3}{2}}^{-}(p)\rangle &=&\lambda^{-}_{\frac{3}{2}} U^{-}_\mu(p,s) \, ,  \\
\langle 0| J_{\mu\nu} (0)|P_{\frac{1}{2}}^{-}(p)\rangle &=&g^{-}_{\frac{1}{2}}p_\mu p_\nu U^{-}(p,s) \, , \nonumber\\
\langle 0| J_{\mu\nu} (0)|P_{\frac{3}{2}}^{+}(p)\rangle &=&f^{+}_{\frac{3}{2}} \left[p_\mu U^{+}_{\nu}(p,s)+p_\nu U^{+}_{\mu}(p,s)\right] \, , \nonumber\\
\langle 0| J_{\mu\nu} (0)|P_{\frac{5}{2}}^{-}(p)\rangle &=&\sqrt{2}\lambda^{-}_{\frac{5}{2}} U^{-}_{\mu\nu}(p,s) \, ,
\end{eqnarray}
the spinors $U^\pm(p,s)$ satisfy the Dirac equations  $(\not\!\!p-M_{\pm})U^{\pm}(p)=0$, while the spinors $U^{\pm}_\mu(p,s)$ and $U^{\pm}_{\mu\nu}(p,s)$ satisfy the Rarita-Schwinger equations $(\not\!\!p-M_{\pm})U^{\pm}_\mu(p)=0$ and $(\not\!\!p-M_{\pm})U^{\pm}_{\mu\nu}(p)=0$,  and the relations $\gamma^\mu U^{\pm}_\mu(p,s)=0$,
$p^\mu U^{\pm}_\mu(p,s)=0$, $\gamma^\mu U^{\pm}_{\mu\nu}(p,s)=0$,
$p^\mu U^{\pm}_{\mu\nu}(p,s)=0$, $ U^{\pm}_{\mu\nu}(p,s)= U^{\pm}_{\nu\mu}(p,s)$, respectively.   The currents $J(0)$, $J_\mu(0)$ and $J_{\mu\nu}(0)$ also couple potentially to the ${\frac{1}{2}}^+$, ${\frac{1}{2}}^-$, ${\frac{3}{2}}^+$ and ${\frac{1}{2}}^+$, ${\frac{3}{2}}^-$, ${\frac{5}{2}}^+$  hidden-charm  pentaquark molecular
 states $P_{\frac{1}{2}}^{+}$, $P_{\frac{1}{2}}^{-}$, $P_{\frac{3}{2}}^{+}$ and $P_{\frac{1}{2}}^{+}$, $P_{\frac{3}{2}}^{-}$, $P_{\frac{5}{2}}^{+}$, respectively,
 \begin{eqnarray}
\langle 0| J (0)|P_{\frac{1}{2}}^{+}(p)\rangle &=&\lambda^{+}_{\frac{1}{2}}i\gamma_5 U^{+}(p,s) \, ,  \\
\langle 0| J_{\mu} (0)|P_{\frac{1}{2}}^{-}(p)\rangle &=&f^{-}_{\frac{1}{2}}p_\mu i\gamma_5 U^{-}(p,s) \, , \nonumber\\
\langle 0| J_{\mu} (0)|P_{\frac{3}{2}}^{+}(p)\rangle &=&\lambda^{+}_{\frac{3}{2}}i\gamma_5 U^{+}_{\mu}(p,s) \, , \\
\langle 0| J_{\mu\nu} (0)|P_{\frac{1}{2}}^{+}(p)\rangle &=&g^{+}_{\frac{1}{2}}p_\mu p_\nu i\gamma_5 U^{+}(p,s) \, , \nonumber\\
\langle 0| J_{\mu\nu} (0)|P_{\frac{3}{2}}^{-}(p)\rangle &=&f^{-}_{\frac{3}{2}} i\gamma_5\left[p_\mu U^{-}_{\nu}(p,s)+p_\nu U^{-}_{\mu}(p,s)\right] \, , \nonumber\\
\langle 0| J_{\mu\nu} (0)|P_{\frac{5}{2}}^{+}(p)\rangle &=&\sqrt{2}\lambda^{+}_{\frac{5}{2}}i\gamma_5 U^{+}_{\mu\nu}(p,s) \, ,
\end{eqnarray}
because multiplying $i \gamma_{5}$ to the currents $J(x)$, $J_{\mu}(x)$  and $J_{\mu\nu}(x)$ changes their parity \cite{Chung82,Bagan93,Oka96,WangHbaryon}.
The $\lambda^{\pm}_{\frac{1}{2}/\frac{3}{2}/\frac{5}{2}}$, $f^{\pm}_{\frac{1}{2}/\frac{3}{2}}$ and $g^{\pm}_{\frac{1}{2}}$ are the pole residues or current-pentaquark-molecule  coupling constants.

In this article, we refer to a five-quark state with fractional spin  as a pentaquark molecular state
if its dominant component  is of the color singlet-singlet type, in other words,  the meson-baryon type, the meson and baryon are not necessary to be the physical states, they just have the same quantum numbers as the constituents of the interpolating currents. If the constituents are in relative S-wave, P-wave, D-wave or F-wave,  the $\bar{D}\Sigma_c$, $\bar{D}\Sigma_c^*$, $\bar{D}^{*}\Sigma_c$ and $\bar{D}^{*}\Sigma_c^*$ pentaquark molecular states maybe have the spin-parity $J^P={\frac{1}{2}}^\pm$,  ${\frac{3}{2}}^\pm$, ${\frac{5}{2}}^\pm$, etc, the relevant (not all the) spin-parity are listed in Table 2.

In general, we expect to solve the eigenequation of the  QCD Hamiltonian and obtain the eigenstates and eigenvalues for the five-quark systems. By analyzing the eigenvalues and substructures of the eigenstates, we can distinguish the diquark-diquark-antiquark type pentaquark states and meson-baryon type pentaquark molecular states. However, at the present time, it is a very difficult work to solve eigenequation of the  QCD Hamiltonian for the five-quark systems.

\begin{table}
\begin{center}
\begin{tabular}{|c|c|c|c|c|}\hline\hline
                        &S-wave              &P-wave                                &D-wave                              &F-wave\\ \hline

$\bar{D}\Sigma_c$       &${\frac{1}{2}}^-$   &${\frac{1}{2}}^+$                     &                                    &\\  \hline

$\bar{D}\Sigma_c^*$     &${\frac{3}{2}}^-$   &${\frac{1}{2}}^+,\,{\frac{3}{2}}^+$   &${\frac{1}{2}}^-$                   &\\  \hline

$\bar{D}^{*}\Sigma_c$   &${\frac{3}{2}}^-$   &${\frac{1}{2}}^+,\,{\frac{3}{2}}^+$   &${\frac{1}{2}}^-$                   &  \\  \hline

$\bar{D}^{*}\Sigma_c^*$ &${\frac{5}{2}}^-$   &${\frac{3}{2}}^+,\,{\frac{5}{2}}^+$   &${\frac{1}{2}}^-,\,{\frac{3}{2}}^-$ &${\frac{1}{2}}^+$    \\  \hline
\hline
\end{tabular}
\end{center}
\caption{ The relevant (not all the) spin-parity of the pentaquark molecular states.}
\end{table}

 At the phenomenological side, we  insert  a complete set  of intermediate pentaquark molecular states with the
same quantum numbers as the current operators $J(x)$,
$i\gamma_5 J(x)$, $J_\mu(x)$,
$i\gamma_5 J_\mu(x)$, $J_{\mu\nu}(x)$ and
$i\gamma_5 J_{\mu\nu}(x)$ into the correlation functions
$\Pi(p)$, $\Pi_{\mu\nu}(p)$ and $\Pi_{\mu\nu\alpha\beta}(p)$ to obtain the hadronic representation
\cite{SVZ79,PRT85}, because the scattering meson-baryon states can only contribute a finite width to the pentaquark molecular states to modify the dispersion relation.
 After isolating the pole terms of the lowest
 hidden-charm  pentaquark molecular states, we obtain the
following results:
\begin{eqnarray}
  \Pi(p) & = & {\lambda^{-}_{\frac{1}{2}}}^2  {\!\not\!{p}+ M_{-} \over M_{-}^{2}-p^{2}  } +  {\lambda^{+}_{\frac{1}{2}}}^2  {\!\not\!{p}- M_{+} \over M_{+}^{2}-p^{2}  } +\cdots  \, ,\\
   \Pi_{\mu\nu}(p) & = & {\lambda^{-}_{\frac{3}{2}}}^2  {\!\not\!{p}+ M_{-} \over M_{-}^{2}-p^{2}  } \left(- g_{\mu\nu}+\frac{\gamma_\mu\gamma_\nu}{3}+\frac{2p_\mu p_\nu}{3p^2}-\frac{p_\mu\gamma_\nu-p_\nu \gamma_\mu}{3\sqrt{p^2}}
\right)\nonumber\\
&&+  {\lambda^{+}_{\frac{3}{2}}}^2  {\!\not\!{p}- M_{+} \over M_{+}^{2}-p^{2}  } \left(- g_{\mu\nu}+\frac{\gamma_\mu\gamma_\nu}{3}+\frac{2p_\mu p_\nu}{3p^2}-\frac{p_\mu\gamma_\nu-p_\nu \gamma_\mu}{3\sqrt{p^2}}
\right)   \nonumber \\
& &+ {f^{+}_{\frac{1}{2}}}^2  {\!\not\!{p}+ M_{+} \over M_{+}^{2}-p^{2}  } p_\mu p_\nu+  {f^{-}_{\frac{1}{2}}}^2  {\!\not\!{p}- M_{-} \over M_{-}^{2}-p^{2}  } p_\mu p_\nu  +\cdots  \, ,\\
\Pi_{\mu\nu\alpha\beta}(p) & = & 2{\lambda^{-}_{\frac{5}{2}}}^2  {\!\not\!{p}+ M_{-} \over M_{-}^{2}-p^{2}  } \left[\frac{ \widetilde{g}_{\mu\alpha}\widetilde{g}_{\nu\beta}+\widetilde{g}_{\mu\beta}\widetilde{g}_{\nu\alpha}}{2}-\frac{\widetilde{g}_{\mu\nu}\widetilde{g}_{\alpha\beta}}{5}-\frac{1}{10}\left( \gamma_{\mu}\gamma_{\alpha}+\frac{\gamma_{\mu}p_{\alpha}-\gamma_{\alpha}p_{\mu}}{\sqrt{p^2}}-\frac{p_{\mu}p_{\alpha}}{p^2}\right)\widetilde{g}_{\nu\beta}\right.\nonumber\\
&&\left.-\frac{1}{10}\left( \gamma_{\nu}\gamma_{\alpha}+\frac{\gamma_{\nu}p_{\alpha}-\gamma_{\alpha}p_{\nu}}{\sqrt{p^2}}-\frac{p_{\nu}p_{\alpha}}{p^2}\right)\widetilde{g}_{\mu\beta}
+\cdots\right]\nonumber\\
&&+2 {\lambda^{+}_{\frac{5}{2}}}^2  {\!\not\!{p}- M_{+} \over M_{+}^{2}-p^{2}  } \left[\frac{ \widetilde{g}_{\mu\alpha}\widetilde{g}_{\nu\beta}+\widetilde{g}_{\mu\beta}\widetilde{g}_{\nu\alpha}}{2}
-\frac{\widetilde{g}_{\mu\nu}\widetilde{g}_{\alpha\beta}}{5}-\frac{1}{10}\left( \gamma_{\mu}\gamma_{\alpha}+\frac{\gamma_{\mu}p_{\alpha}-\gamma_{\alpha}p_{\mu}}{\sqrt{p^2}}-\frac{p_{\mu}p_{\alpha}}{p^2}\right)\widetilde{g}_{\nu\beta}\right.\nonumber\\
&&\left.
-\frac{1}{10}\left( \gamma_{\nu}\gamma_{\alpha}+\frac{\gamma_{\nu}p_{\alpha}-\gamma_{\alpha}p_{\nu}}{\sqrt{p^2}}-\frac{p_{\nu}p_{\alpha}}{p^2}\right)\widetilde{g}_{\mu\beta}
 +\cdots\right]   \nonumber\\
 && +{f^{+}_{\frac{3}{2}}}^2  {\!\not\!{p}+ M_{+} \over M_{+}^{2}-p^{2}  } \left[ p_\mu p_\alpha \left(- g_{\nu\beta}+\frac{\gamma_\nu\gamma_\beta}{3}+\frac{2p_\nu p_\beta}{3p^2}-\frac{p_\nu\gamma_\beta-p_\beta \gamma_\nu}{3\sqrt{p^2}}
\right)+\cdots \right]\nonumber\\
&&+  {f^{-}_{\frac{3}{2}}}^2  {\!\not\!{p}- M_{-} \over M_{-}^{2}-p^{2}  } \left[ p_\mu p_\alpha \left(- g_{\nu\beta}+\frac{\gamma_\nu\gamma_\beta}{3}+\frac{2p_\nu p_\beta}{3p^2}-\frac{p_\nu\gamma_\beta-p_\beta \gamma_\nu}{3\sqrt{p^2}}
\right)+\cdots \right]   \nonumber \\
& &+ {g^{-}_{\frac{1}{2}}}^2  {\!\not\!{p}+ M_{-} \over M_{-}^{2}-p^{2}  } p_\mu p_\nu p_\alpha p_\beta+  {g^{+}_{\frac{1}{2}}}^2  {\!\not\!{p}- M_{+} \over M_{+}^{2}-p^{2}  } p_\mu p_\nu p_\alpha p_\beta  +\cdots \, ,
\end{eqnarray}
where $\widetilde{g}_{\mu\nu}=g_{\mu\nu}-\frac{p_{\mu}p_{\nu}}{p^2}$.
In calculations, we have used the following summations of the Rarita-Schwinger  spinors \cite{HuangShiZhong},
\begin{eqnarray}
\sum_s U \overline{U}&=&\left(\!\not\!{p}+M_{\pm}\right) \,  ,  \\
\sum_s U_\mu \overline{U}_\nu&=&\left(\!\not\!{p}+M_{\pm}\right)\left( -g_{\mu\nu}+\frac{\gamma_\mu\gamma_\nu}{3}+\frac{2p_\mu p_\nu}{3p^2}-\frac{p_\mu
\gamma_\nu-p_\nu \gamma_\mu}{3\sqrt{p^2}} \right) \,  ,  \\
\sum_s U_{\mu\nu}\overline {U}_{\alpha\beta}&=&\left(\!\not\!{p}+M_{\pm}\right)\left\{\frac{\widetilde{g}_{\mu\alpha}\widetilde{g}_{\nu\beta}+\widetilde{g}_{\mu\beta}\widetilde{g}_{\nu\alpha}}{2} -\frac{\widetilde{g}_{\mu\nu}\widetilde{g}_{\alpha\beta}}{5}-\frac{1}{10}\left( \gamma_{\mu}\gamma_{\alpha}+\frac{\gamma_{\mu}p_{\alpha}-\gamma_{\alpha}p_{\mu}}{\sqrt{p^2}}-\frac{p_{\mu}p_{\alpha}}{p^2}\right)\widetilde{g}_{\nu\beta}\right. \nonumber\\
&&-\frac{1}{10}\left( \gamma_{\nu}\gamma_{\alpha}+\frac{\gamma_{\nu}p_{\alpha}-\gamma_{\alpha}p_{\nu}}{\sqrt{p^2}}-\frac{p_{\nu}p_{\alpha}}{p^2}\right)\widetilde{g}_{\mu\beta}
-\frac{1}{10}\left( \gamma_{\mu}\gamma_{\beta}+\frac{\gamma_{\mu}p_{\beta}-\gamma_{\beta}p_{\mu}}{\sqrt{p^2}}-\frac{p_{\mu}p_{\beta}}{p^2}\right)\widetilde{g}_{\nu\alpha}\nonumber\\
&&\left.-\frac{1}{10}\left( \gamma_{\nu}\gamma_{\beta}+\frac{\gamma_{\nu}p_{\beta}-\gamma_{\beta}p_{\nu}}{\sqrt{p^2}}-\frac{p_{\nu}p_{\beta}}{p^2}\right)\widetilde{g}_{\mu\alpha} \right\} \, ,
\end{eqnarray}
and $p^2=M^2_{\pm}$ on the mass-shell.

In this article, we choose the structures $\!\not\!{p}$,  $1$,   $\!\not\!{p}g_{\mu\nu}$, $g_{\mu\nu}$ and $\!\not\!{p}\left(g_{\mu\alpha}g_{\nu\beta}+g_{\mu\beta}g_{\nu\alpha}\right)$, $g_{\mu\alpha}g_{\nu\beta}+g_{\mu\beta}g_{\nu\alpha}$ for the correlation functions $\Pi(p)$, $\Pi_{\mu\nu}(p)$ and $\Pi_{\mu\nu\alpha\beta}(p)$ respectively to study the $J^P={\frac{1}{2}}^\mp$, ${\frac{3}{2}}^\mp$ and ${\frac{5}{2}}^\mp$ pentaquark molecular states,
\begin{eqnarray}
\Pi(p)&=&\Pi_{\frac{1}{2}}^1(p^2)\!\not\!{p}+\Pi_{\frac{1}{2}}^0(p^2)\, , \nonumber\\
\Pi_{\mu\nu}(p)&=&-\Pi_{\frac{3}{2}}^1(p^2)\!\not\!{p}\,g_{\mu\nu}-\Pi_{\frac{3}{2}}^0(p^2)\,g_{\mu\nu}+\cdots\, , \nonumber\\
\Pi_{\mu\nu\alpha\beta}(p)&=&\Pi_{\frac{5}{2}}^1(p^2)\!\not\!{p}\left( g_{\mu\alpha}g_{\nu\beta}+g_{\mu\beta}g_{\nu\alpha}\right)+\Pi_{\frac{5}{2}}^0(p^2)\,\left( g_{\mu\alpha}g_{\nu\beta}+g_{\mu\beta}g_{\nu\alpha}\right)+ \cdots \, .
\end{eqnarray}

If we choose the structures  $\!\not\!{p} p_{\mu} p_{\nu}$ and $p_{\mu} p_{\nu}$ in the correlation function $\Pi_{\mu\nu}(p)$, both the $J^P={\frac{1}{2}}^\pm$ and ${\frac{3}{2}}^\pm$ pentaquark molecular states have contributions. If we choose the  structures  $\!\not\!{p}\left( p_{\mu} p_{\alpha}g_{\nu\beta}+p_{\nu}p_{\beta}g_{\mu\alpha}\right)$ and $\left( p_{\mu} p_{\alpha}g_{\nu\beta}+p_{\nu}p_{\beta}g_{\mu\alpha}\right)$  in the correlation function $\Pi_{\mu\nu\alpha\beta}(p)$, both the $J^P={\frac{3}{2}}^\pm$ and ${\frac{5}{2}}^\pm$ pentaquark molecular states have contributions.
On the other hand, if we choose the structures  $\!\not\!{p} p_{\mu} p_{\nu}p_{\alpha}p_{\beta}$ and $ p_{\mu} p_{\nu}p_{\alpha}p_{\beta}$  in the correlation function $\Pi_{\mu\nu\alpha\beta}(p)$, all the $J^P={\frac{1}{2}}^\pm$,  ${\frac{3}{2}}^\pm$ and ${\frac{5}{2}}^\pm$ pentaquark molecular states have contributions. We can distinguish those contributions unambiguously and obtain the QCD sum rules for the $J^P={\frac{1}{2}}^\pm$,  ${\frac{3}{2}}^\pm$ and ${\frac{5}{2}}^\pm$ pentaquark molecular states respectively. However, it is a very difficult work, the QCD sum rules obtained in this way are always failed to work well. In fact, we usually construct a current  to interpolate the baryon or pentaquark states  with the largest  spin.

Now we obtain the spectral densities at phenomenological side through the dispersion relation,
\begin{eqnarray}
\frac{{\rm Im}\Pi_{j}^1(s)}{\pi}&=& {\lambda^{-}_{j}}^2 \delta\left(s-M_{-}^2\right)+{\lambda^{+}_{j}}^2 \delta\left(s-M_{+}^2\right) =\, \rho^1_{j,H}(s) \, , \\
\frac{{\rm Im}\Pi^0_{j}(s)}{\pi}&=&M_{-}{\lambda^{-}_{j}}^2 \delta\left(s-M_{-}^2\right)-M_{+}{\lambda^{+}_{j}}^2 \delta\left(s-M_{+}^2\right)
=\rho^0_{j,H}(s) \, ,
\end{eqnarray}
where $j=\frac{1}{2}$, $\frac{3}{2}$, $\frac{5}{2}$, the subscript $H$ denotes  the hadron side,
then we introduce the  weight functions $\sqrt{s}\exp\left(-\frac{s}{T^2}\right)$ and $\exp\left(-\frac{s}{T^2}\right)$ to obtain the QCD sum rules at the phenomenological side (or the hadron side),
\begin{eqnarray}
\int_{4m_c^2}^{s_0}ds \left[\sqrt{s}\rho^1_{j,H}(s)+\rho^0_{j,H}(s)\right]\exp\left( -\frac{s}{T^2}\right)
&=&2M_{-}{\lambda^{-}_{j}}^2\exp\left( -\frac{M_{-}^2}{T^2}\right) \, ,\\
\int_{4m_c^2}^{s_0}ds \left[\sqrt{s}\rho^1_{j,H}(s)-\rho^0_{j,H}(s)\right]\exp\left( -\frac{s}{T^2}\right)
&=&2M_{+}{\lambda^{+}_{j}}^2\exp\left( -\frac{M_{+}^2}{T^2}\right) \, ,
\end{eqnarray}
where the $s_0$ are the continuum threshold parameters and the $T^2$ are the Borel parameters.
We separate the  contributions  of the negative parity pentaquark molecular states from that of the positive parity pentaquark molecular states unambiguously.

In the following,  we briefly outline  the operator product expansion for the correlation functions $\Pi(p)$, $\Pi_{\mu\nu}(p)$ and $\Pi_{\mu\nu\alpha\beta}(p)$ in perturbative QCD.  We contract the $u$, $d$ and $c$ quark fields in the correlation functions
$\Pi(p)$, $\Pi_{\mu\nu}(p)$ and $\Pi_{\mu\nu\alpha\beta}(p)$  with Wick theorem, and obtain the results:
\begin{eqnarray}\label{Pi12}
\Pi^{\bar{D}\Sigma_c}(p)&=&-i\, \varepsilon^{ijk} \varepsilon^{i^{\prime}j^{\prime}k^{\prime}}
 \int d^4x e^{ip\cdot x}\, \gamma^\alpha \gamma_5 C_{k^{\prime}k}(x)\gamma_5\gamma^\beta\nonumber\\
&&\Big\{   -Tr\left[i\gamma_5 C_{m^\prime m}(-x) i\gamma_5  U_{mm^\prime}(x)\right] \,Tr\left[\gamma_\alpha D_{jj^\prime}(x) \gamma_\beta C U^{T}_{ii^\prime}(x)C\right]  \nonumber\\
&& +  Tr \left[i\gamma_5 C_{m^\prime m}(-x) i\gamma_5  U_{mi^\prime}(x) \gamma_\beta C D^T_{jj^\prime}(x)C \gamma_\alpha  U_{im^\prime}(x)\right]    \Big\} \, ,
\end{eqnarray}

\begin{eqnarray}
\Pi^{\bar{D}\Sigma_c^*}_{\mu\nu}(p)&=&i\, \varepsilon^{ijk} \varepsilon^{i^{\prime}j^{\prime}k^{\prime}}
 \int d^4x e^{ip\cdot x} \, C_{k^{\prime}k}(x)\nonumber\\
&&\Big\{   -Tr\left[i\gamma_5 C_{m^\prime m}(-x) i\gamma_5  U_{mm^\prime}(x)\right] \,Tr\left[\gamma_\mu D_{jj^\prime}(x) \gamma_\nu C U^{T}_{ii^\prime}(x)C\right]  \nonumber\\
&& +  Tr \left[i\gamma_5 C_{m^\prime m}(-x) i\gamma_5  U_{mi^\prime}(x) \gamma_\nu C D^T_{jj^\prime}(x)C \gamma_\mu  U_{im^\prime}(x)\right]     \Big\} \, ,
\end{eqnarray}

\begin{eqnarray}
\Pi^{\bar{D}^*\Sigma_c}_{\mu\nu}(p)&=&-i\, \varepsilon^{ijk} \varepsilon^{i^{\prime}j^{\prime}k^{\prime}}
 \int d^4x e^{ip\cdot x}\, \gamma^\alpha \gamma_5 C_{k^{\prime}k}(x)\gamma_5\gamma^\beta\nonumber\\
&&\Big\{   -Tr\left[\gamma_\nu C_{m^\prime m}(-x) \gamma_\mu  U_{mm^\prime}(x)\right] \,Tr\left[\gamma_\alpha D_{jj^\prime}(x) \gamma_\beta C U^{T}_{ii^\prime}(x)C\right]  \nonumber\\
&& +  Tr \left[\gamma_\nu C_{m^\prime m}(-x) \gamma_\mu  U_{mi^\prime}(x) \gamma_\beta C D^T_{jj^\prime}(x)C \gamma_\alpha  U_{im^\prime}(x)\right]    \Big\} \, ,
\end{eqnarray}

\begin{eqnarray}\label{Pi52}
\Pi^{\bar{D}^*\Sigma_c^*}_{\mu\nu\alpha\beta}(p)&=&i\, \varepsilon^{ijk} \varepsilon^{i^{\prime}j^{\prime}k^{\prime}}
 \int d^4x e^{ip\cdot x}\,  C_{k^{\prime}k}(x)\nonumber\\
&&\Big\{  -Tr\left[\gamma_\alpha C_{m^\prime m}(-x) \gamma_\mu  U_{mm^\prime}(x)\right] \,Tr\left[\gamma_\nu D_{jj^\prime}(x) \gamma_\beta C U^{T}_{ii^\prime}(x)C\right]  \nonumber\\
&&-Tr\left[\gamma_\beta C_{m^\prime m}(-x) \gamma_\mu  U_{mm^\prime}(x)\right] \,Tr\left[\gamma_\nu D_{jj^\prime}(x) \gamma_\alpha C U^{T}_{ii^\prime}(x)C\right]  \nonumber\\
&&-Tr\left[\gamma_\alpha C_{m^\prime m}(-x) \gamma_\nu  U_{mm^\prime}(x)\right] \,Tr\left[\gamma_\mu D_{jj^\prime}(x) \gamma_\beta C U^{T}_{ii^\prime}(x)C\right]  \nonumber\\
&&-Tr\left[\gamma_\beta C_{m^\prime m}(-x) \gamma_\nu  U_{mm^\prime}(x)\right] \,Tr\left[\gamma_\mu D_{jj^\prime}(x) \gamma_\alpha C U^{T}_{ii^\prime}(x)C\right]  \nonumber\\
&& +  Tr \left[\gamma_\alpha C_{m^\prime m}(-x) \gamma_\mu  U_{mi^\prime}(x) \gamma_\beta C D^T_{jj^\prime}(x)C \gamma_\nu  U_{im^\prime}(x)\right]    \nonumber\\
&& +  Tr \left[\gamma_\beta C_{m^\prime m}(-x) \gamma_\mu  U_{mi^\prime}(x) \gamma_\alpha C D^T_{jj^\prime}(x)C \gamma_\nu  U_{im^\prime}(x)\right]    \nonumber\\
&& +  Tr \left[\gamma_\alpha C_{m^\prime m}(-x) \gamma_\nu  U_{mi^\prime}(x) \gamma_\beta C D^T_{jj^\prime}(x)C \gamma_\mu  U_{im^\prime}(x)\right]    \nonumber\\
&& +  Tr \left[\gamma_\beta C_{m^\prime m}(-x) \gamma_\nu  U_{mi^\prime}(x) \gamma_\alpha C D^T_{jj^\prime}(x)C \gamma_\mu  U_{im^\prime}(x)\right]    \Big\} \, ,
\end{eqnarray}
where
the $U_{ij}(x)$, $D_{ij}(x)$ and $C_{ij}(x)$ are the full $u$, $d$ and $c$ quark propagators respectively ($S_{ij}(x)=U_{ij}(x),\,D_{ij}(x)$),
 \begin{eqnarray}\label{Lpropagator}
S_{ij}(x)&=& \frac{i\delta_{ij}\!\not\!{x}}{ 2\pi^2x^4}-\frac{\delta_{ij}\langle
\bar{q}q\rangle}{12} -\frac{\delta_{ij}x^2\langle \bar{q}g_s\sigma Gq\rangle}{192} -\frac{ig_sG^{a}_{\alpha\beta}t^a_{ij}(\!\not\!{x}
\sigma^{\alpha\beta}+\sigma^{\alpha\beta} \!\not\!{x})}{32\pi^2x^2} \nonumber\\
&&  -\frac{1}{8}\langle\bar{q}_j\sigma^{\mu\nu}q_i \rangle \sigma_{\mu\nu}+\cdots \, ,
\end{eqnarray}
\begin{eqnarray}
C_{ij}(x)&=&\frac{i}{(2\pi)^4}\int d^4k e^{-ik \cdot x} \left\{
\frac{\delta_{ij}}{\!\not\!{k}-m_c}
-\frac{g_sG^n_{\alpha\beta}t^n_{ij}}{4}\frac{\sigma^{\alpha\beta}(\!\not\!{k}+m_c)+(\!\not\!{k}+m_c)
\sigma^{\alpha\beta}}{(k^2-m_c^2)^2}\right.\nonumber\\
&&\left. -\frac{g_s^2 (t^at^b)_{ij} G^a_{\alpha\beta}G^b_{\mu\nu}(f^{\alpha\beta\mu\nu}+f^{\alpha\mu\beta\nu}+f^{\alpha\mu\nu\beta}) }{4(k^2-m_c^2)^5}+\cdots\right\} \, ,\nonumber\\
f^{\alpha\beta\mu\nu}&=&(\!\not\!{k}+m_c)\gamma^\alpha(\!\not\!{k}+m_c)\gamma^\beta(\!\not\!{k}+m_c)\gamma^\mu(\!\not\!{k}+m_c)\gamma^\nu(\!\not\!{k}+m_c)\, ,
\end{eqnarray}
and  $t^n=\frac{\lambda^n}{2}$, the $\lambda^n$ is the Gell-Mann matrix   \cite{PRT85,Pascual-1984}, then compute  the integrals both in the coordinate and momentum spaces to obtain the correlation functions $\Pi(p)$, $\Pi_{\mu\nu}(p)$ and $\Pi_{\mu\nu\alpha\beta}(p)$ therefore the QCD spectral densities $\rho^1_{\frac{1}{2}/\frac{3}{2}/\frac{5}{2},QCD}(s)$ and $\rho^0_{\frac{1}{2}/\frac{3}{2}/\frac{5}{2},QCD}(s)$ through the dispersion  relation.
In Eq.\eqref{Lpropagator}, we retain the term $\langle\bar{q}_j\sigma_{\mu\nu}q_i \rangle$  comes from the Fierz re-arrangement of the $\langle q_i \bar{q}_j\rangle$ to  absorb the gluons  emitted from other  quark lines to extract the mixed condensate  $\langle\bar{q}g_s\sigma G q\rangle$.

From Eqs.\eqref{Pi12}-\eqref{Pi52}, we can see that there are two type contributions, one contains two Tr's, one contains one Tr.  The terms with two Tr's have both  factorizable contributions and non-factorizable contributions, while the terms with one  Tr have only non-factorizable contributions. At the leading order, the perturbative terms have only factorizable contributions. The  non-factorizable contributions play a important role in determining the pentaquark molecular states. If there are only factorizable contributions of the terms in the two Tr's, the intermediate scattering baryon-meson states will dominate the QCD sum rules. On the other hand, if we take into account  both the  factorizable contributions and non-factorizable contributions, the intermediate  baryon-meson loops only  contribute  a finite imaginary part to modify the dispersion relation at the hadron side,
\begin{eqnarray}
\Pi_{\frac{1}{2}}(p) &=&-\frac{\!\not\!{p} +M_{-}}{ p^2-M_{-}^2+i\sqrt{p^2}\Gamma_{-}(p^2)}{\lambda^{-}_{\frac{1}{2}}}^2 -\frac{\!\not\!{p} -M_{+}}{ p^2-M_{+}^2+i\sqrt{p^2}\Gamma_{+}(p^2)}{\lambda^{+}_{\frac{1}{2}}}^2 +\cdots \, , \nonumber\\
\Pi_{\frac{3}{2}}(p) &=&-\frac{\!\not\!{p} +M_{-}}{ p^2-M_{-}^2+i\sqrt{p^2}\Gamma_{-}(p^2)}{\lambda^{-}_{\frac{3}{2}}}^2 -\frac{\!\not\!{p} -M_{+}}{ p^2-M_{+}^2+i\sqrt{p^2}\Gamma_{+}(p^2)}{\lambda^{+}_{\frac{3}{2}}}^2 +\cdots \, , \nonumber\\
\Pi_{\frac{5}{2}}(p) &=&-\frac{\!\not\!{p} +M_{-}}{ p^2-M_{-}^2+i\sqrt{p^2}\Gamma_{-}(p^2)}{\lambda^{-}_{\frac{5}{2}}}^2 -\frac{\!\not\!{p} -M_{+}}{ p^2-M_{+}^2+i\sqrt{p^2}\Gamma_{+}(p^2)}{\lambda^{+}_{\frac{5}{2}}}^2 +\cdots \, .
 \end{eqnarray}
In calculations. we observe that   the zero width approximation will  not impair the  predictive  ability    significantly even for large widths \cite{Wang-Octet}, the scattering baryon-meson states can be neglected safely.

Furthermore, from Eqs.\eqref{Pi12}-\eqref{Pi52}, we can see that there are two heavy quark propagators and three light quark propagators in the correlation functions, if each heavy line emits a gluon and each light quark line contributes  a quark pair, we obtain a operator $GG\bar{u}u\bar{u}u\bar{d}d$, which is of dimension 13, we should take into account the vacuum condensates at least
up to dimension 13.
In this article, we carry out the operator product expansion to the vacuum condensates  up to dimension-$13$ and assume vacuum saturation for the
 higher dimensional vacuum condensates. We take the truncations $n\leq 13$ and $k\leq 1$ in a consistent way,
the operators of the orders $\mathcal{O}( \alpha_s^{k})$ with $k> 1$ are  discarded.  In previous QCD sum rules for the pentaquark molecular states, the operator product expansion was carried out up to the vacuum condensates of the dimension $8$ or $6$ \cite{HXChen-PRL,HXChen-EPJC,Azizi-PRD}, the vacuum condensates $\langle\bar{q} q\rangle^3$, $\langle\bar{q}g_s\sigma Gq\rangle^2$, $\langle\bar{q} q\rangle^2\langle\bar{q}g_s\sigma Gq\rangle $ and $\langle\bar{q} q\rangle \langle\bar{q}g_s\sigma Gq\rangle^2 $ were  discarded.
 The vacuum condensates  $\langle\bar{q} q\rangle^2\langle\bar{q}g_s\sigma Gq\rangle $ and $\langle\bar{q} q\rangle \langle\bar{q}g_s\sigma Gq\rangle^2 $, which come from the Feynman diagrams shown in Figs.1-2,   play an important role in determining the Borel windows, as  there appear terms of the orders $\mathcal{O}\left(\frac{1}{T^2}\right)$, $\mathcal{O}\left(\frac{1}{T^4}\right)$, $\mathcal{O}\left(\frac{1}{T^6}\right)$ in the QCD spectral densities, which  manifest themselves at small values of the Borel parameter $T^2$, we have to choose large values of the $T^2$ to warrant convergence of the operator product expansion and appearance of the Borel platforms. In the Borel windows, the  vacuum condensates $\langle\bar{q} q\rangle^2\langle\bar{q}g_s\sigma Gq\rangle $ and $\langle\bar{q} q\rangle \langle\bar{q}g_s\sigma Gq\rangle^2 $ play a less important role.
  Although the vacuum  condensates  $\langle \bar{q}q\rangle\langle \frac{\alpha_s}{\pi}GG\rangle$,
$\langle \bar{q}q\rangle^2\langle \frac{\alpha_s}{\pi}GG\rangle$ and $\langle \bar{q}q\rangle^3\langle \frac{\alpha_s}{\pi}GG\rangle$ are the vacuum expectations
of the operators of the order
$\mathcal{O}(\alpha_s)$, and they are neglected due to the small contributions of the  gluon condensates in the QCD sum rules for the multiquark states  \cite{Wang-tetraquark-QCDSR,Wang-molecule-QCDSR,Wang-EPJC-update}.

\begin{figure}
 \centering
 \includegraphics[totalheight=3.0cm,width=10cm]{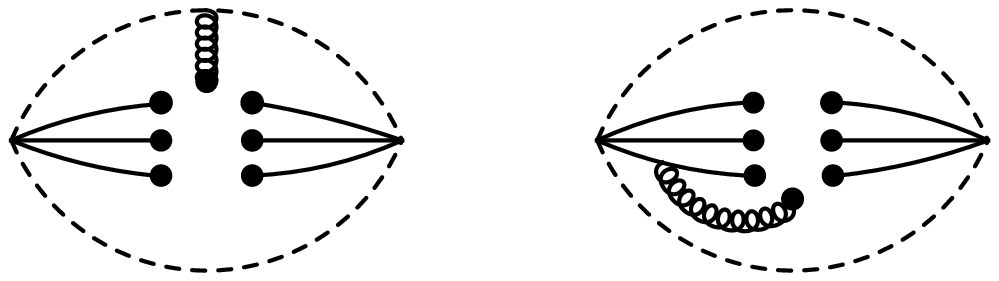}
    \caption{The diagrams contribute  to the  mixed condensate $\langle\bar{q} q\rangle^2\langle\bar{q}g_s\sigma Gq\rangle $ of dimension $11$. Other
diagrams obtained by interchanging of the heavy quark lines (dashed lines) or light quark lines (solid lines) are implied. }
\end{figure}

\begin{figure}
 \centering
 \includegraphics[totalheight=3.0cm,width=15cm]{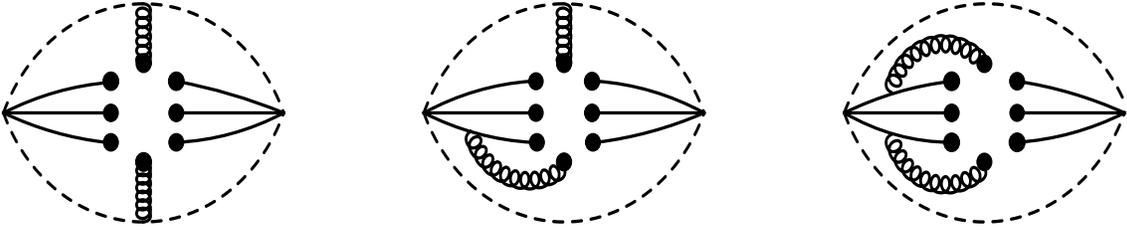}
    \caption{The diagrams contribute  to the  mixed condensate $\langle\bar{q} q\rangle \langle\bar{q}g_s\sigma Gq\rangle^2 $ of dimension $13$. Other
diagrams obtained by interchanging of the heavy quark lines (dashed lines) or light quark lines (solid lines) are implied. }
\end{figure}

 Once the analytical QCD spectral densities $\rho^1_{j,QCD}(s)$ and $\rho^0_{j,QCD}(s)$ are obtained,  we can take the
quark-hadron duality below the continuum thresholds  $s_0$ and introduce the weight functions $\sqrt{s}\exp\left(-\frac{s}{T^2}\right)$ and $\exp\left(-\frac{s}{T^2}\right)$ to obtain  the following QCD sum rules:
\begin{eqnarray}\label{QCDN}
2M_{-}{\lambda^{-}_{j}}^2\exp\left( -\frac{M_{-}^2}{T^2}\right)
&=& \int_{4m_c^2}^{s_0}ds \left[\sqrt{s}\rho^1_{j,QCD}(s)+\rho^0_{j,QCD}(s)\right]\exp\left( -\frac{s}{T^2}\right)\, ,
\end{eqnarray}
\begin{eqnarray}\label{QCDP}
2M_{+}{\lambda^{+}_{j}}^2\exp\left( -\frac{M_{+}^2}{T^2}\right) &=&\int_{4m_c^2}^{s_0}ds \left[\sqrt{s}\rho^1_{j,QCD}(s)-\rho^0_{j,QCD}(s)\right]\exp\left( -\frac{s}{T^2}\right)\, ,
\end{eqnarray}
where $j=\frac{1}{2}$, $\frac{3}{2}$, $\frac{5}{2}$, $\rho^{0}_{j,QCD}(s)=m_c\,\overline{\rho}^{0}_{j,QCD}(s)$,
\begin{eqnarray}
\rho^{1}_{j,QCD}(s)&=&\rho_0^{1}(s)+\rho_3^{1}(s)+\rho_4^{1}(s)+\rho_5^{1}(s)+\rho_6^{1}(s)+\rho_8^{1}(s)+\rho_9^{1}(s)+\rho_{10}^{1}(s) \nonumber\\
&&+\rho_{11}^{1}(s)+\rho_{13}^{1}(s)\, , \nonumber\\
\overline{\rho}^{0}_{j,QCD}(s)&=&\rho_0^{0}(s)+\rho_3^{0}(s)+\rho_4^{0}(s)+\rho_5^{0}(s)+\rho_6^{0}(s)+\rho_8^{0}(s)+\rho_9^{0}(s)+\rho_{10}^{0}(s) \nonumber\\
&&+\rho_{11}^{0}(s)+\rho_{13}^{0}(s)\, ,
\end{eqnarray}
the explicit expressions of the  QCD spectral densities $\rho_i^{1/0}(s)$  with $i=0$, $3$, $4$, $5$, $6$, $8$, $9$, $10$, $11$ and $13$ are given  in the appendix.

We differentiate   Eqs.\eqref{QCDN}-\eqref{QCDP} with respect to  $\tau=\frac{1}{T^2}$, then eliminate the
 pole residues $\lambda^{\pm}_{\frac{1}{2}/\frac{3}{2}/\frac{5}{2}}$ and obtain the QCD sum rules for
 the masses of the pentaquark molecular  states,
 \begin{eqnarray}\label{QCDSR-M}
 M^2_{-} &=& \frac{-\frac{d}{d \tau}\int_{4m_c^2}^{s_0}ds \,\left[\sqrt{s}\,\rho^1_{QCD}(s)+\,\rho^0_{QCD}(s)\right]\exp\left(- \tau s\right)}{\int_{4m_c^2}^{s_0}ds \left[\sqrt{s}\,\rho_{QCD}^1(s)+\,\rho^0_{QCD}(s)\right]\exp\left( -\tau s\right)}\, ,
 \end{eqnarray}

 \begin{eqnarray}
 M^2_{+} &=& \frac{-\frac{d}{d \tau}\int_{4m_c^2}^{s_0}ds \,\left[\sqrt{s}\,\rho^1_{QCD}(s)-\,\rho^0_{QCD}(s)\right]\exp\left(- \tau s\right)}{\int_{4m_c^2}^{s_0}ds \left[\sqrt{s}\,\rho_{QCD}^1(s)-\,\rho^0_{QCD}(s)\right]\exp\left( -\tau s\right)}\, ,
\end{eqnarray}
where $\rho_{QCD}^1(s)=\rho_{j,QCD}^1(s)$ and $\rho^0_{QCD}(s)=\rho^0_{j,QCD}(s)$.
In numerical calculations, we observe that the masses $M_{+}$ of the $\bar{D}\Sigma_c$, $\bar{D}\Sigma_c^*$, $\bar{D}^{*}\Sigma_c$ and
$  \bar{D}^{*}\Sigma_c^*$  pentaquark molecular states with the $J^{P}={\frac{1}{2}}^+$, ${\frac{3}{2}}^+$, ${\frac{3}{2}}^+$ and ${\frac{5}{2}}^+$ are about $4.96\,\rm{GeV}$, $4.60\,\rm{GeV}$, $5.31\,\rm{GeV}$ and $4.71\,\rm{GeV}$ respectively, which are much larger than the corresponding  $\bar{D}+\Sigma_c$,  $\bar{D}+\Sigma_c^*$, $\bar{D}^{*}+\Sigma_c$ and
$  \bar{D}^{*}+\Sigma_c^*$  threshold holds
$4.318\,\rm{GeV}$, $4.382\,\rm{GeV}$, $4.460\,\rm{GeV}$ and $4.524\,\rm{GeV}$ respectively.  In this article, we would not pay attention to the pentaquark molecular states with positive parity, as they may be resonance states or virtual states.

\section{Numerical results and discussions}
We take  the standard values of the vacuum condensates $\langle
\bar{q}q \rangle=-(0.24\pm 0.01\, \rm{GeV})^3$,   $\langle
\bar{q}g_s\sigma G q \rangle=m_0^2\langle \bar{q}q \rangle$,
$m_0^2=(0.8 \pm 0.1)\,\rm{GeV}^2$,   $\langle \frac{\alpha_s
GG}{\pi}\rangle=(0.33\,\rm{GeV})^4 $    at the energy scale  $\mu=1\, \rm{GeV}$
\cite{SVZ79,PRT85,ColangeloReview}, and choose the $\overline{MS}$ mass  $m_{c}(m_c)=(1.28\pm0.03)\,\rm{GeV}$
 from the Particle Data Group \cite{PDG}.
Furthermore, we take into account the energy-scale dependence of  the input parameters,
\begin{eqnarray}
\langle\bar{q}q \rangle(\mu)&=&\langle\bar{q}q \rangle({\rm 1 GeV})\left[\frac{\alpha_{s}({\rm 1 GeV})}{\alpha_{s}(\mu)}\right]^{\frac{12}{25}}\, , \nonumber\\
 \langle\bar{q}g_s \sigma Gq \rangle(\mu)&=&\langle\bar{q}g_s \sigma Gq \rangle({\rm 1 GeV})\left[\frac{\alpha_{s}({\rm 1 GeV})}{\alpha_{s}(\mu)}\right]^{\frac{2}{25}}\, , \nonumber\\
m_c(\mu)&=&m_c(m_c)\left[\frac{\alpha_{s}(\mu)}{\alpha_{s}(m_c)}\right]^{\frac{12}{25}} \, ,\nonumber\\
\alpha_s(\mu)&=&\frac{1}{b_0t}\left[1-\frac{b_1}{b_0^2}\frac{\log t}{t} +\frac{b_1^2(\log^2{t}-\log{t}-1)+b_0b_2}{b_0^4t^2}\right]\, ,
\end{eqnarray}
   where $t=\log \frac{\mu^2}{\Lambda^2}$, $b_0=\frac{33-2n_f}{12\pi}$, $b_1=\frac{153-19n_f}{24\pi^2}$,
   $b_2=\frac{2857-\frac{5033}{9}n_f+\frac{325}{27}n_f^2}{128\pi^3}$,
   $\Lambda=210\,\rm{MeV}$, $292\,\rm{MeV}$  and  $332\,\rm{MeV}$ for the flavors
   $n_f=5$, $4$ and $3$, respectively  \cite{PDG,Narison-mix,Narison-Book}.

In this article, we study the pentaquark molecular states in three cases,\\
{\bf (I)}. We evolve  the input parameters to the  energy scale  $ \mu =\sqrt{M_{P}^2-(2{\mathbb{M}}_c)^2}$ to extract the masses $M_{P}$ with the truncation of the operator product expansion $D=13$;\\
{\bf (II)}. We evolve the input parameters except for $m_c(m_c)$ to the  energy scale  $ \mu =1\,\rm{GeV}$ to extract the masses $M_{P}$ with the truncation of the operator product expansion $D=10$; \\
{\bf (III)}. We evolve the input parameters except for $m_c(m_c)$ to the  energy scale  $ \mu =1\,\rm{GeV}$ to extract the masses $M_{P}$ with the truncation of the operator product expansion $D=13$. \\

Now we take a short digression to discuss the energy scale formula, $ \mu =\sqrt{M_{P}^2-(2{\mathbb{M}}_Q)^2}$. In the heavy quark limit,
the $Q$-quark serves as a static well potential and  can combine with a  light quark $q$  to form a heavy diquark  in  color antitriplet, or combine with a light diquark in  color antitriplet to form a baryon state in color singlet.
 The $\overline{Q}$-quark serves  as another static well potential, and can combine with a light diquark $\varepsilon^{ijk}q^i\,q^{\prime j}$ to form a heavy triquark in color triplet, or combine with a light quark $q$  to form a heavy meson  in  color singlet,
\begin{eqnarray}
q^j+Q^k &\to & \varepsilon^{ijk}\, q^j\,Q^k\, , \nonumber\\
\varepsilon^{ijk}q^i\,q^{\prime j}+Q^k &\to & \varepsilon^{ijk}\, q^i\,q^{\prime j}Q^k\, , \nonumber\\
\varepsilon^{ijl}q^i\,q^{\prime j} +\overline{Q}^k&\to & \varepsilon^{lkm}\varepsilon^{ijl}\,q^i \,q^{\prime j}\, \overline{Q}^k\, ,\nonumber\\
q^j+\overline{Q}^k &\to & \delta^{jk}\,  \overline{Q}\,q\, ,
\end{eqnarray}
where the $i$, $j$, $k$, $l$, $m$ are color indexes.
 Then
\begin{eqnarray}
 \varepsilon^{ijk}\, q^j\,Q^k+\varepsilon^{imn} \bar{q}^{\prime m}\,\overline{Q}^n &\to &  {\rm compact \,\,\, tetraquark \,\,\, states}\, , \nonumber\\
  \varepsilon^{lkm}\varepsilon^{ijl}\,q^i \,q^{\prime j}\, \overline{Q}^k+\varepsilon^{mnb} q^{\prime \prime n}\,Q^b &\to &
  {\rm compact \,\,\, pentaquark \,\,\, states}\, , \nonumber\\
  \overline{Q}q+ \overline{q}^\prime Q &\to & {\rm tetraquark   \,\,\, molecular \,\,\, states}\, , \nonumber\\
  \overline{Q}q+ \varepsilon^{ijk}\, q^{\prime i}\,q^{\prime \prime j}Q^k&\to & {\rm pentaquark   \,\,\, molecular \,\,\, states}\, .
\end{eqnarray}
The two heavy quarks $Q$ and $\bar{Q}$ stabilize the four-quark systems $q\bar{q}^{\prime}Q\bar{Q}$ or the five quark systems $qq^\prime q^{\prime\prime} Q \overline{Q}$ , just as in the case of the $(\mu^-e^+)(\mu^+ e^-)$ molecule in QED \cite{Brodsky-2014}.
The tetraquark (molecular) states $q\bar{q}^\prime Q \overline{Q}$ ($X,\,Y,\,Z$) and pentaquark (molecular) states $qq^\prime q^{\prime\prime} Q \overline{Q}$ ($P$)
are characterized by the effective heavy quark masses ${\mathbb{M}}_Q$  (or constituent quark masses not as robust) and the virtuality
$V=\sqrt{M^2_{X/Y/Z/P}-(2{\mathbb{M}}_Q)^2}$  (or bound energy not as robust). The QCD sum rules have three typical energy scales $\mu^2$, $T^2$, $V^2$. It is natural to take the energy  scales of the QCD spectral densities to be $\mu=V$.

The effective $Q$-quark  masses ${\mathbb{M}}_Q$   embody  the net effects of the complex dynamics, appear as parameters and their values are fitted by the QCD sum rules. The   ${\mathbb{M}}_Q$ have uncertainties, the optimal values in the diquark-antidiquark (diquark-diquark-antiquark) systems are not necessary the optimal  values in the  meson-meson (meson-baryon) systems. In the  multiquark states consist of color singlet constituents, irrespective of the meson-meson type or meson-baryon type multiquark states, or in the  multiquark states consist of color (anti)triplet  constituents, irrespective of the diquark-antidiquark type or diquark-diquark-antiquark type multiquark states, the effective $Q$-quark  masses   ${\mathbb{M}}_Q$ should have   universal values.

We fit  the  effective $Q$-quark masses ${\mathbb{M}}_{Q}$ to reproduce the experimental
masses of the $Z_c(3900)$ and $Z_b(10610)$ in the  scenario of  tetraquark  states or molecular states \cite{Wang-tetraquark-QCDSR,Wang-molecule-QCDSR}, as there are controversies concerning the
tetraquark and molecule assignments,
 then use the  energy scale formula $\mu=\sqrt{M^2_{X/Y/Z/P}-(2{\mathbb{M}}_Q)^2}$  to study the
 hidden-charm (hidden-bottom) tetraquark states and  hidden-charm (hidden-bottom) pentaquark states or hidden-charm (hidden-bottom) tetraquark molecular states and  hidden-charm (hidden-bottom) pentaquark molecular states.

   In Ref.\cite{Wang-molecule-QCDSR}, we obtain the optimal value ${\mathbb{M}}_c= 1.84\,\rm{GeV}$ for the tetraquark molecular states. Later, we re-checked the numerical calculations and corrected a small
error involving the mixed condensates and obtained the updated value ${\mathbb{M}}_c= 1.85\,\rm{GeV}$ \cite{Wang-CPC-4390}.

 In the case ({\bf I}), we choose the  Borel parameters $T^2$ and continuum threshold
parameters $s_0$  to satisfy   the  following four criteria:

$\bf C1.$ Pole dominance at the phenomenological side;

$\bf C2.$ Convergence of the operator product expansion;

$\bf C3.$ Appearance of the Borel platforms;

$\bf C4.$ Satisfying the energy scale formula,\\
by try and error.

In the cases ({\bf II}) and ({\bf III}), we choose the  Borel parameters $T^2$ and continuum threshold
parameters $s_0$  to satisfy   the  three criteria,  $\bf C1$, $\bf C2$ and $\bf C3$.

Now we write down the contributions of the different terms in the operator product expansion,
\begin{eqnarray}\label{Dn}
D(n)&=& \frac{  \int_{4m_c^2}^{s_0} ds\,\rho_{n}(s)\,\exp\left(-\frac{s}{T^2}\right)}{\int_{4m_c^2}^{s_0} ds \,\rho(s)\,\exp\left(-\frac{s}{T^2}\right)}\, ,
\end{eqnarray}
where the $\rho_{n}(s)$ are the QCD spectral densities for the vacuum condensates of dimension $n$, and the total spectral densities
$\rho(s)=\sqrt{s}\rho^1_{QCD}(s)+ \rho^0_{QCD}(s)$.
 There is another definition for the $D(n)$,
\begin{eqnarray}
D(n)&=& \frac{  \int_{4m_c^2}^{\infty} ds\,\rho_{n}(s)\,\exp\left(-\frac{s}{T^2}\right)}{\int_{4m_c^2}^{\infty} ds \,\rho(s)\,\exp\left(-\frac{s}{T^2}\right)}\, ,
\end{eqnarray}
which enhance the contributions of the vacuum condensates of low dimension and lead to smaller Borel parameters. Such a definition  only warrants the operator product expansion is convergent if all the hadron states are taken into account on the phenomenological side. In this article, we prefer the definition shown in Eq.\eqref{Dn} as we only take into account the ground state contributions.

 The contributions of the perturbative terms $D(0)$ are usually small for the multiquark  states,
    we approximate the continuum contributions as $\rho(s)\Theta(s-s_0)$, and define
    the pole contributions ($\rm{PC}$) or ground state contributions as
   \begin{eqnarray}
{\rm PC}&=& \frac{  \int_{4m_c^2}^{s_0} ds\,\left[\sqrt{s}\rho_{QCD}^1(s)+  \rho_{QCD}^0(s)\right]\,
\exp\left(-\frac{s}{T^2}\right)}{\int_{4m_c^2}^{\infty} ds \,\left[\sqrt{s}\rho_{QCD}^1(s)+ \rho_{QCD}^0(s)\right]\,\exp\left(-\frac{s}{T^2}\right)}\, .
\end{eqnarray}

In Ref.\cite{WangHbaryon},  we separate the contributions of the positive parity and negative parity baryon states explicitly, and study the    heavy, doubly-heavy and triply-heavy baryon states  with the QCD sum rules in a systematic way. In calculations, we observe that the continuum threshold parameters $\sqrt{s_0}=M_{\rm{gr}}+ (0.6-0.8)\,\rm{GeV}$  can reproduce  the masses of the observed heavy and doubly-heavy baryon states \cite{PDG},  where the subscript $\rm{gr}$ denotes the ground  baryon states.
The pentaquark states and pentaquark molecular states are another type baryon states considering  the fractional spins  $1\over 2$, $3\over 2$, $5\over 2$,
 we can take the continuum threshold parameters as $\sqrt{s_0}< M_{P}+0.8\,\rm{GeV}$.

The resulting Borel parameters or Borel windows $T^2$, continuum threshold parameters $s_0$, optimal  energy scales of the QCD spectral densities and pole contributions of the ground state pentaquark molecular states are shown  explicitly in Table 3. From the table, we can see that the pole dominance  or the $\bf C1$ is satisfied in the cases ({\bf I}) and ({\bf II}), while in the case ({\bf III}) the pole contributions are very small, less than $25\%$.

In the QCD sum rules for the multiquark states, we usually choose the same pole contributions as $(40-60)\%$ \cite{Wang1508-EPJC,WangHuang1508,Wang-tetraquark-QCDSR,Wang-molecule-QCDSR,Wang-EPJC-update}, which satisfy the pole dominance,  the resulting Borel windows are small,  $T^2_{max}-T^2_{min}\approx 0.4\,\rm{GeV}^2$. If we enlarge or narrow the pole contributions, the Borel windows are changed, the corresponding predictions are also changed. In Refs.\cite{Wang1508-EPJC,WangHuang1508,Wang-tetraquark-QCDSR,Wang-molecule-QCDSR,Wang-EPJC-update}, we study the tetraquark states, tetraquark molecular states and pentaquark states with the QCD sum rules in a consistent way by choosing the pole contributions $(40-60)\%$, and obtain satisfactory results  in assigning the exotic states. In the present work, we choose the pole contributions $(40-60)\%$ in the case ({\bf I}) and expect to obtain reliable predictions.

\begin{figure}
 \centering
\includegraphics[totalheight=5cm,width=7cm]{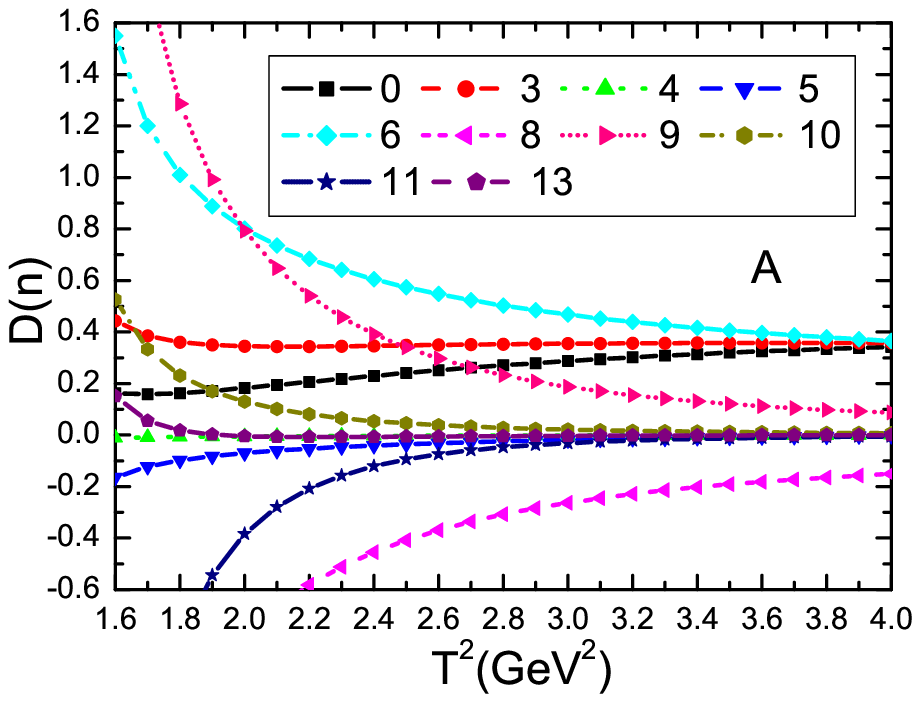}
\includegraphics[totalheight=5cm,width=7cm]{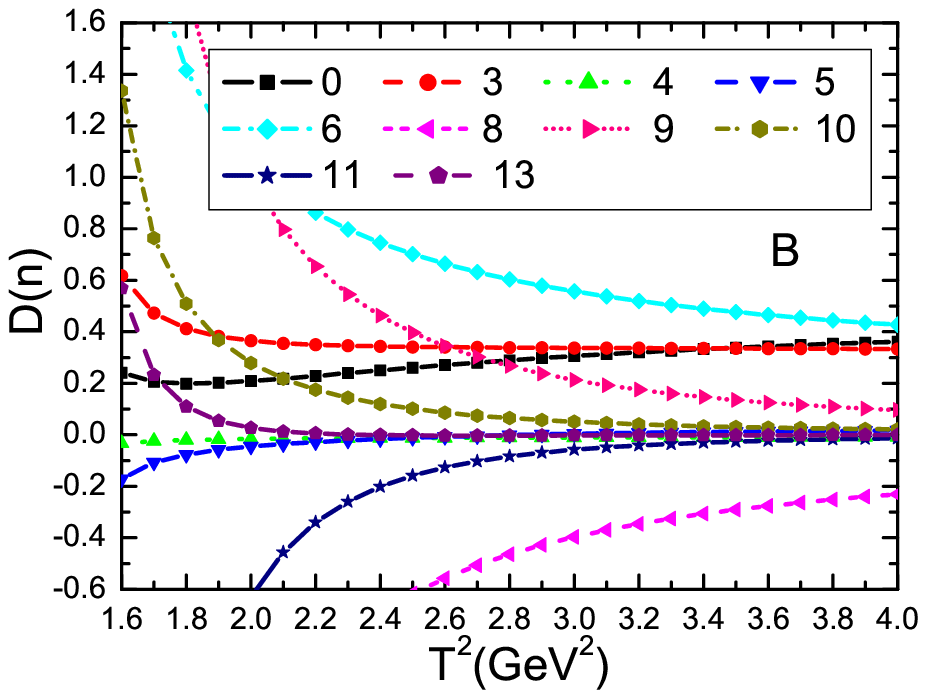}
\includegraphics[totalheight=5cm,width=7cm]{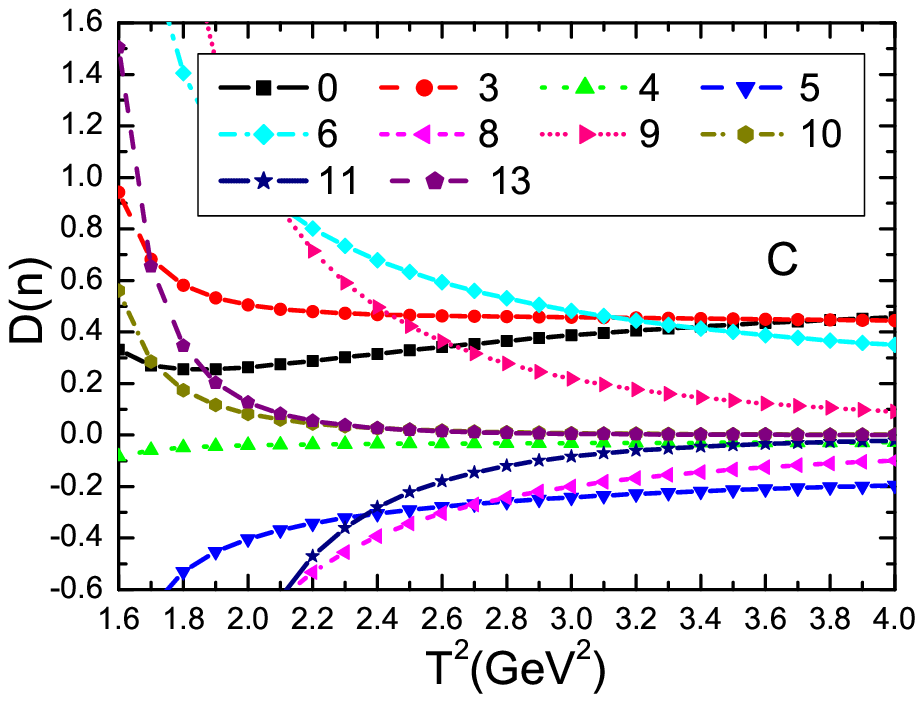}
\includegraphics[totalheight=5cm,width=7cm]{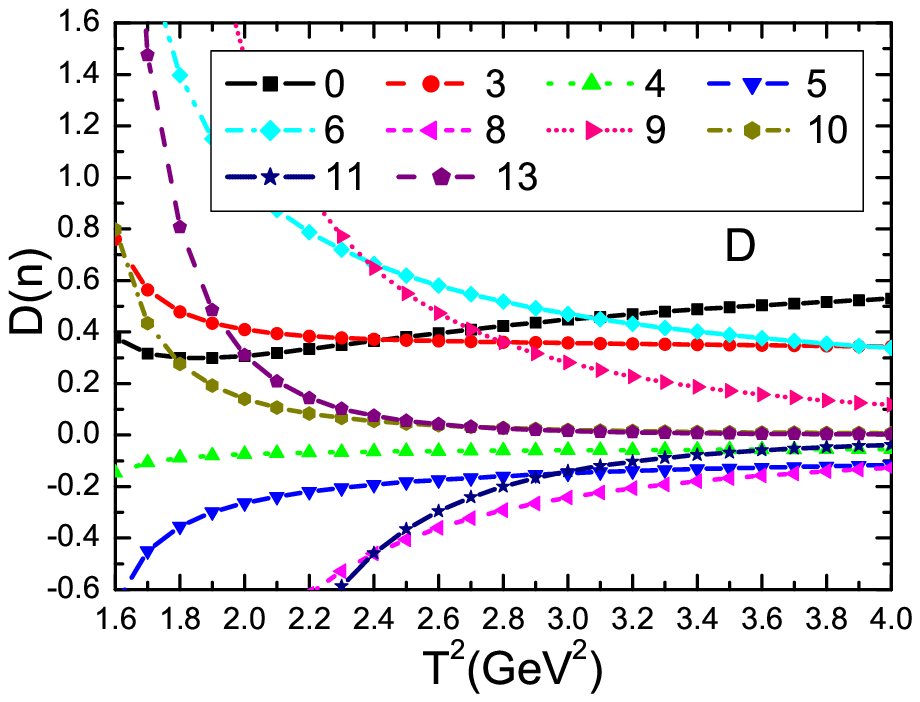}
 \caption{ The contributions  of the vacuum condensates of dimension $n=0$, $3$, $4$, $\cdots$ with variations of the Borel parameter $T^2$ for central values of other input parameters in the case ({\bf I}),  where  the $A$, $B$, $C$ and $D$ denote the pentaquark molecular  states $\bar{D}\Sigma_c$, $\bar{D}\Sigma_c^*$, $\bar{D}^{*}\Sigma_c$ and $ \bar{D}^{*}\Sigma_c^*$, respectively.   }\label{FrDSigma}
\end{figure}

In Figs.\ref{FrDSigma}-\ref{FrDSigmaMix},  we plot the contributions of the vacuum condensates of dimension $n$ with variations of the Borel parameters $T^2$ for the central values of  other input parameters shown in Table 3  in the cases ({\bf I}), ({\bf II}) and ({\bf III}), respectively. From the figures, we can see that the contributions $D(n)$ change quickly with variations of the Borel parameters  at the regions $T^2\leq3.0\,\rm{GeV}^2$, $2.6\,\rm{GeV}^2$ and $3.3\,\rm{GeV}^2$ in the cases ({\bf I}), ({\bf II}) and ({\bf III}), respectively, which cannot lead to stable QCD sum rules, and the operator product expansion is not convergent,  we should choose (much) larger Borel parameters $T^2$. In Fig.\ref{FR-Dimension}, we plot the absolute  contributions of the vacuum condensates of dimension $n$ for the central values of the input parameters shown in Table 3 in the cases ({\bf I}), ({\bf II}) and ({\bf III}), respectively. From the figure, we can see that the contributions of the perturbative terms $D(0)$ are not  the dominant contributions, the contributions of the vacuum condensates of dimensions $6$ and $8$ are very large. If we take the contributions of the  vacuum condensates of dimension $6$ as milestones, the contributions of the vacuum condensates $|D(n)|$ decrease quickly with increase of the dimensions $n$,
    the operator product expansion is well convergent. The convergent behaviors have relation  $({\bf I})>({\bf II})>({\bf III})$.

In calculations, we observe that in the case ({\bf II}), we take into account the vacuum condensates up to dimension 10, not to dimension 13, there are no terms associated with  $\frac{1}{T^2}$, $\frac{1}{T^4}$, $\frac{1}{T^6}$, which warrant those terms manifest themselves at low $T^2$ and appearance of the Borel platforms,
the predicted masses increase  monotonously  with  increase of the Borel parameters. We choose small Borel windows $T^2_{max}-T^2_{min}=0.4\,\rm{GeV}^2$,   and obtain the Borel platforms  by requiring the uncertainties $\frac{\delta M_{P}}{M_{P}} $ induced by the Borel parameters are about $1\%$.

We take into account  all uncertainties  of the input   parameters,
and obtain  the masses and pole residues of
 the $J^P={1\over 2}^{-}$, ${\frac{3}{2}}^-$  and ${\frac{5}{2}}^-$  hidden-charm pentaquark molecular states, which are shown explicitly in Table 4 and Figs.7-12. From Table 4, we can see that the  criterion  $\bf C4$ is  satisfied in the case ({\bf I}).

 In Figs.\ref{massDSigma}-\ref{residueDSigmaMix}, we plot the masses and pole residues at much larger ranges of the Borel parameters than the Borel windows.
 From Figs.\ref{massDSigma}-\ref{residueDSigma}, we can see that the predicted masses and pole residues in the case ({\bf I}) decrease monotonously and quickly  with
 increase of the Borel parameters at the region $T^2\leq 2.0\,\rm{GeV}^2$, then reach small platforms and increase slowly with increase of the Borel parameters.
 From Figs.\ref{massDSigmaNiel}-\ref{residueDSigmaNiel}, we can see that the predicted masses and pole residues in the case ({\bf II}) increase monotonously and quickly  with increase of the Borel parameters at the region $T^2<2.6\,\rm{GeV}^2$, then  increase slowly with increase of the Borel parameters.
 From Figs.\ref{massDSigmaMix}-\ref{residueDSigmaMix}, we can see that the predicted masses and pole residues in the case ({\bf III}) decrease monotonously and quickly  with
  increase of the Borel parameters at the region $T^2<3.0\,\rm{GeV}^2$, then  decrease very slowly with increase of the Borel parameters. In all the three cases, we can define Borel platforms by requiring the uncertainties $\frac{\delta M_{P}}{M_{P}} $ induced by the Borel parameters are about $1\%$, the criterion  $\bf C3$ can be satisfied.  The flatness of the platforms   have relation  $({\bf III})>({\bf I})>({\bf II})$.

 In summary, in the case ({\bf I}), the criteria $\bf C1$, $\bf C2$, $\bf C3$, $\bf C4$ can be satisfied; in the case ({\bf II}), the criteria $\bf C1$, $\bf C2$, $\bf C3$  can be satisfied; in the case ({\bf III}), the criteria  $\bf C2$, $\bf C3$ can be satisfied. While the convergent behaviors have relation  $({\bf I})>({\bf II})>({\bf III})$ and the flatness   of the platforms  have relation  $({\bf III})>({\bf I})>({\bf II})$.

 In the case ({\bf III}), if we choose small Borel parameters, the pole contributions can be enhanced, however, the convergence of the operator product expansion breaks down. On the other hand, if we choose larger continuum threshold parameters to enhance the pole contributions, we can obtain much  larger masses than the total masses of the two constituents, which correspond to virtual states or resonances, not meson-baryon bound states. The masses extracted from the continuum state dominated QCD sum rules are not robust, the case ({\bf III}) are not preferred.

Compared to the QCD sum rules in the case  ({\bf II}), the QCD sum rules in the case ({\bf I}) have better convergent behaviors in the operator product expansion and more flat Borel platforms. We do not prefer  the case ({\bf II}) as they lead to  two energy scales, $\mu=m_c$ and $\mu=1\,\rm{GeV}$,  in the QCD spectral densities, just like in the case of the $ssq\bar{q}c$ pentaquark states \cite{WangZhangPcssss}.

In this article, we prefer the QCD sum rules in the case ({\bf I}), which  support assigning the $P_c(4312)$ to be the $\bar{D}\Sigma_c$ pentaquark molecular state with $J^P={\frac{1}{2}}^-$, assigning the $P_c(4380)$ to be the $\bar{D}\Sigma_c^*$ pentaquark molecular state with $J^P={\frac{3}{2}}^-$,  assigning the $P_c(4440/4457)$ to be the $\bar{D}^{*}\Sigma_c$ pentaquark molecular state with $J^P={\frac{3}{2}}^-$ or the $\bar{D}^{*}\Sigma_c^*$ pentaquark molecular state with $J^P={\frac{5}{2}}^-$, see Table 4. As the mass alone cannot identify a hadron,  more experimental data are still needed to determine the $P_c(4312)$, $P_c(4380)$, $P_c(4440)$ and $P_c(4457)$ unambiguously. In other words, the QCD sum rules indicate that there maybe exist the $\bar{D}\Sigma_c$, $\bar{D}\Sigma_c^*$, $\bar{D}^{*}\Sigma_c$  and   $\bar{D}^{*}\Sigma_c^*$  pentaquark molecular states with the $J^P={\frac{1}{2}}^-$, ${\frac{3}{2}}^-$, ${\frac{3}{2}}^-$ and ${\frac{5}{2}}^-$, respectively, which lie in the corresponding  $\bar{D}\Sigma_c$, $\bar{D}\Sigma_c^*$, $\bar{D}^{*}\Sigma_c$  and   $\bar{D}^{*}\Sigma_c^*$ thresholds, respectively, see Table 4.
We have to study the two-body strong decays of the pentaquark molecular states $\bar{D}\Sigma_c$, $\bar{D}\Sigma_c^*$, $\bar{D}^{*}\Sigma_c$,   $\bar{D}^{*}\Sigma_c^* \to J/\psi p$ with the three-point QCD sum rules to assign the $P_c(4312)$, $P_c(4380)$, $P_c(4440)$ and $P_c(4457)$   in a more robust way, as we need more parameters beyond the masses to assign the $P_c(4312)$, $P_c(4380)$, $P_c(4440)$ and $P_c(4457)$  unambiguously. However, it is a difficult work to deal with the tensor (or spinor) structures in the three-point QCD sum rules for the hadronic coupling constants involving  the pentaquark molecular states with the spin $\geq \frac{3}{2}$. It is our next work.

In Fig.\ref{massDSigmaDimension}, we plot the masses of the pentaquark molecular states  with variations of the Borel parameter $T^2$ for the central values of other input parameters in Table 3  in the case ({\bf I}) with truncations of the operator product expansion $D=8$, $9$, $10$, $11$ and $13$, respectively. From the figure, we can see that the predicted
masses change significantly outside of the Borel windows,   the higher dimensional vacuum condensates play an important role in determining the Borel windows;
the regions between the two perpendicular lines are the Borel windows. Even in the Borel windows, the predicted masses change considerably  with the truncations of the operator product expansion, we should truncate the operator product expansion in a consistent way.

\begin{table}
\begin{center}
\begin{tabular}{|c|c|c|c|c|c|c|c|}\hline\hline
                                     &$J^P$                &$D$      &$\mu(\rm GeV)$    &$T^2 (\rm{GeV}^2)$  &$s_0 (\rm{GeV}^2)$   &pole     \\ \hline
$\bar{D}^0\,\Sigma_c^{+}(2455)$      &${\frac{1}{2}}^-$    &13       &2.2               &$3.1-3.5$           &$25.0\pm1.0$         &$(41-62)\%$  \\
                                     &                     &10       &1.0               &$2.7-3.1$           &$24.5\pm1.0$         &$(38-63)\%$     \\
                                     &                     &13       &1.0               &$3.4-4.2$           &$21.5\pm1.0$         &$(7-24)\%$    \\ \hline

$\bar{D}^0\,\Sigma_c^{*+}(2520)$     &${\frac{3}{2}}^-$    &13       &2.4               &$3.3-3.7$           &$25.5\pm1.0$         &$(39-59)\%$  \\
                                     &                     &10       &1.0               &$2.8-3.2$           &$25.5\pm1.0$         &$(40-64)\%$     \\
                                     &                     &13       &1.0               &$3.5-4.3$           &$22.0\pm1.0$         &$(7-23)\%$    \\ \hline

$\bar{D}^{*0}\,\Sigma_c^+(2455)$     &${\frac{3}{2}}^-$    &13       &2.5               &$3.3-3.7$           &$26.5\pm1.0$         &$(40-60)\%$  \\
                                     &                     &10       &1.0               &$2.8-3.2$           &$26.5\pm1.0$         &$(40-65)\%$     \\
                                     &                     &13       &1.0               &$3.7-4.6$           &$23.0\pm1.0$         &$(5-19)\%$    \\  \hline

$\bar{D}^{*0}\,\Sigma_c^{*+}(2520)$  &${\frac{5}{2}}^-$    &13       &2.6               &$3.4-3.8$           &$27.0\pm1.0$         &$(39-59)\%$  \\
                                     &                     &10       &1.0               &$3.0-3.4$           &$27.5\pm1.0$         &$(39-62)\%$     \\
                                     &                     &13       &1.0               &$3.5-4.4$           &$23.5\pm1.0$         &$(7-25)\%$    \\  \hline  \hline
\end{tabular}
\end{center}
\caption{ The truncations of the operator product expansion $D$, optimal energy scales $\mu$, Borel parameters $T^2$, continuum threshold parameters $s_0$ and
 pole contributions (pole)    for the hidden-charm   pentaquark molecular states, the energy scale $\mu=1\,\rm{GeV}$ denotes the input parameters except for the $m_c(m_c)$ are taken at $1\,\rm{GeV}$.}
\end{table}

\begin{table}
\begin{center}
\begin{tabular}{|c|c|c|c|c|c|c|c|}\hline\hline
                                  &$J^P$                &$D$      &$\mu(\rm GeV)$ &$M (\rm{GeV})$          &$\lambda (10^{-3}\rm{GeV}^6)$ & Thresholds (MeV)     \\  \hline
$\bar{D}^0\,\Sigma_c^+(2455)$     &${\frac{1}{2}}^-$    &13       &2.2            &$4.32^{+0.11}_{-0.11}$  &$1.95^{+0.37}_{-0.33}  $      &4318       \\
                                  &                     &10       &1.0            &$4.30^{+0.14}_{-0.17}$  &$1.00^{+0.27}_{-0.24} $       &            \\
                                  &                     &13       &1.0            &$4.30^{+0.09}_{-0.10}$  &$0.77^{+0.19}_{-0.15} $       &           \\ \hline

$\bar{D}^0\,\Sigma_c^+(2520)$     &${\frac{3}{2}}^-$    &13       &2.4            &$4.39^{+0.10}_{-0.11}$  &$1.23^{+0.21}_{-0.20}  $      &4382       \\
                                  &                     &10       &1.0            &$4.39^{+0.14}_{-0.17}$  &$0.64^{+0.16}_{-0.15} $       &            \\
                                  &                     &13       &1.0            &$4.38^{+0.10}_{-0.09}$  &$0.47^{+0.11}_{-0.09} $       &           \\ \hline

$\bar{D}^{*0}\,\Sigma_c^+(2455)$  &${\frac{3}{2}}^-$    &13       &2.5            &$4.46^{+0.11}_{-0.12}$  &$2.31^{+0.41}_{-0.38} $       &4460           \\
                                  &                     &10       &1.0            &$4.45^{+0.16}_{-0.20}$  &$1.15^{+0.31}_{-0.29} $       &         \\
                                  &                     &13       &1.0            &$4.48^{+0.11}_{-0.10}$  &$0.88^{+0.19}_{-0.18} $       &        \\ \hline

$\bar{D}^{*0}\,\Sigma_c^+(2520)$  &${\frac{5}{2}}^-$    &13       &2.6            &$4.50^{+0.12}_{-0.12}$  &$1.74^{+0.31}_{-0.28} $       &4524    \\
                                  &                     &10       &1.0            &$4.51^{+0.16}_{-0.19}$  &$0.93^{+0.25}_{-0.22} $       &        \\
                                  &                     &13       &1.0            &$4.51^{+0.11}_{-0.10}$  &$0.66^{+0.15}_{-0.12} $       &       \\ \hline  \hline
\end{tabular}
\end{center}
\caption{ The predicted masses and pole residues of the hidden-charm   pentaquark molecular states.}
\end{table}

\begin{figure}
 \centering
 \includegraphics[totalheight=5cm,width=7cm]{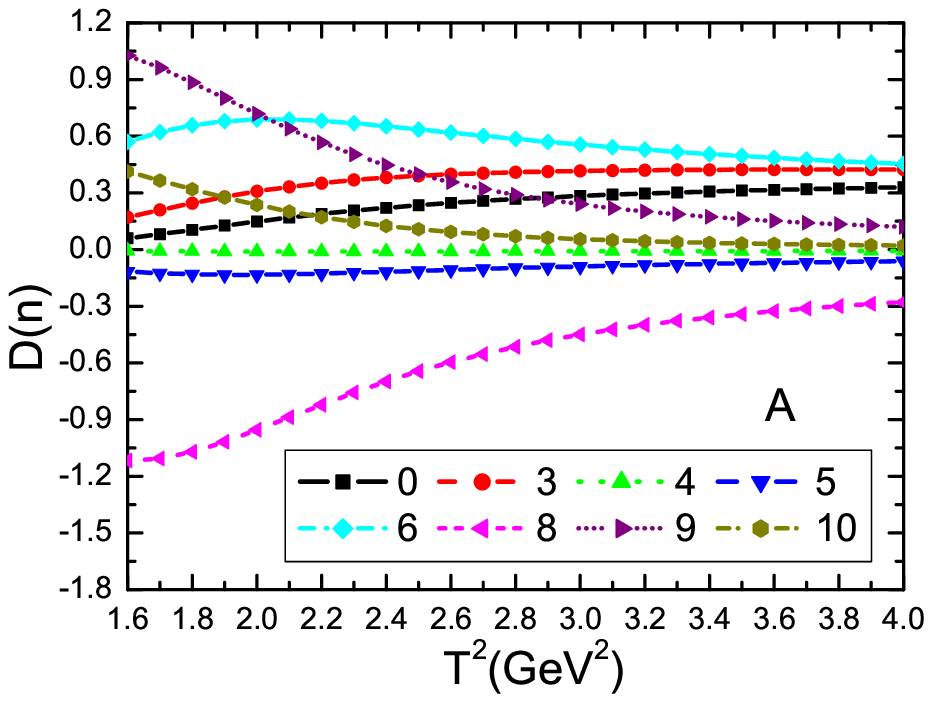}
 \includegraphics[totalheight=5cm,width=7cm]{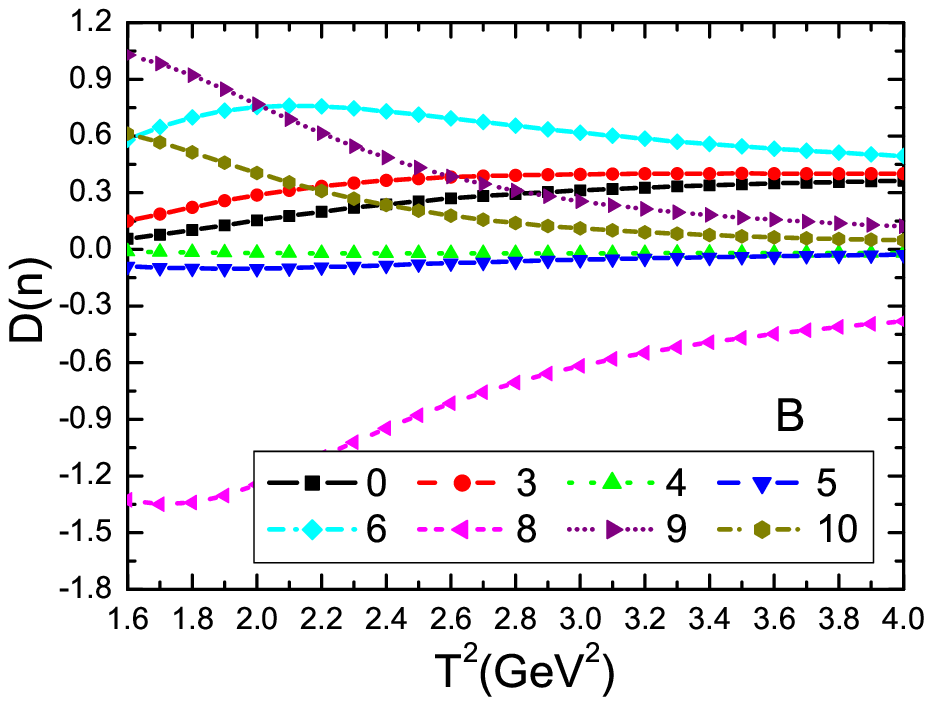}
 \includegraphics[totalheight=5cm,width=7cm]{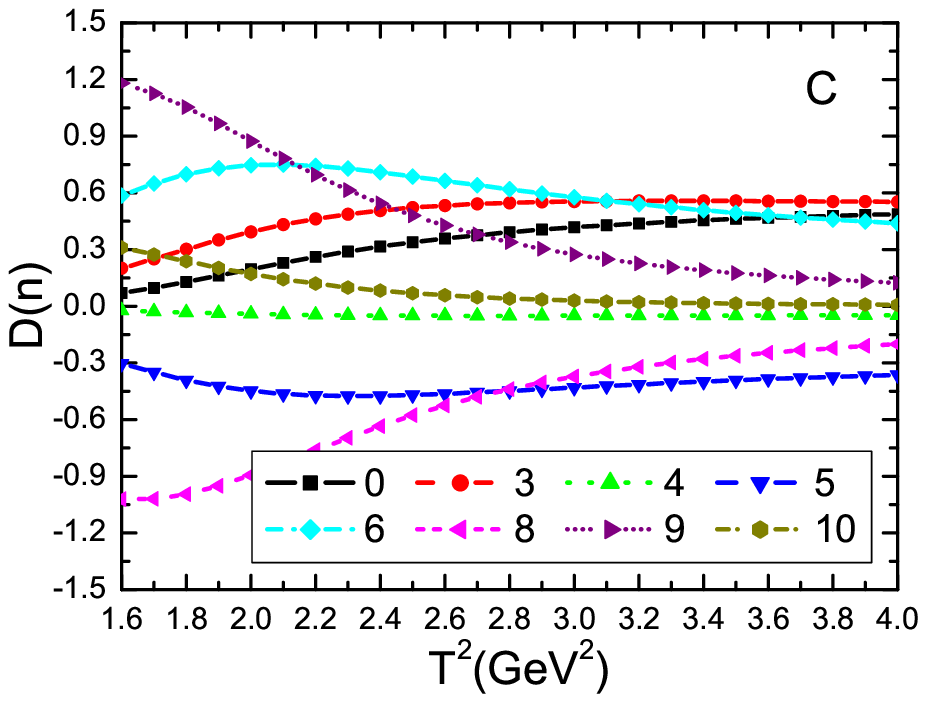}
 \includegraphics[totalheight=5cm,width=7cm]{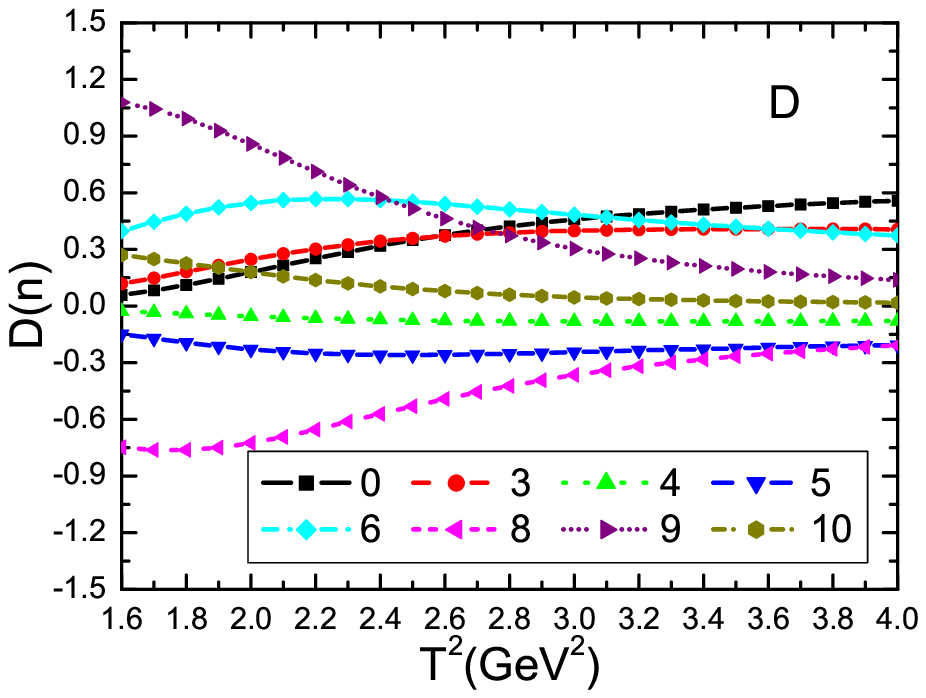}
  \caption{ The contributions  of the vacuum condensates of dimension $n=0$, $3$, $4$, $\cdots$ with variations of the Borel parameter $T^2$ for central values of  other  input parameters in the case ({\bf II}),  where  the $A$, $B$, $C$ and $D$ denote the pentaquark molecular  states $\bar{D}\Sigma_c$, $\bar{D}\Sigma_c^*$, $\bar{D}^{*}\Sigma_c$ and $ \bar{D}^{*}\Sigma_c^*$, respectively.   }
\end{figure}

\begin{figure}
 \centering
 \includegraphics[totalheight=5cm,width=7cm]{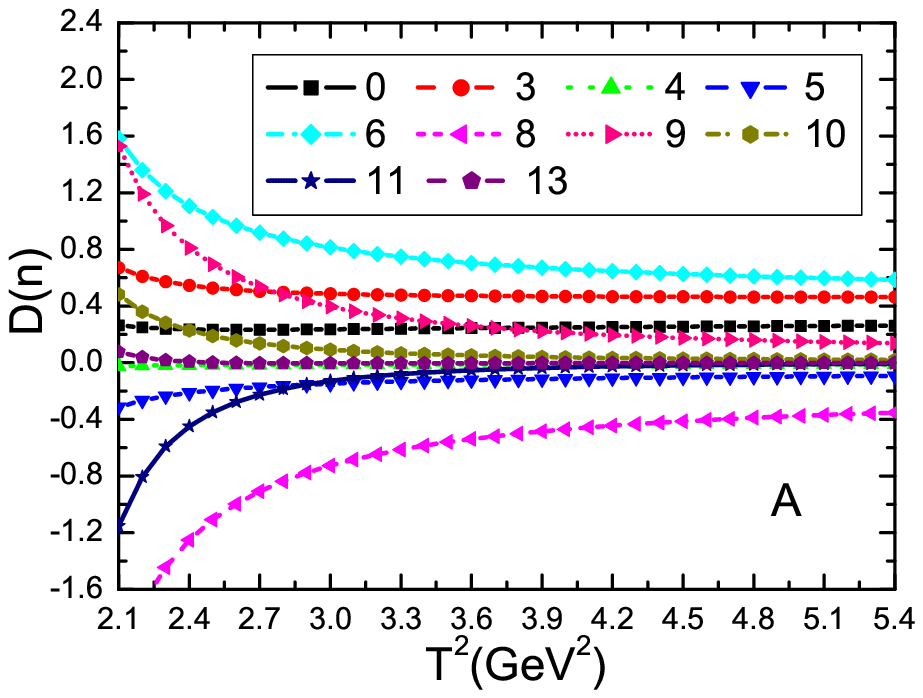}
 \includegraphics[totalheight=5cm,width=7cm]{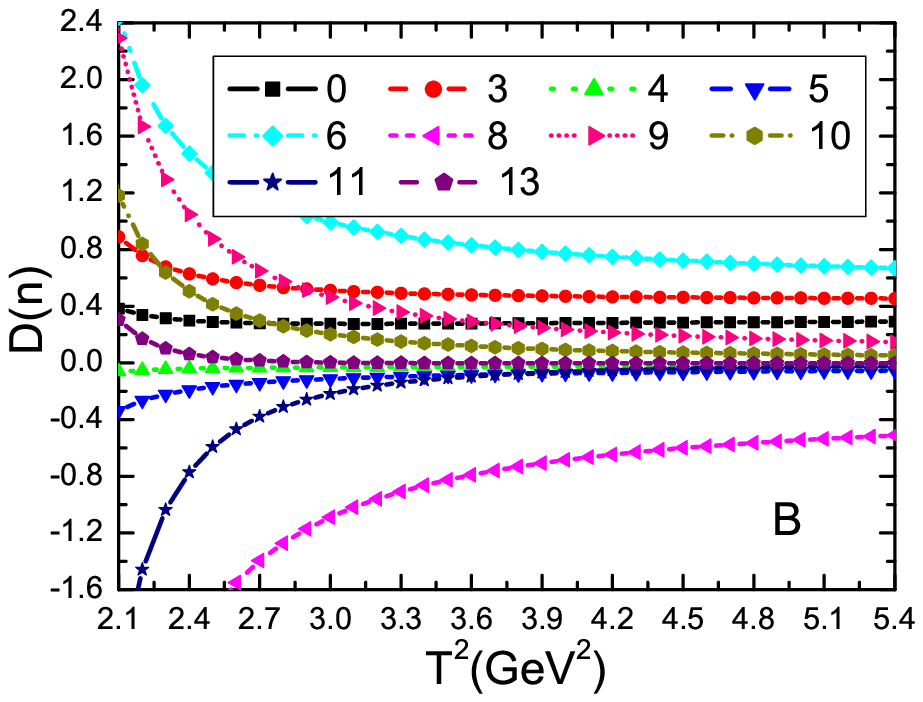}
 \includegraphics[totalheight=5cm,width=7cm]{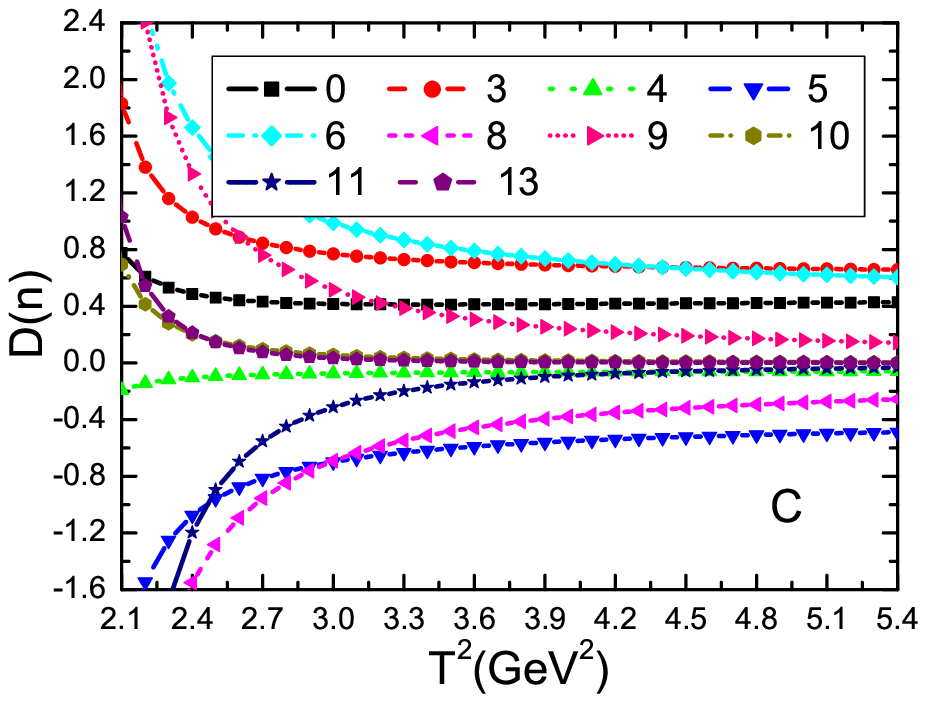}
 \includegraphics[totalheight=5cm,width=7cm]{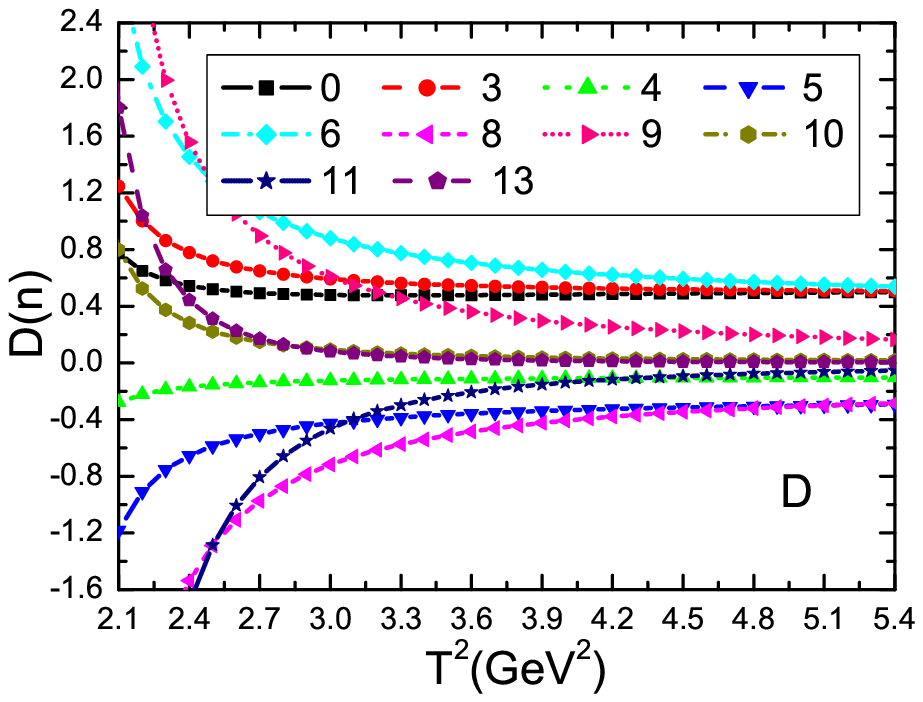}
  \caption{ The contributions  of the vacuum condensates of dimension $n=0$, $3$, $4$, $\cdots$ with variations of the Borel parameter $T^2$ for central values of  other  input parameters in the case ({\bf III}),  where  the $A$, $B$, $C$ and $D$ denote the pentaquark molecular  states $\bar{D}\Sigma_c$, $\bar{D}\Sigma_c^*$, $\bar{D}^{*}\Sigma_c$ and $ \bar{D}^{*}\Sigma_c^*$, respectively.   }\label{FrDSigmaMix}
\end{figure}

\begin{figure}
 \centering
  \includegraphics[totalheight=5cm,width=7cm]{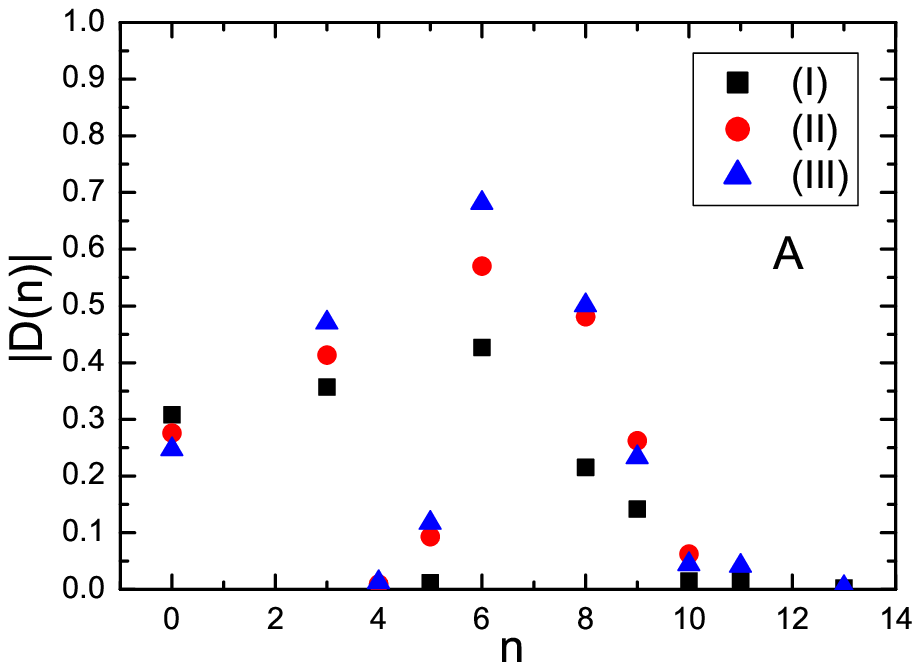}
 \includegraphics[totalheight=5cm,width=7cm]{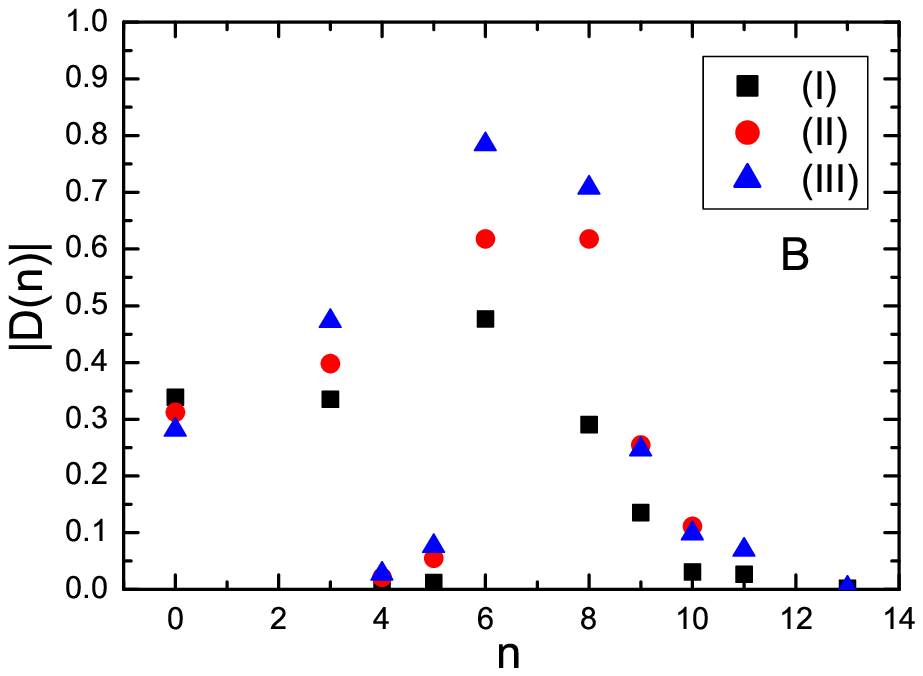}
 \includegraphics[totalheight=5cm,width=7cm]{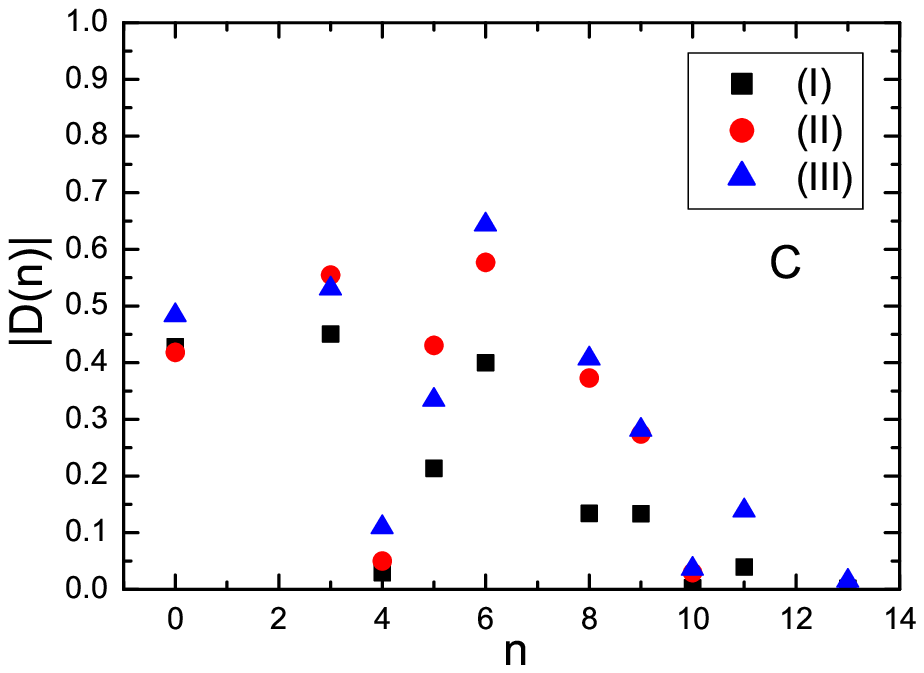}
 \includegraphics[totalheight=5cm,width=7cm]{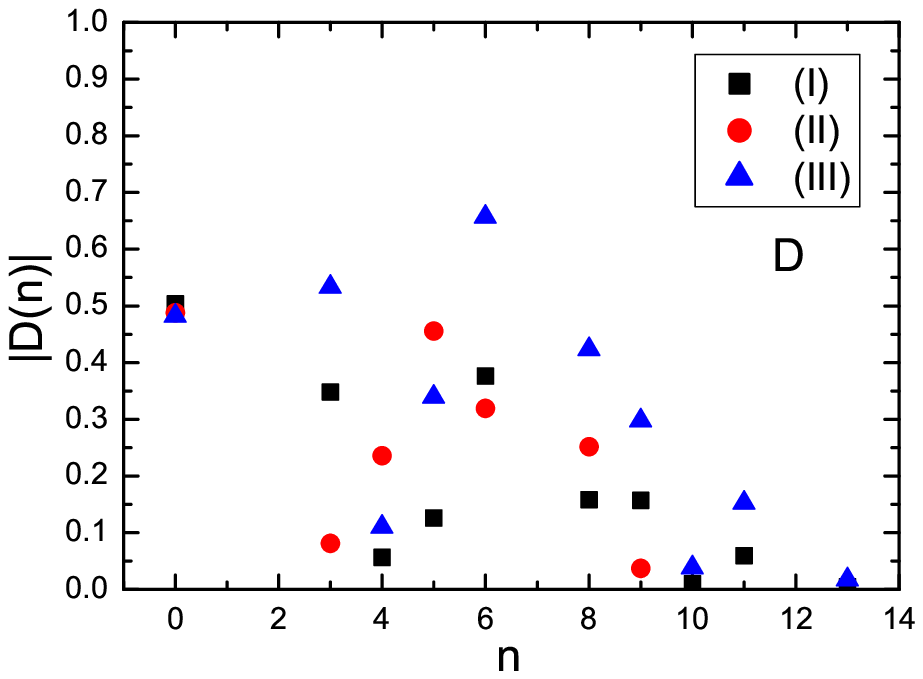}
 \caption{ The absolute contributions  of the vacuum condensates of dimension $n$ for central values of the input parameters in the Borel windows in the cases ({\bf I}), ({\bf II}) and ({\bf III}), where  the $A$, $B$, $C$ and $D$  denote the pentaquark molecular  states  $\bar{D}\Sigma_c$,  $\bar{D}\Sigma_c^*$, $\bar{D}^{*}\Sigma_c$ and $ \bar{D}^{*}\Sigma_c^*$, respectively.   }\label{FR-Dimension}
\end{figure}

\begin{figure}
 \centering
  \includegraphics[totalheight=5cm,width=7cm]{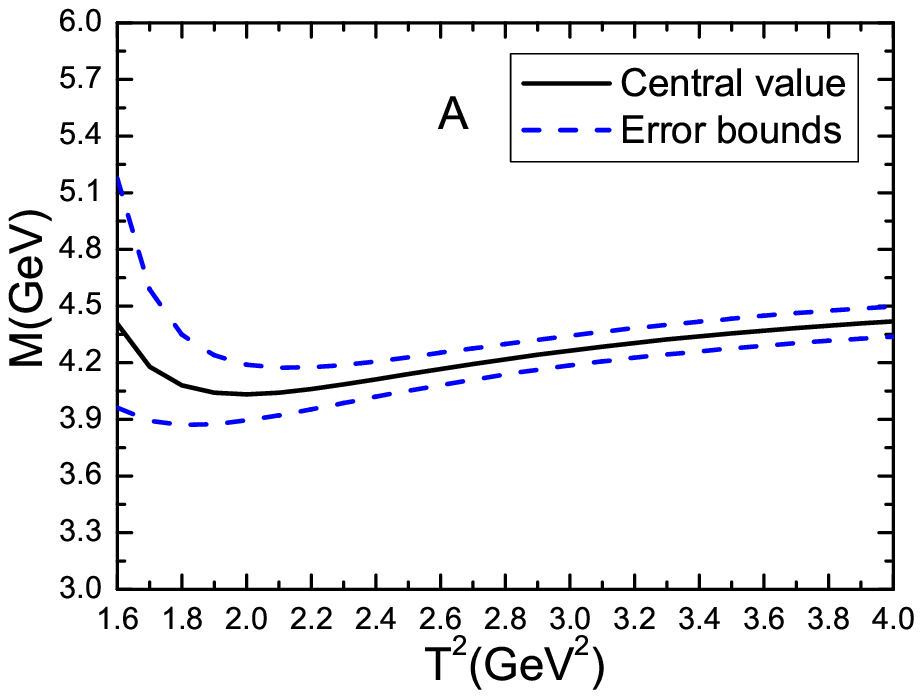}
 \includegraphics[totalheight=5cm,width=7cm]{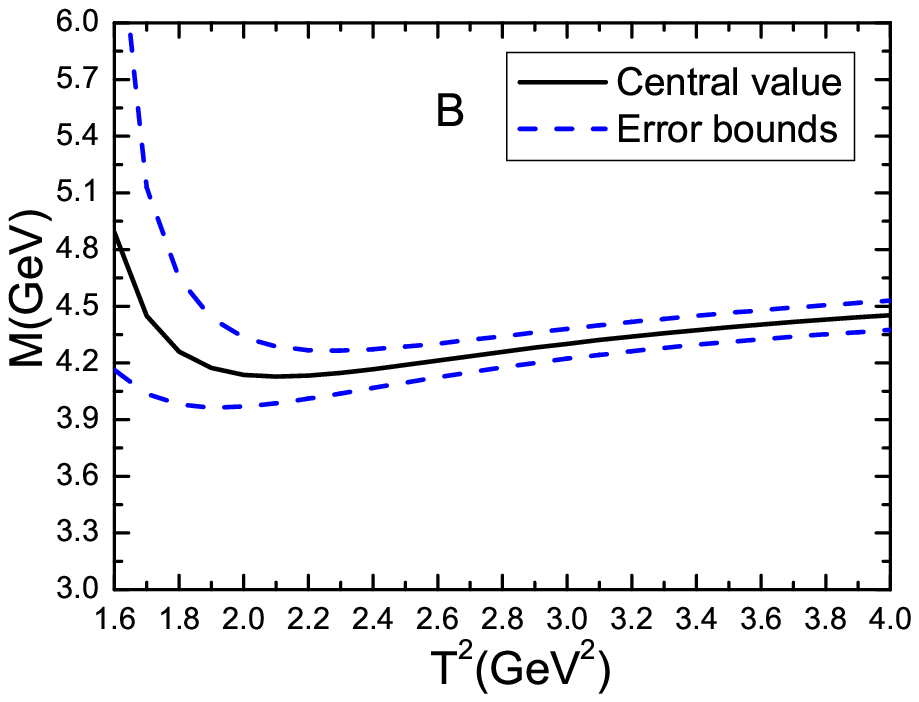}
 \includegraphics[totalheight=5cm,width=7cm]{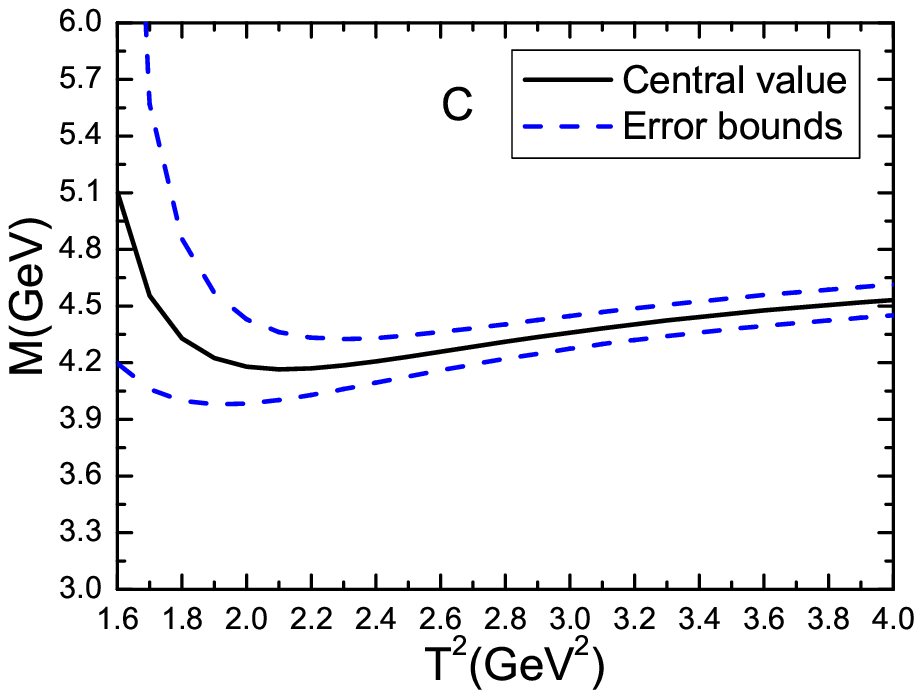}
 \includegraphics[totalheight=5cm,width=7cm]{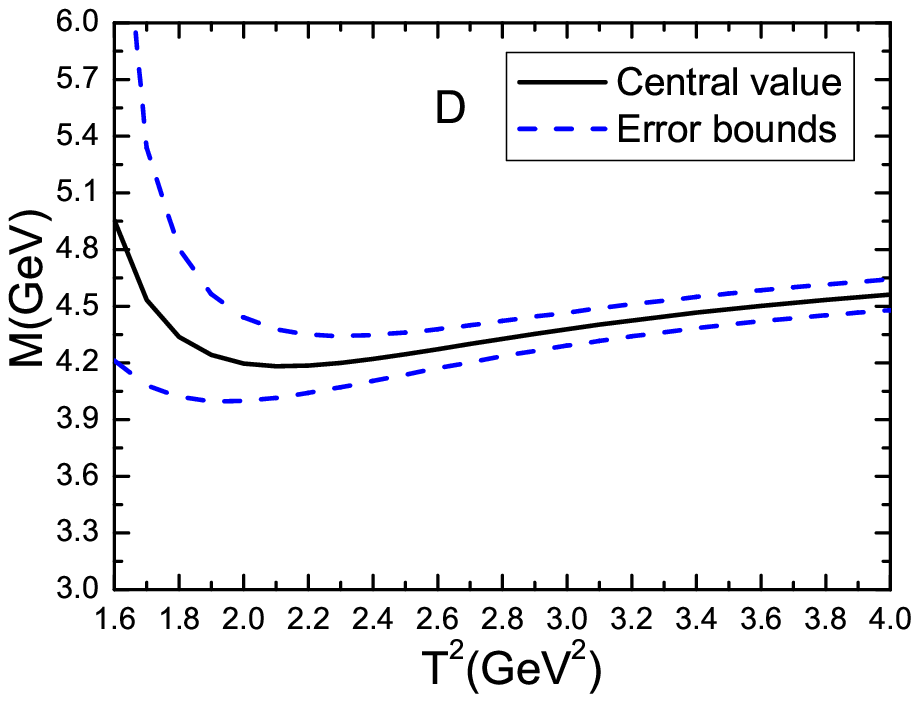}
  \caption{ The masses of the pentaquark molecular states  with variations of the Borel parameter $T^2$  in the case ({\bf I}),  where  the $A$, $B$, $C$ and $D$  denote the pentaquark molecular  states  $\bar{D}\Sigma_c$, $\bar{D}\Sigma_c^*$, $\bar{D}^{*}\Sigma_c$ and $ \bar{D}^{*}\Sigma_c^*$, respectively.   }\label{massDSigma}
\end{figure}

\begin{figure}
 \centering
 \includegraphics[totalheight=5cm,width=7cm]{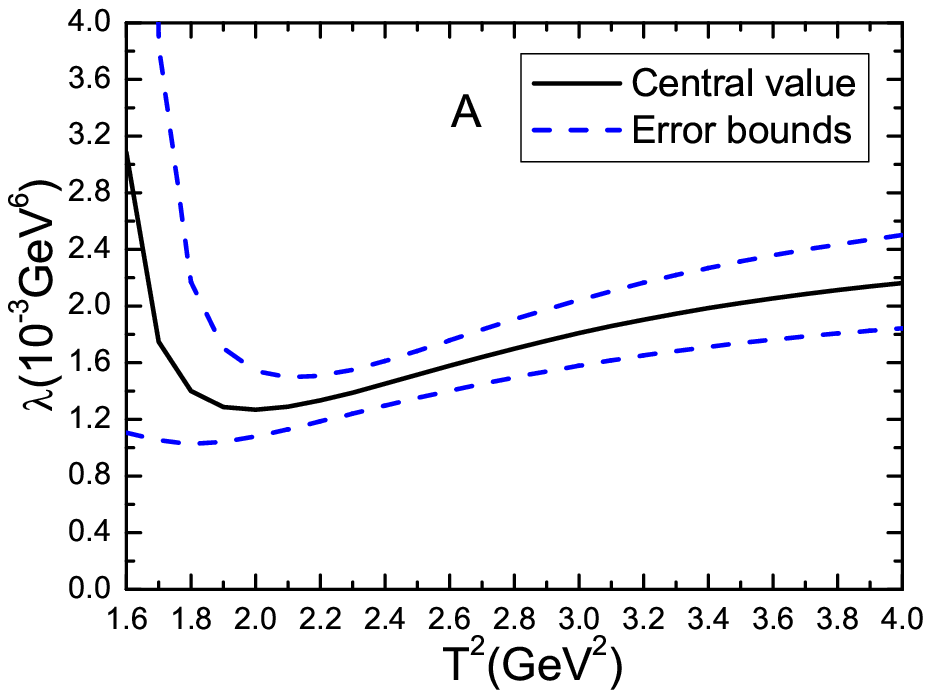}
 \includegraphics[totalheight=5cm,width=7cm]{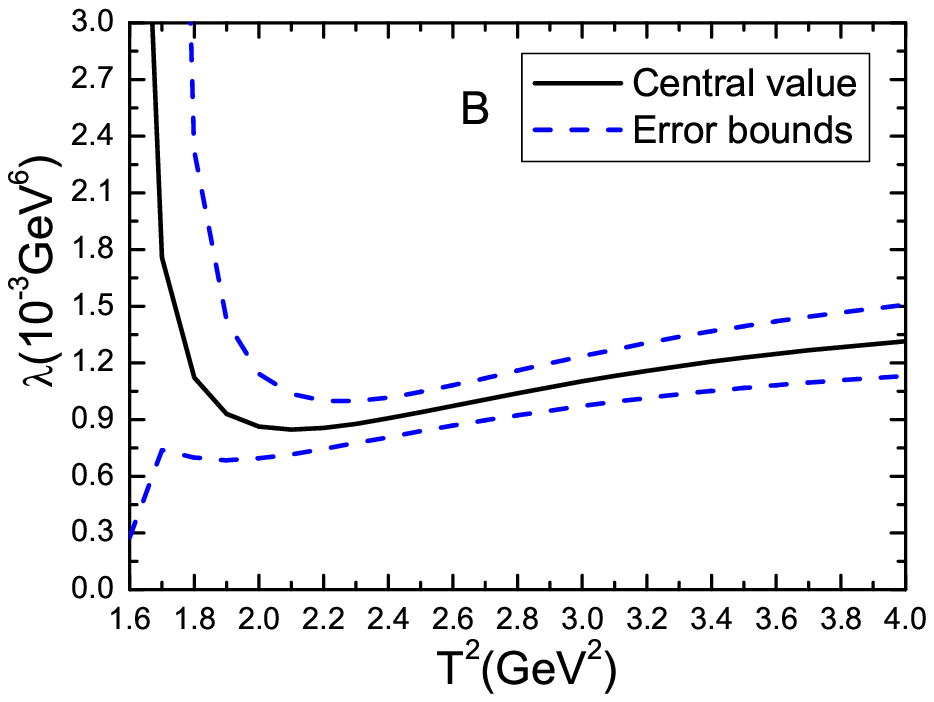}
 \includegraphics[totalheight=5cm,width=7cm]{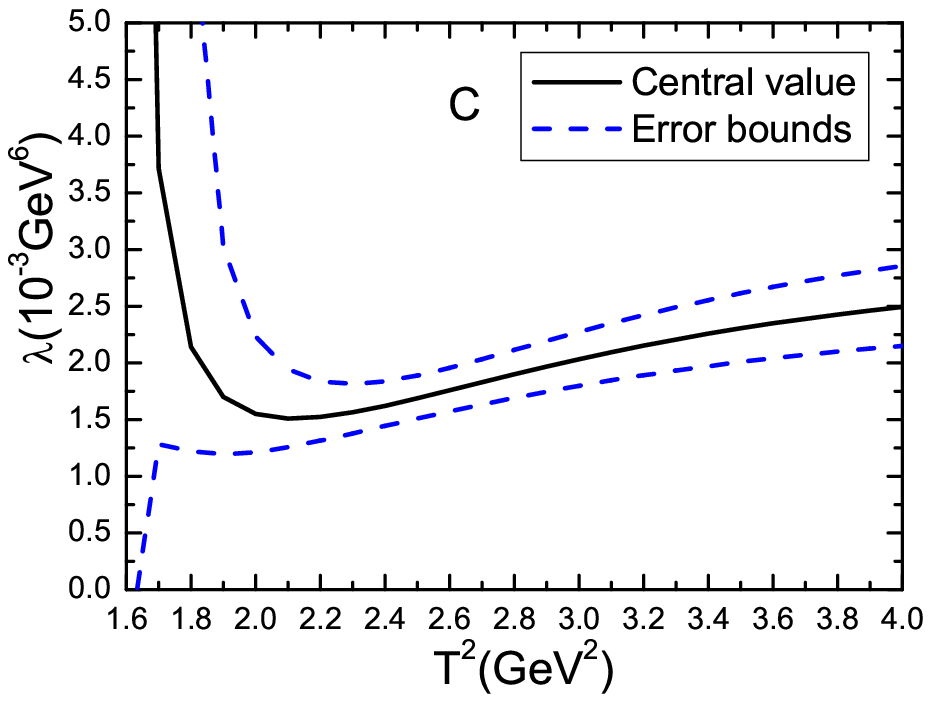}
 \includegraphics[totalheight=5cm,width=7cm]{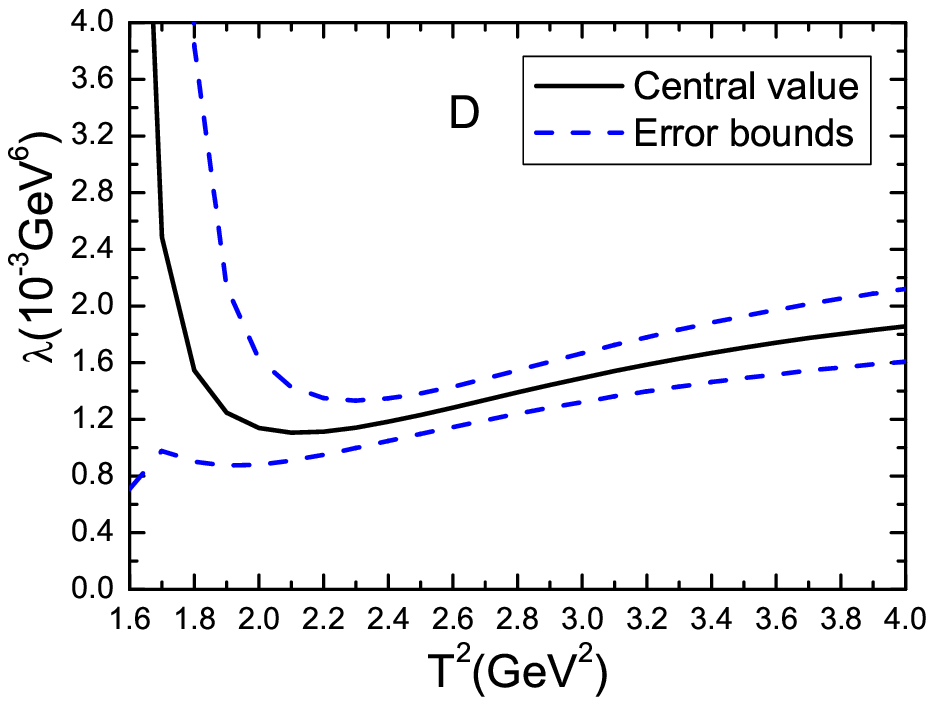}
\caption{ The pole residues of the pentaquark molecular states  with variations of the Borel parameter $T^2$  in the case ({\bf I}),  where  the $A$, $B$, $C$ and $D$  denote the pentaquark molecular  states  $\bar{D}\Sigma_c$, $\bar{D}\Sigma_c^*$, $\bar{D}^{*}\Sigma_c$ and $ \bar{D}^{*}\Sigma_c^*$, respectively.   }\label{residueDSigma}
\end{figure}

\begin{figure}
 \centering
\includegraphics[totalheight=5cm,width=7cm]{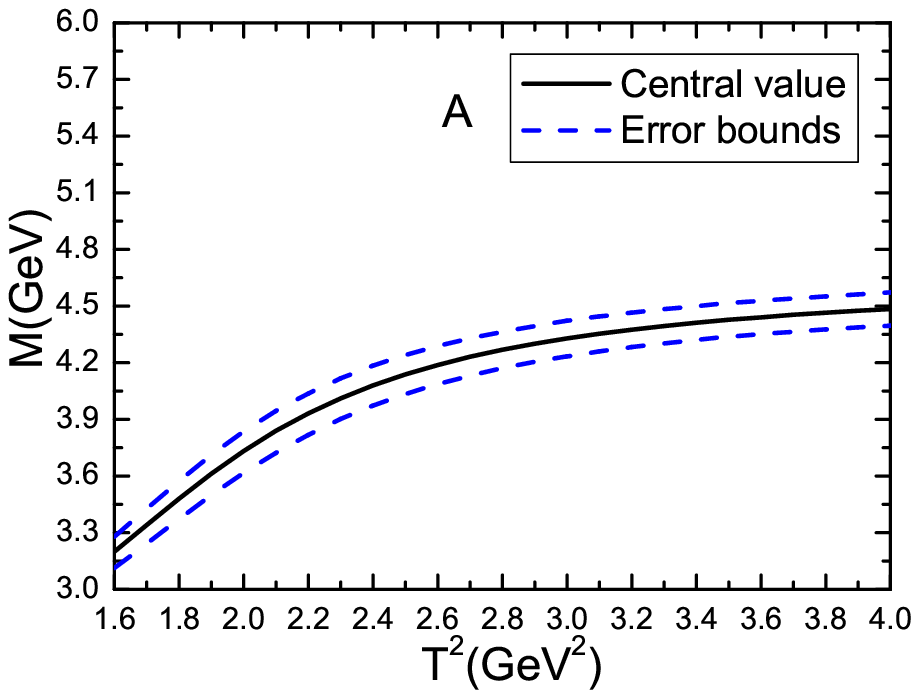}
\includegraphics[totalheight=5cm,width=7cm]{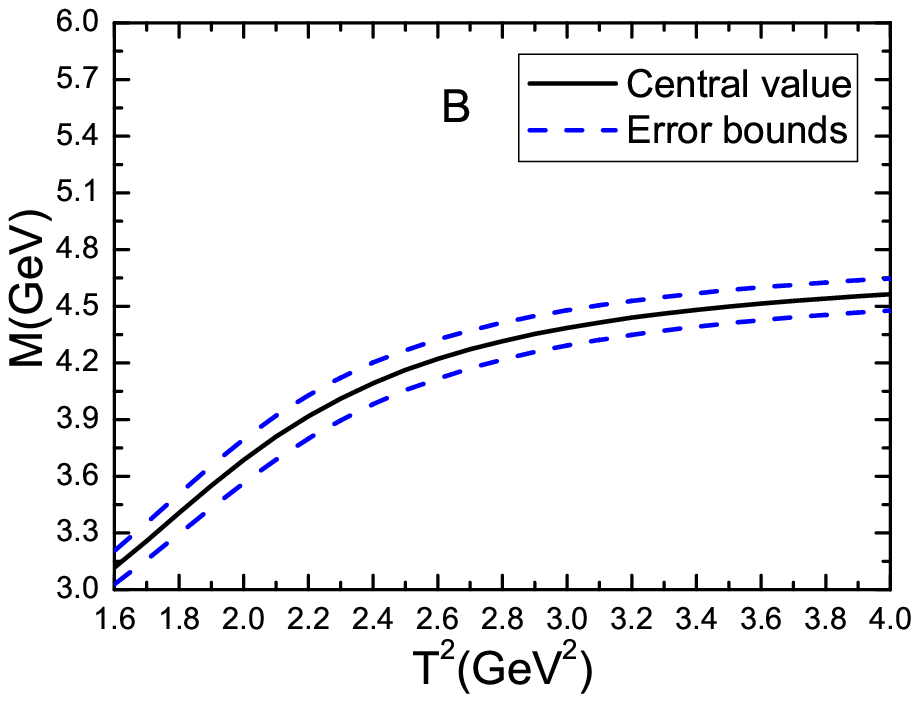}
\includegraphics[totalheight=5cm,width=7cm]{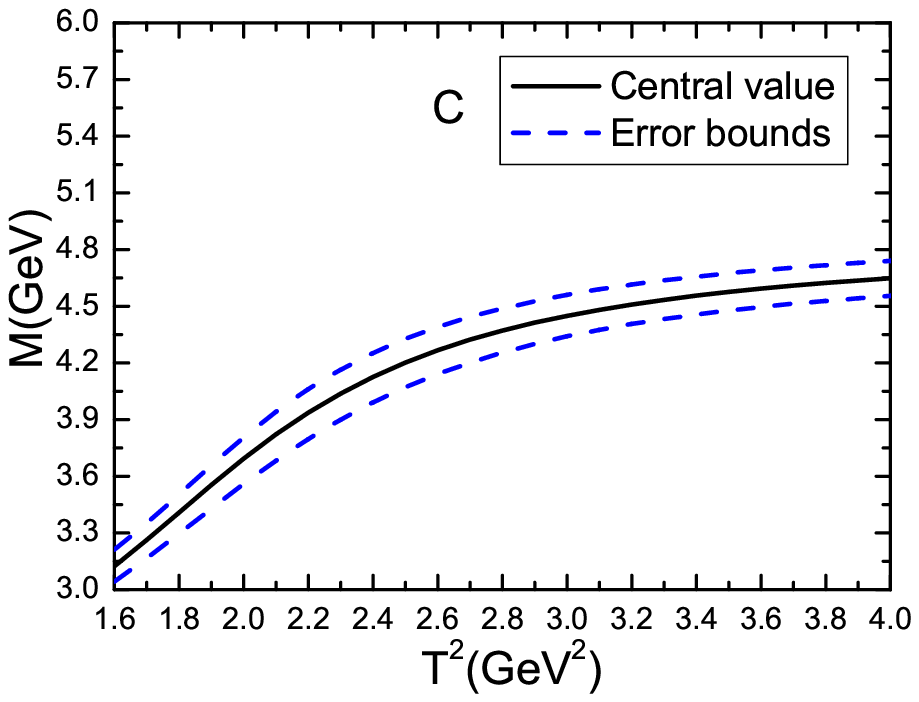}
\includegraphics[totalheight=5cm,width=7cm]{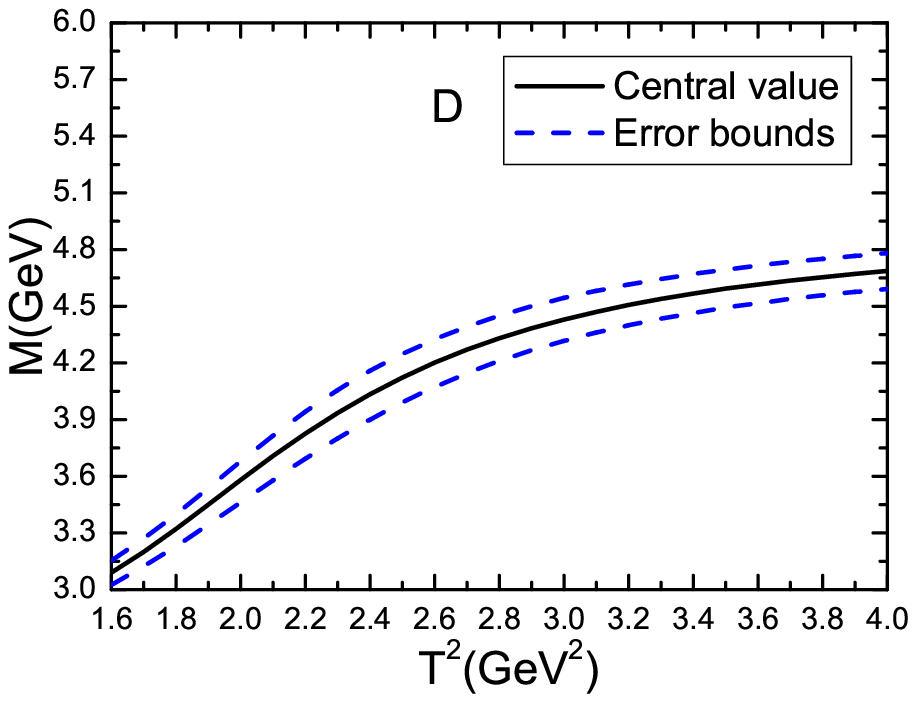}
\caption{ The masses of the pentaquark molecular states  with variations of the Borel parameter $T^2$  in the case ({\bf II}),  where  the $A$, $B$, $C$ and $D$  denote the pentaquark molecular  states  $\bar{D}\Sigma_c$, $\bar{D}\Sigma_c^*$, $\bar{D}^{*}\Sigma_c$ and $ \bar{D}^{*}\Sigma_c^*$, respectively.   }\label{massDSigmaNiel}
\end{figure}

\begin{figure}
 \centering
 \includegraphics[totalheight=5cm,width=7cm]{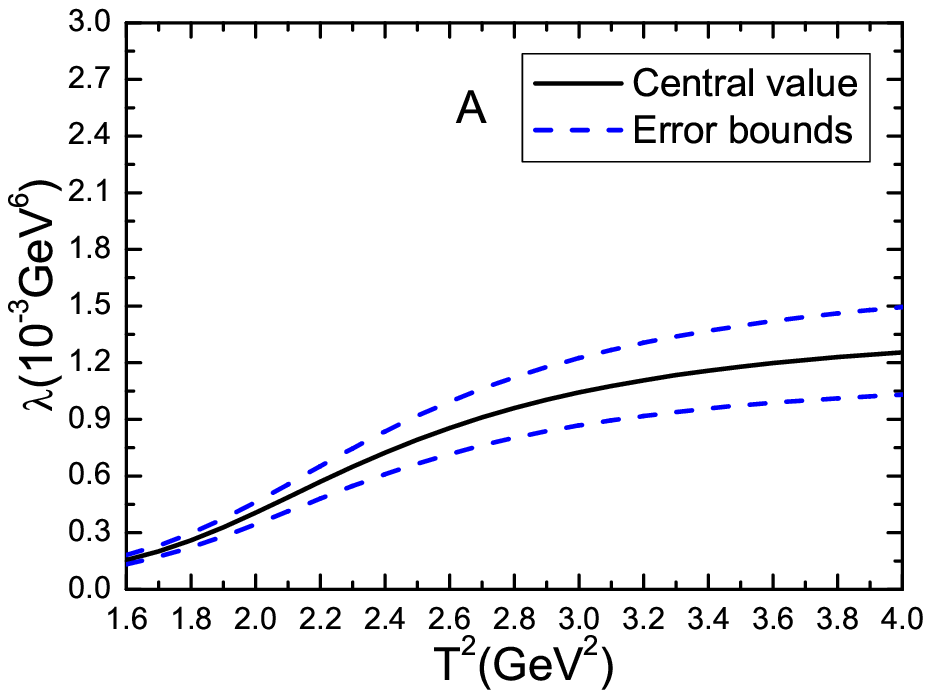}
 \includegraphics[totalheight=5cm,width=7cm]{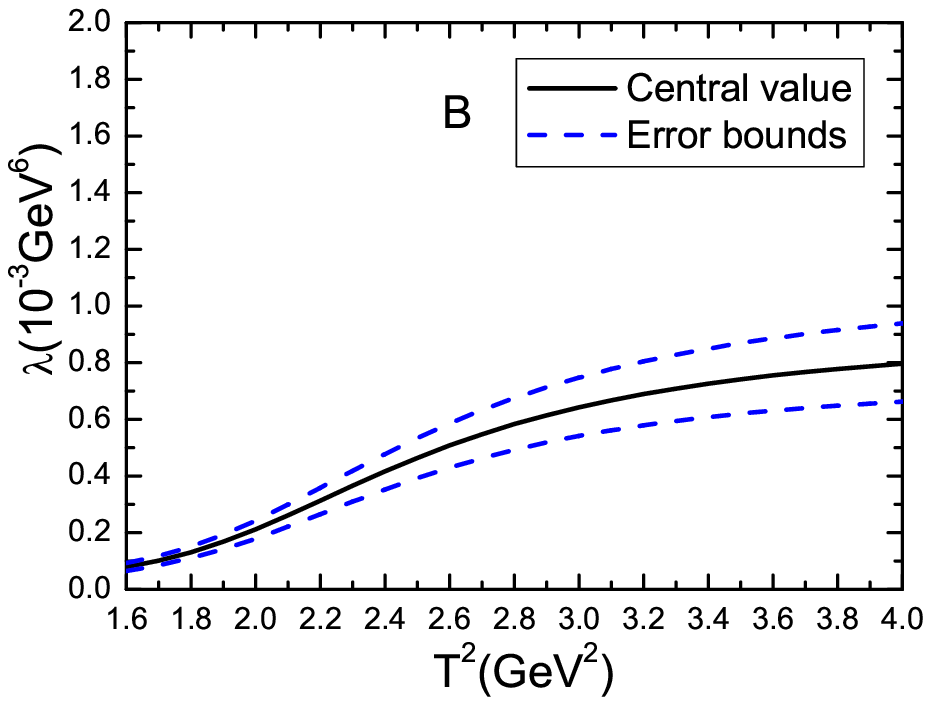}
 \includegraphics[totalheight=5cm,width=7cm]{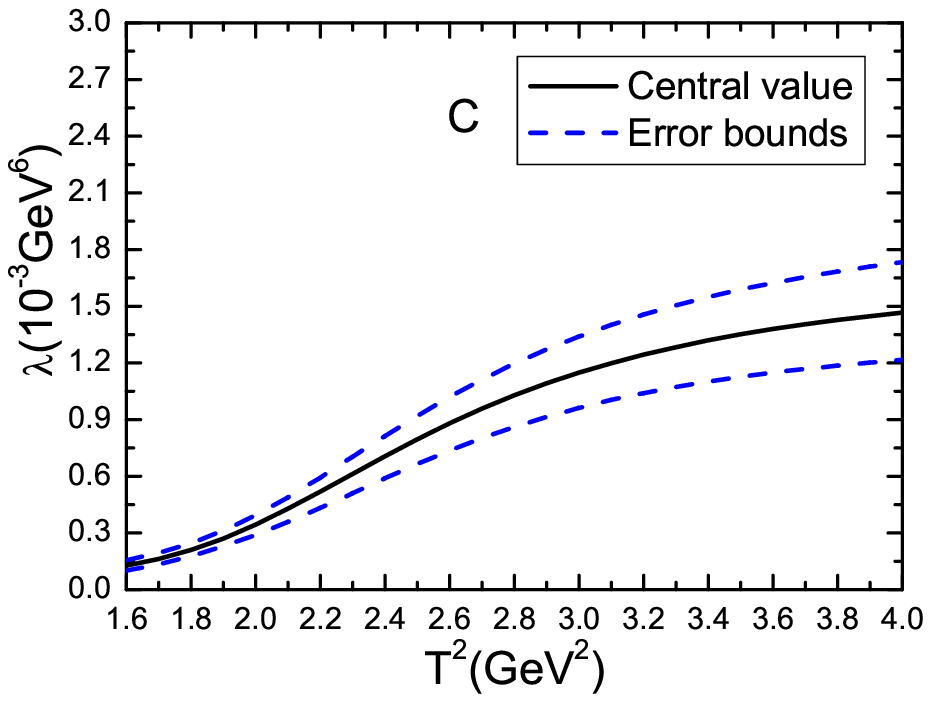}
 \includegraphics[totalheight=5cm,width=7cm]{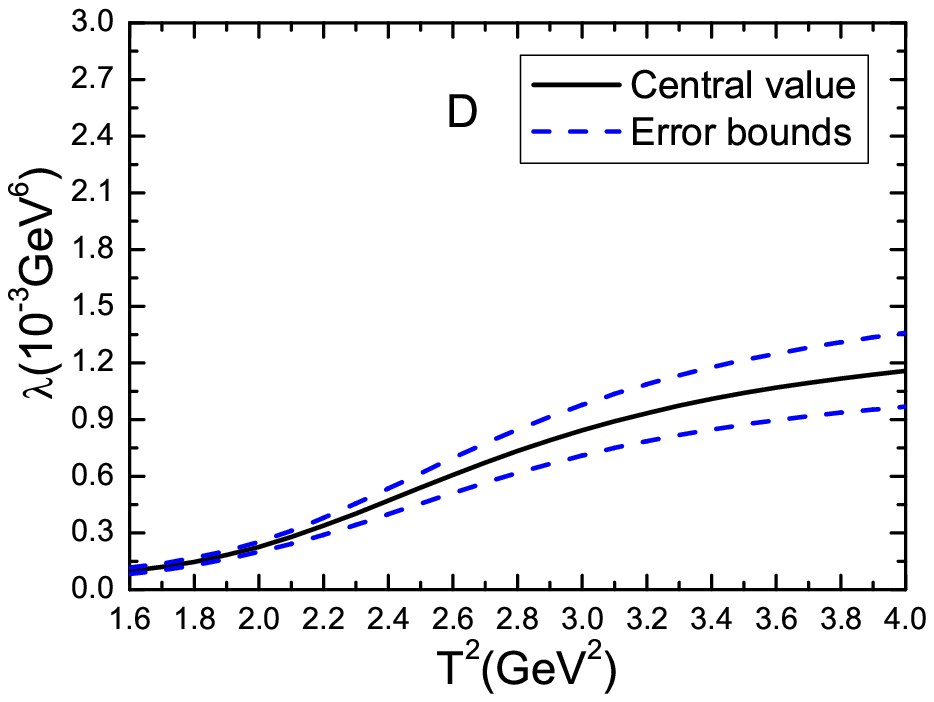}
\caption{ The pole residues of the pentaquark molecular states  with variations of the Borel parameter $T^2$  in the case ({\bf II}),  where  the $A$, $B$, $C$ and $D$  denote the pentaquark molecular  states  $\bar{D}\Sigma_c$, $\bar{D}\Sigma_c^*$, $\bar{D}^{*}\Sigma_c$ and $ \bar{D}^{*}\Sigma_c^*$, respectively.   }\label{residueDSigmaNiel}
\end{figure}

\begin{figure}
 \centering
 \includegraphics[totalheight=5cm,width=7cm]{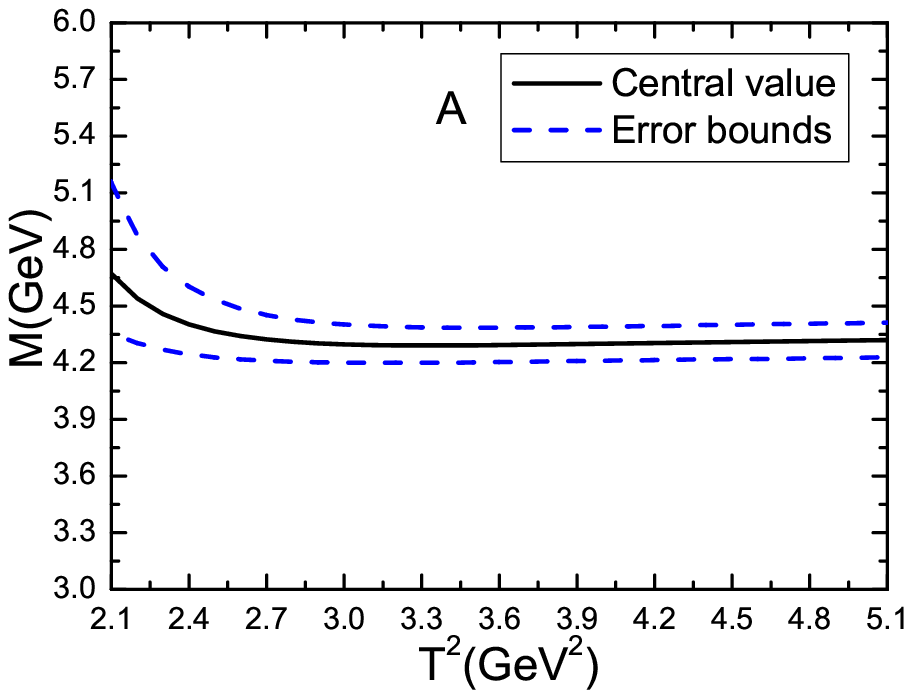}
 \includegraphics[totalheight=5cm,width=7cm]{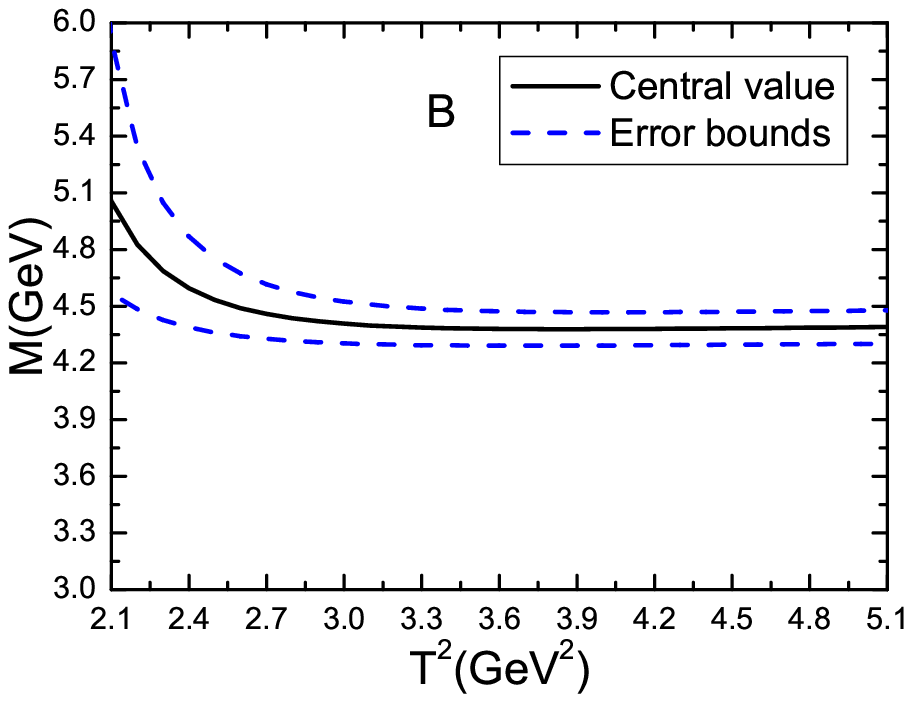}
 \includegraphics[totalheight=5cm,width=7cm]{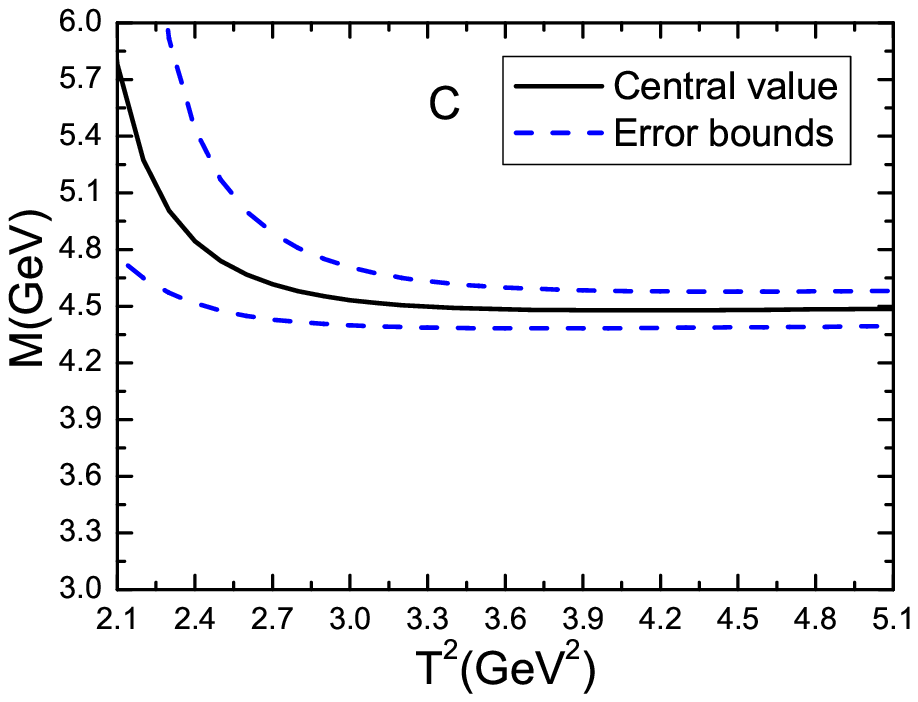}
 \includegraphics[totalheight=5cm,width=7cm]{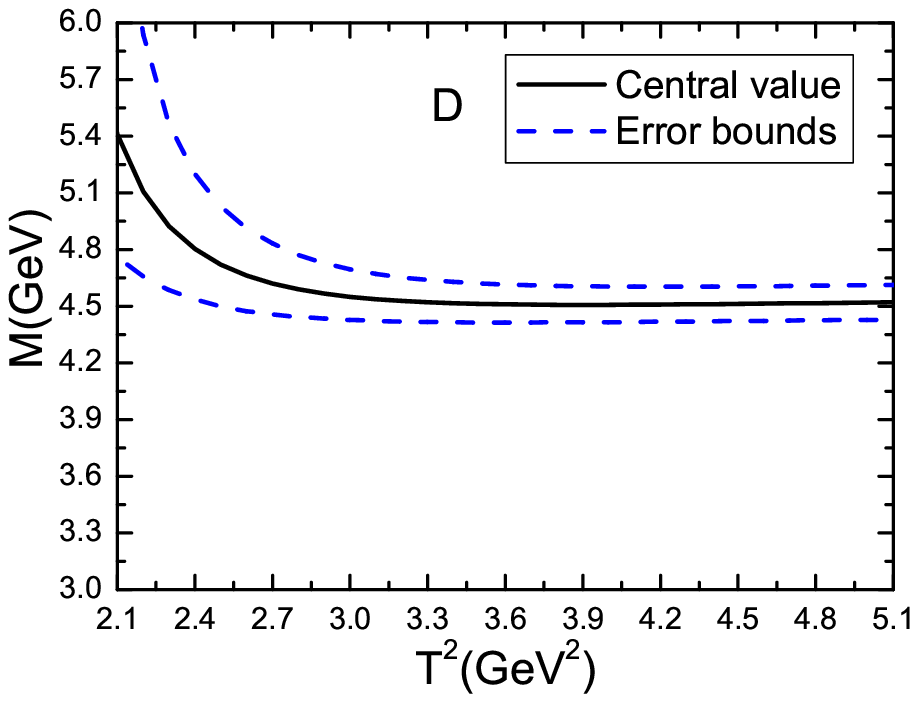}
\caption{ The masses of the pentaquark molecular states  with variations of the Borel parameter $T^2$  in the case ({\bf III}),  where  the $A$, $B$, $C$ and $D$  denote the pentaquark molecular  states  $\bar{D}\Sigma_c$, $\bar{D}\Sigma_c^*$, $\bar{D}^{*}\Sigma_c$ and $ \bar{D}^{*}\Sigma_c^*$, respectively.   }\label{massDSigmaMix}
\end{figure}

\begin{figure}
 \centering
 \includegraphics[totalheight=5cm,width=7cm]{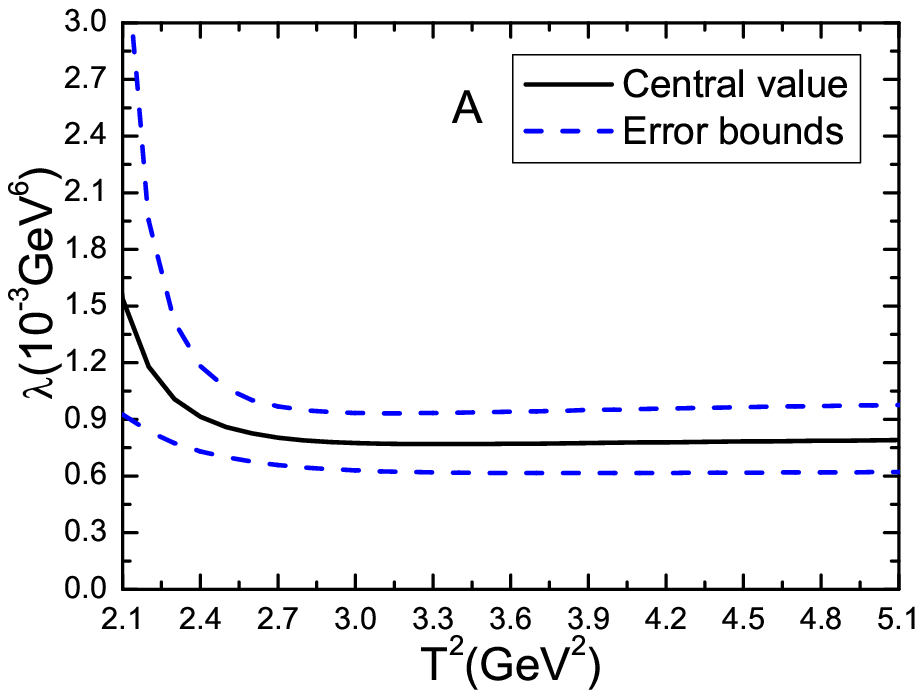}
\includegraphics[totalheight=5cm,width=7cm]{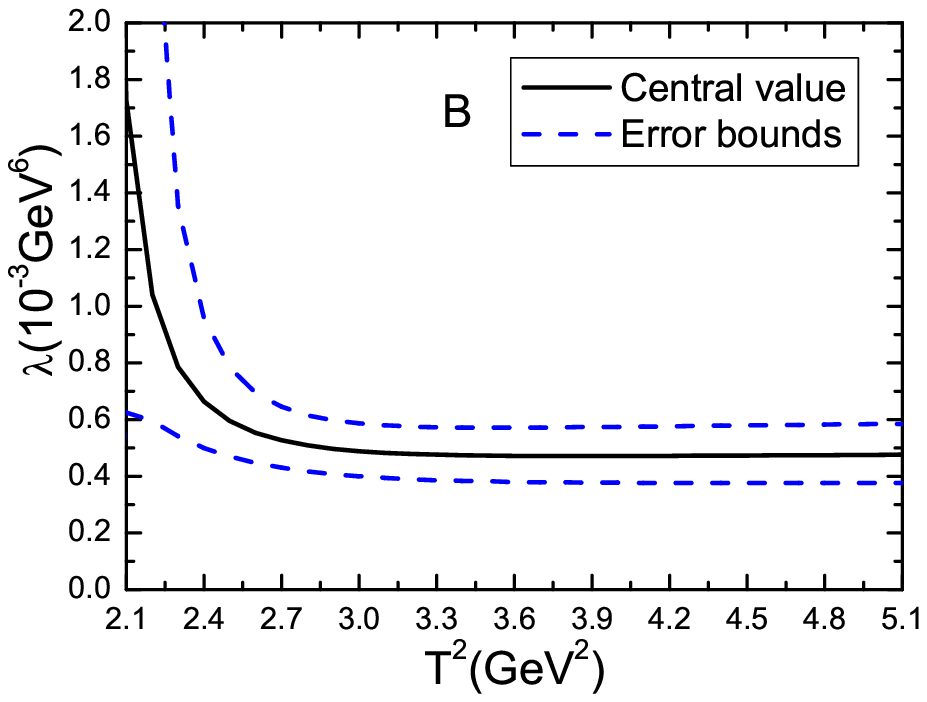}
\includegraphics[totalheight=5cm,width=7cm]{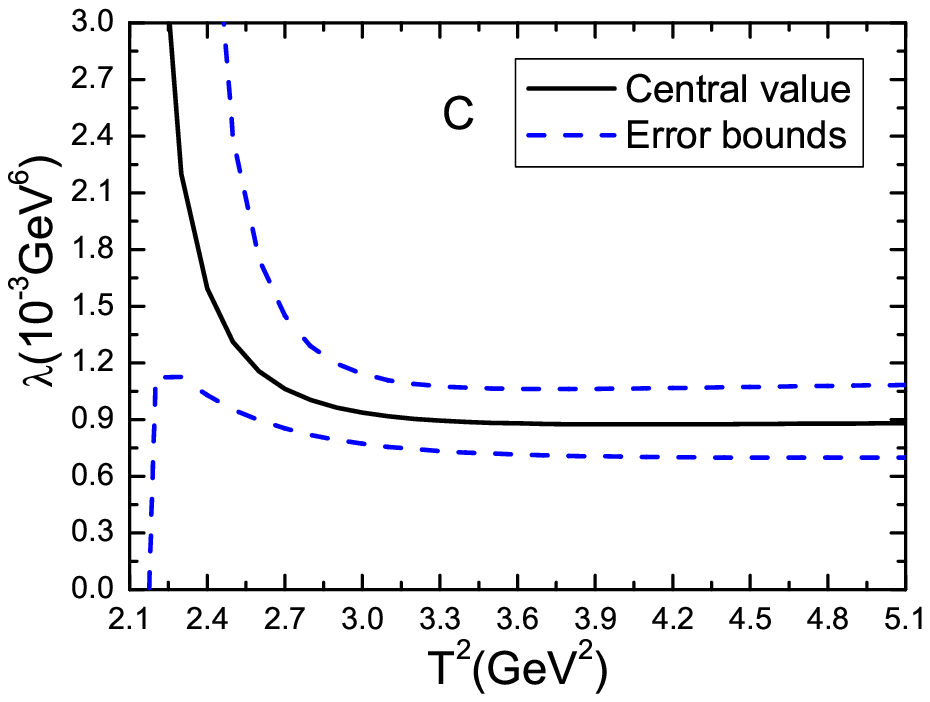}
\includegraphics[totalheight=5cm,width=7cm]{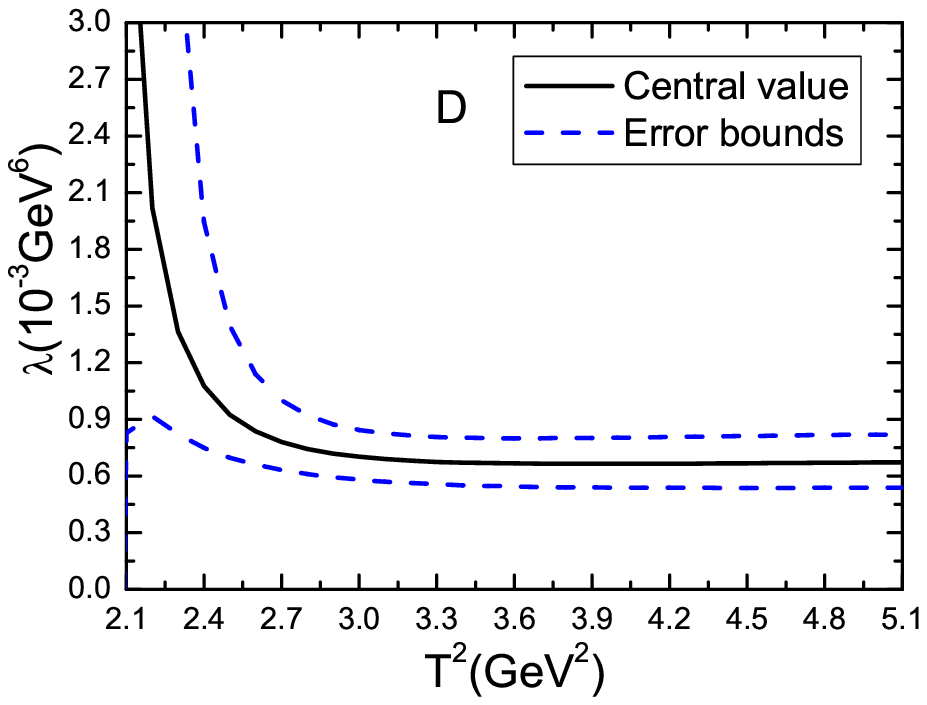}
\caption{ The pole residues of the pentaquark molecular states  with variations of the Borel parameter $T^2$  in the case ({\bf III}),  where  the $A$, $B$, $C$ and $D$  denote the pentaquark molecular  states  $\bar{D}\Sigma_c$, $\bar{D}\Sigma_c^*$, $\bar{D}^{*}\Sigma_c$ and $ \bar{D}^{*}\Sigma_c^*$, respectively.   }\label{residueDSigmaMix}
\end{figure}

\begin{figure}
 \centering
 \includegraphics[totalheight=5cm,width=7cm]{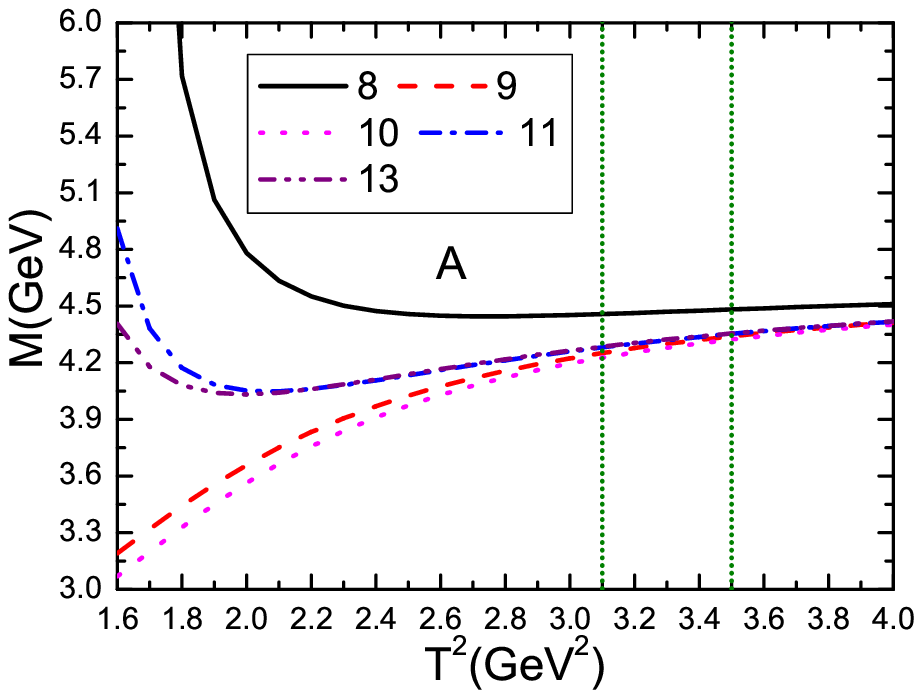}
 \includegraphics[totalheight=5cm,width=7cm]{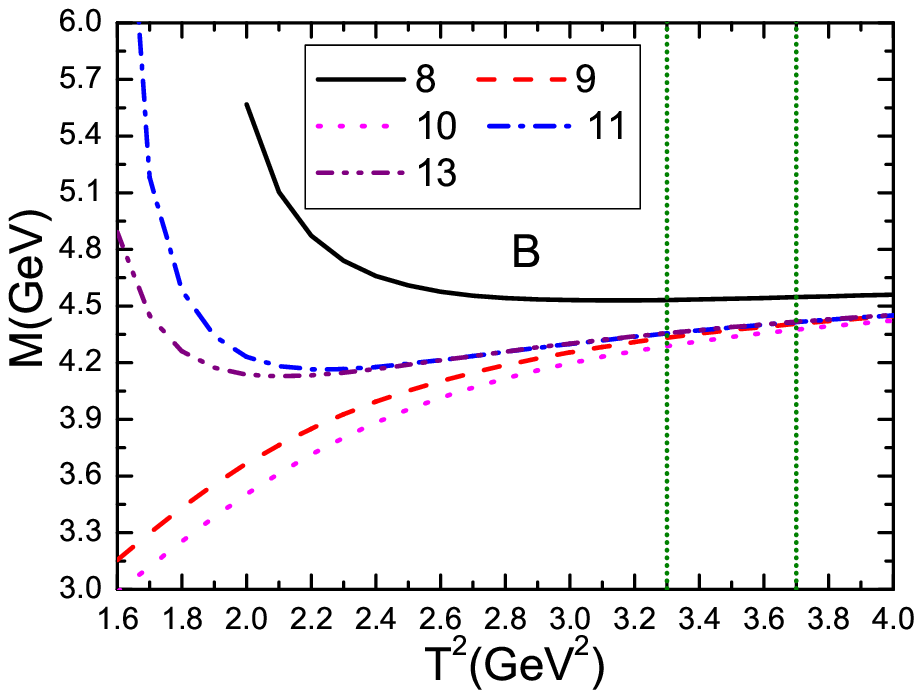}
 \includegraphics[totalheight=5cm,width=7cm]{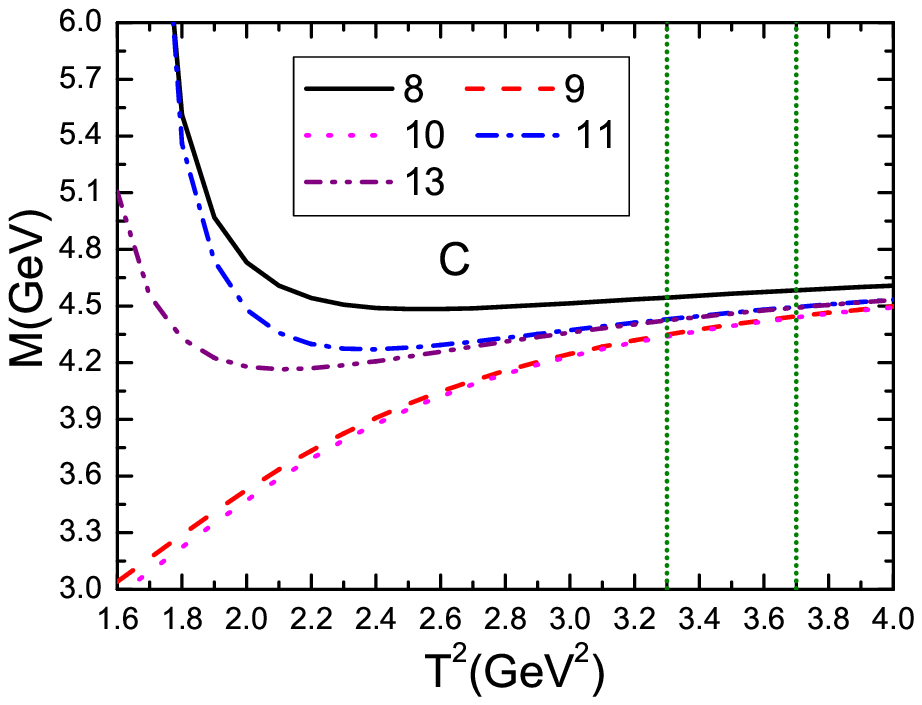}
 \includegraphics[totalheight=5cm,width=7cm]{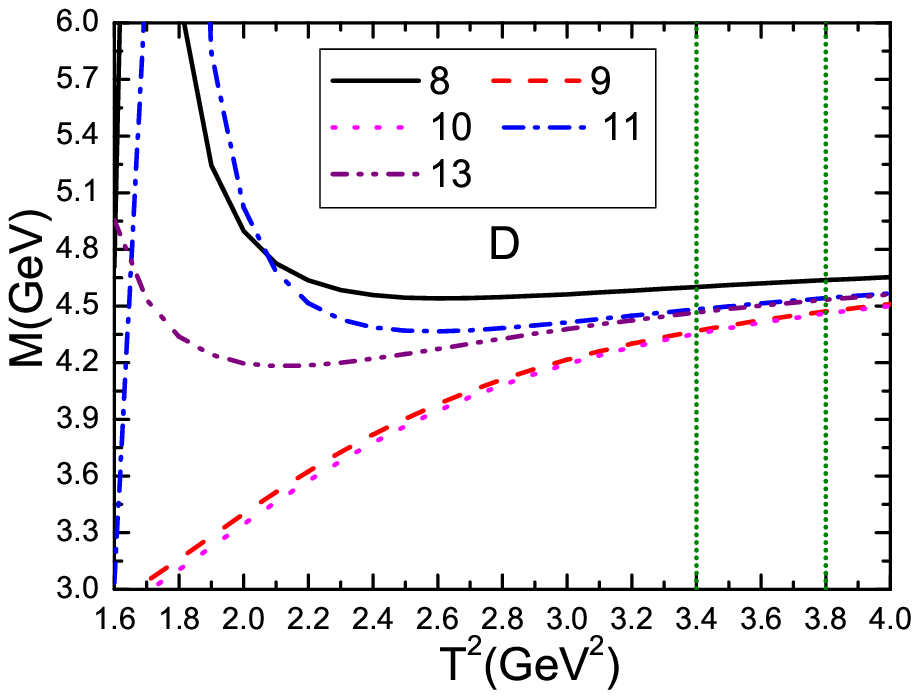}
  \caption{ The masses of the pentaquark molecular states  with variations of the Borel parameter $T^2$  in the case ({\bf I}) with truncations of the operator product expansion $D=8$, $9$, $10$, $11$ and $13$,  where  the $A$, $B$, $C$ and $D$  denote the pentaquark molecular  states  $\bar{D}\Sigma_c$, $\bar{D}\Sigma_c^*$, $\bar{D}^{*}\Sigma_c$ and $ \bar{D}^{*}\Sigma_c^*$, respectively.   }\label{massDSigmaDimension}
\end{figure}

Now we discuss the possible uncertainties originate from the energy scales   in the case ({\bf I}). In calculations, we observe that the predicted masses $M$ decrease monotonously and slowly with the increase of the energy scales $\mu$.    If we  choose the same continuum threshold parameters $s_0$ as that shown in Table 3, and take the uncertainties $\delta \mu=\pm0.2\,\rm{GeV}$ and vary the Borel parameters $T^2$ to retain the same pole contributions as that shown in Table 3, we obtain the uncertainties $\delta M={}^{+0.00}_{-0.02}\,\rm{GeV}$, ${}^{+0.00}_{-0.02}\,\rm{GeV}$, ${}^{+0.00}_{-0.01}\,\rm{GeV}$ and ${}^{+0.00}_{-0.01}\,\rm{GeV}$ for the $\bar{D}\Sigma_c$, $\bar{D}\Sigma_c^*$,   $\bar{D}^{*}\Sigma_c$  and   $\bar{D}^{*}\Sigma_c^*$  pentaquark molecular states, respectively. In fact, if we take the uncertainties $\delta \mu=\pm0.2\,\rm{GeV}$ and vary both the Borel parameters $T^2$ and continuum threshold parameters $s_0$ to retain the same pole contributions as that shown in Table 3, we can obtain the tiny uncertainties $\delta M\approx 0$, so the uncertainties $\delta M$ originate from the $\delta\mu$ near the optimal energy scales shown in Table 3 can be neglected.

We can define the QCD side of the QCD sum rules as
\begin{eqnarray}
\Pi(\mu)&=&\int_{4m_c^2(\mu)}^{s_0} ds\,\left[\sqrt{s}\rho_{QCD}^1(s,\mu)+  \rho_{QCD}^0(s,\mu)\right]\,
\exp\left(-\frac{s}{T^2}\right)\, .
\end{eqnarray}
The $\Pi(\mu)$ evolves with the   renormalization group equation, we can take into account
the energy-scale dependence according to the following equation,
\begin{eqnarray}
\Pi(\mu)&=&\left[ \frac{\alpha_s(\mu_0)}{\alpha_s(\mu)}\right]^{\gamma}\Pi(\mu_0) \, ,
\end{eqnarray}
 where the $\gamma$ is  the anomalous dimension of the correlation function, and we expect that the energy scale dependence can be factorized out and absorbed into the pole residue, the predicted mass $M$ is energy scale independent, see Eq.\eqref{QCDSR-M}.  The anomalous dimensions $\gamma$ for the QCD sum rules involving the massive quarks are unknown up to now \cite{Ioffe-PPNP}. We have to perform the following routine to take into account the energy scale dependence.
\begin{eqnarray}
\Pi(\mu)&=&\Pi\left(m_c(\mu),\langle\bar{q}q\rangle(\mu),\langle\bar{q}g_s\sigma Gq\rangle(\mu) \right) \nonumber\\
&=&\Pi\left(m_c(\mu_0)\left[\frac{\alpha_{s}(\mu)}{\alpha_{s}(\mu_0)}\right]^{\frac{12}{25}},\,
\langle\bar{q}q \rangle(\mu_0)\left[\frac{\alpha_{s}(\mu_0)}{\alpha_{s}(\mu)}\right]^{\frac{12}{25}},\,
\langle\bar{q}g_s \sigma Gq \rangle(\mu_0)\left[\frac{\alpha_{s}(\mu_0)}{\alpha_{s}(\mu)}\right]^{\frac{2}{25}} \right)\, ,
\end{eqnarray}
and evolve the $c$-quark mass and vacuum condensates to the optimal energy scales $\mu=2.2,\rm{GeV}$, $2.4\,\rm{GeV}$, $2.5\,\rm{GeV}$ and $2.6\,\rm{GeV}$, respectively.
In the operator product expansion, the energy scale $\mu$ separates the regions of short and long distances, the interactions at momenta $p^2>\mu^2$ are included in the Wilson's coefficients, while the effects at
$p^2<\mu^2$ are absorbed into the vacuum condensates, which are energy scale dependent and can be evolved to arbitrary energy scales according to the renormalization group equation. The scale $\mu$ (normalization point) should be large
enough in order to justify the calculations of the Wilson's coefficients in QCD perturbation theory. In this article, the energy scales $\mu=2.2,\rm{GeV}$, $2.4\,\rm{GeV}$, $2.5\,\rm{GeV}$ and $2.6\,\rm{GeV}$ are suitable. We obtain the masses $M$ of the pentaquark molecular states through a  fraction, the energy scale dependence in the numerator and denominator are canceled out to some extent, the $\delta M$ induced by the $\delta\mu$ near the optimal energy scales are very small.

\section{Conclusion}
In this article, we  study  the $\bar{D}\Sigma_c$, $\bar{D}\Sigma_c^*$,  $\bar{D}^{*}\Sigma_c$  and   $\bar{D}^{*}\Sigma_c^*$  pentaquark molecular states
with the QCD sum rules by carrying out the operator product expansion   up to   the vacuum condensates of dimension $13$ in a consistent way.
In calculations,   we separate  the contributions of the negative parity and positive parity pentaquark  molecular  states  unambiguously,
and  study the masses and pole residues of the hidden-charm pentaquark molecular states
 with the QCD sum rules in details. Special attentions are payed to the operator product expansion, as the predicted masses change remarkably  with the truncations of the operator product expansion, we should truncate the operator product expansion in a consistent way.  The present calculations support assigning the $P_c(4312)$ to be the $\bar{D}\Sigma_c$ pentaquark molecular state with $J^P={\frac{1}{2}}^-$, assigning the $P_c(4380)$ to be the $\bar{D}\Sigma_c^*$ pentaquark molecular state with $J^P={\frac{3}{2}}^-$,  assigning the $P_c(4440/4457)$ to be the $\bar{D}^{*}\Sigma_c$ pentaquark molecular state with $J^P={\frac{3}{2}}^-$ or the $\bar{D}^{*}\Sigma_c^*$ pentaquark molecular state with $J^P={\frac{5}{2}}^-$. The QCD sum rules indicate that there maybe exist the $\bar{D}\Sigma_c$,  $\bar{D}\Sigma_c^*$,  $\bar{D}^{*}\Sigma_c$  and   $\bar{D}^{*}\Sigma_c^*$  pentaquark molecular states with the $J^P={\frac{1}{2}}^-$, ${\frac{3}{2}}^-$, ${\frac{3}{2}}^-$ and ${\frac{5}{2}}^-$, respectively.

\section*{Appendix}
The explicit expressions of the QCD spectral densities: \\ \\
{\bf  For the $\bar{D}\Sigma_c$ pentaquark molecular states},

\begin{eqnarray}
\rho_0^1(s)&=&\frac{11}{1966080\pi^8}\int dydz\,yz\left(1-y-z\right)^4\left(s-\overline{m}_c^2\right)^4\left(8s-3\overline{m}_c^2\right) \, ,
\end{eqnarray}

\begin{eqnarray}
\rho_3^1(s)&=&-\frac{19m_c \langle\bar{q}q\rangle}{12288\pi^6} \int dydz\, z\left(1-y-z\right)^2\left(s-\overline{m}_c^2\right)^3 \, ,
\end{eqnarray}

\begin{eqnarray}
\rho_4^1(s)&=&-\frac{11m_c^2}{589824\pi^6}\langle\frac{\alpha_{s}GG}{\pi}\rangle \int dydz\, \left(\frac{z}{y^2}+\frac{y}{z^2}\right)\left(1-y-z\right)^4
\left(s-\overline{m}_c^2\right)\left(5s-3\overline{m}_c^2\right) \nonumber\\
&&+\frac{1}{3072\pi^6}\langle\frac{\alpha_{s}GG}{\pi}\rangle \int dydz\, z\left(1-y-z\right)^3\left(s-\overline{m}_c^2\right)^2\left(2s-\overline{m}_c^2\right)
\nonumber\\
&&-\frac{1}{65536\pi^6}\langle\frac{\alpha_{s}GG}{\pi}\rangle \int dydz\, y\left(1-y-z\right)^3
\left(s-\overline{m}_c^2\right)^2\left(2s-\overline{m}_c^2\right) \nonumber\\
&&-\frac{1}{16384\pi^6}\langle\frac{\alpha_{s}GG}{\pi}\rangle \int dydz\, yz\left(1-y-z\right)^2
\left(s-\overline{m}_c^2\right)^2\left(2s-\overline{m}_c^2\right) \nonumber\\
&&-\frac{1}{786432\pi^6}\langle\frac{\alpha_{s}GG}{\pi}\rangle \int dydz\,
\left(1-y-z\right)^4\left(s-\overline{m}_c^2\right)^2\left(2s-\overline{m}_c^2\right)\, ,
\end{eqnarray}

\begin{eqnarray}
\rho_5^1(s)&=&\frac{11m_c \langle\bar{q}g_{s}\sigma Gq\rangle}{4096\pi^6}\int dydz\, z\left(1-y-z\right)\left(s-\overline{m}_c^2\right)^2
 \nonumber\\
&&-\frac{41m_c\langle\bar{q}g_{s}\sigma Gq\rangle}{16384\pi^6} \int dydz\, \frac{z\left(1-y-z\right)^2}{y}\left(s-\overline{m}_c^2\right)^2
 \nonumber\\
&&+\frac{3m_c\langle\bar{q}g_{s}\sigma Gq\rangle}{32768\pi^6} \int dydz\,\left(1-y-z\right)^2\left(s-\overline{m}_c^2\right)^2\, ,
\end{eqnarray}

\begin{eqnarray}
\rho_6^1(s)&=&\frac{\langle\bar{q}q\rangle^2}{192\pi^4} \int dydz\, yz\left(1-y-z\right)\left(s-\overline{m}_c^2\right)\left(5s-3\overline{m}_c^2\right)\, ,
\end{eqnarray}

\begin{eqnarray}
\rho_8^1(s)&=&-\frac{17\langle\bar{q}q\rangle \langle\bar{q}g_{s}\sigma Gq\rangle}{3072\pi^4} \int dydz\, yz \left(4s-3\overline{m}_c^2\right) \nonumber\\
&&+\frac{3\langle\bar{q}q\rangle \langle\bar{q}g_{s}\sigma Gq\rangle}{1024\pi^4}  \int dydz\, y\left(1-y-z\right)\left(4s-3\overline{m}_c^2\right) \nonumber\\
&&-\frac{5\langle\bar{q}q\rangle \langle\bar{q}g_{s}\sigma Gq\rangle}{3072\pi^4}  \int dydz\, z\left(1-y-z\right)\left(4s-3\overline{m}_c^2\right)\, ,
\end{eqnarray}

\begin{eqnarray}
\rho_{9}^1(s)&=&-\frac{11m_c\langle\bar{q}q\rangle^3}{1152\pi^2} \int dy \, ,
\end{eqnarray}

\begin{eqnarray}
\rho_{10}^1(s)&=&\frac{9\langle\bar{q}g_{s}\sigma Gq\rangle^2}{4096\pi^4} \int dy\, y\left(1-y\right)\left[1+\frac{s}{3}\delta\left(s-\widetilde{m}_c^2\right)\right] \nonumber\\
&&-\frac{9\langle\bar{q}g_{s}\sigma Gq\rangle^2}{4096\pi^4} \int dydz\, y \left[1+\frac{s}{3}\delta\left(s-\overline{m}_c^2\right)\right] \nonumber\\
&&+\frac{39\langle\bar{q}g_{s}\sigma Gq\rangle^2}{32768\pi^4} \int dydz\, z \left[1+\frac{s}{3}\delta\left(s-\overline{m}_c^2\right)\right] \nonumber\\
&&-\frac{11\langle\bar{q}g_{s}\sigma Gq\rangle^2}{24576\pi^4}  \int dydz\, \left(1-y-z\right)\left[1+\frac{s}{3}\delta\left(s-\overline{m}_c^2\right)\right] \, ,
\end{eqnarray}

\begin{eqnarray}
\rho_{11}^1(s)&=&\frac{11m_c\langle\bar{q}q\rangle^2 \langle\bar{q}g_{s}\sigma Gq\rangle}{768\pi^2}  \int dy\, \left(1+\frac{s}{2T^2}\right)
\delta\left(s-\widetilde{m}_c^2\right)  \nonumber\\
&&-\frac{5m_c\langle\bar{q}q\rangle^2 \langle\bar{q}g_{s}\sigma Gq\rangle}{576\pi^2} \int dy\, \frac{1-y}{y}\delta\left(s-\widetilde{m}_c^2\right) \nonumber\\
&&-\frac{7m_c\langle\bar{q}q\rangle^2 \langle\bar{q}g_{s}\sigma Gq\rangle}{1536\pi^2} \int dy\,\delta\left(s-\widetilde{m}_c^2\right)\, ,
\end{eqnarray}

\begin{eqnarray}
\rho_{13}^1(s)&=&-\frac{11m_c\langle\bar{q}q\rangle \langle\bar{q}g_{s}\sigma Gq\rangle^2}{3072\pi^2T^2}  \int dy\,
\left(1+\frac{s}{T^2}+\frac{s^2}{2T^4}\right)\delta\left(s-\widetilde{m}_c^2\right) \nonumber\\
&&+\frac{5m_c\langle\bar{q}q\rangle \langle\bar{q}g_{s}\sigma Gq\rangle^2}{1152\pi^2T^2} \int dy\,\frac{1-y}{y}
\left(1+\frac{s}{T^2}\right)\delta\left(s-\widetilde{m}_c^2\right) \nonumber\\
&&+\frac{7m_c\langle\bar{q}q\rangle \langle\bar{q}g_{s}\sigma Gq\rangle^2}{3072\pi^2T^2} \int dy\,
\left(1+\frac{s}{T^2}\right)\delta\left(s-\widetilde{m}_c^2\right) \nonumber\\
&&-\frac{41m_c \langle\bar{q}q\rangle \langle\bar{q}g_{s}\sigma Gq\rangle^2}{18432\pi^2T^2} \int dy\,
\frac{1}{y}\delta\left(s-\widetilde{m}_c^2\right)\, ,
\end{eqnarray}

\begin{eqnarray}
\rho_0^0(s)&=&\frac{11}{3932160\pi^8}\int dydz \,y\left(1-y-z\right)^4\left(s-\overline{m}_c^2\right)^4\left(7s-2\overline{m}_c^2\right) \, ,
\end{eqnarray}

\begin{eqnarray}
\rho_3^0(s)&=&-\frac{m_c \langle\bar{q}q\rangle}{1536\pi^6}\int dydz\, \left(1-y-z\right)^2 \left(s-\overline{m}_c^2\right)^3\, ,
\end{eqnarray}

\begin{eqnarray}
\rho_4^0(s)&=&-\frac{11m_c^2}{589824\pi^6}\langle\frac{\alpha_{s}GG}{\pi}\rangle \int dydz\, \left(\frac{1}{y^2}+\frac{y}{z^3}\right)\left(1-y-z\right)^4
\left(s-\overline{m}_c^2\right)\left(2s-\overline{m}_c^2\right) \nonumber\\
&&+\frac{1}{1179648\pi^6}\langle\frac{\alpha_{s}GG}{\pi}\rangle \int dydz\, \frac{y\left(1-y-z\right)^4}{z^2}\left(s-\overline{m}_c^2\right)^2\left(5s-2\overline{m}_c^2\right)\nonumber\\
&&+\frac{5}{589824\pi^6}\langle\frac{\alpha_{s}GG}{\pi}\rangle \int dydz\,\left( 1+\frac{y}{z}\right) \left(1-y-z\right)^3
\left(s-\overline{m}_c^2\right)^2\left(5s-2\overline{m}_c^2\right)\nonumber\\
&&-\frac{1}{16384\pi^6}\langle\frac{\alpha_{s}GG}{\pi}\rangle \int dydz\, y\left(1-y-z\right)^2
\left(s-\overline{m}_c^2\right)^2\left(5s-2\overline{m}_c^2\right)\nonumber\\
&&-\frac{1}{2359296\pi^6}\langle\frac{\alpha_{s}GG}{\pi}\rangle \int dydz\, \frac{\left(1-y-z\right)^4}{z}
\left(s-\overline{m}_c^2\right)^2\left(5s-2\overline{m}_c^2\right)\, ,
\end{eqnarray}

\begin{eqnarray}
\rho_{5}^0(s)&=&\frac{9m_c \langle\bar{q}g_{s}\sigma Gq\rangle}{8192\pi^6}\int dydz\, \left(1-y-z\right)\left(s-\overline{m}_c^2\right)^2 \nonumber\\
&&-\frac{5m_c\langle\bar{q}g_{s}\sigma Gq\rangle}{4096\pi^6} \int dydz\,
\frac{\left(1-y-z\right)^2}{y}\left(s-\overline{m}_c^2\right)^2 \nonumber\\
&&+\frac{m_c\langle\bar{q}g_{s}\sigma Gq\rangle}{8192\pi^6} \int dydz\, \frac{\left(1-y-z\right)^2}{z}\left(s-\overline{m}_c^2\right)^2\, ,
\end{eqnarray}

\begin{eqnarray}
\rho_{6}^0(s)&=&\frac{19\langle\bar{q}q\rangle^2}{768\pi^4} \int dydz\, y\left(1-y-z\right)
\left(s-\overline{m}_c^2\right)\left(2s-\overline{m}_c^2\right)\, ,
\end{eqnarray}

\begin{eqnarray}
\rho_{8}^0(s)&=&-\frac{41\langle\bar{q}q\rangle \langle\bar{q}g_{s}\sigma Gq\rangle}{3072\pi^4} \int dydz\,
y\left(3s-2\overline{m}_c^2\right) \nonumber\\
&&+\frac{19\langle\bar{q}q\rangle \langle\bar{q}g_{s}\sigma Gq\rangle}{3072\pi^4}  \int dydz\,
\frac{y\left(1-y-z\right)}{z}\left(3s-2\overline{m}_c^2\right)\nonumber\\
&&-\frac{\langle\bar{q}q\rangle \langle\bar{q}g_{s}\sigma Gq\rangle}{512\pi^4}  \int dydz\,
\left(1-y-z\right)\left(3s-2\overline{m}_c^2\right)\, ,
\end{eqnarray}

\begin{eqnarray}
\rho_{9}^0(s)&=&-\frac{11m_c\langle\bar{q}q\rangle^3}{288\pi^2} \int dy\, ,
\end{eqnarray}

\begin{eqnarray}
\rho_{10}^0(s)&=&\frac{11\langle\bar{q}g_{s}\sigma Gq\rangle^2}{3072\pi^4} \int dy\, y
\left[1+\frac{s}{2}\delta\left(s-\widetilde{m}_c^2\right)\right] \nonumber\\
&&-\frac{41\langle\bar{q}g_{s}\sigma Gq\rangle^2}{12288\pi^4}  \int dydz\,
\frac{y}{z}\left[1+\frac{s}{2}\delta\left(s-\overline{m}_c^2\right)\right] \nonumber\\
&&+\frac{7\langle\bar{q}g_{s}\sigma Gq\rangle^2}{6144\pi^4}  \int dydz\,
\left[1+\frac{s}{2}\delta\left(s-\overline{m}_c^2\right)\right] \nonumber\\
&&-\frac{\langle\bar{q}g_{s}\sigma Gq\rangle^2}{2048\pi^4}  \int dydz\, \frac{\left(1-y-z\right)}{z}
\left[1+\frac{s}{2}\delta\left(s-\overline{m}_c^2\right)\right]\, ,
\end{eqnarray}

\begin{eqnarray}
\rho_{11}^0(s)&=&\frac{11m_c\langle\bar{q}q\rangle^2 \langle\bar{q}g_{s}\sigma Gq\rangle}{384\pi^2}  \int dy\,
\left(1+\frac{s}{T^2}\right)\delta\left(s-\widetilde{m}_c^2\right)  \nonumber\\
&&-\frac{61m_c\langle\bar{q}q\rangle^2 \langle\bar{q}g_{s}\sigma Gq\rangle}{4608\pi^2} \int dy\,
\frac{1}{y(1-y)}\delta\left(s-\widetilde{m}_c^2\right) \, ,
\end{eqnarray}

\begin{eqnarray}
\rho_{13}^0(s)&=&-\frac{11m_c\langle\bar{q}q\rangle \langle\bar{q}g_{s}\sigma Gq\rangle^2}{1536\pi^2T^6}  \int dy\,
s^2\delta\left(s-\widetilde{m}_c^2\right) \nonumber\\
&&+\frac{61m_c\langle\bar{q}q\rangle \langle\bar{q}g_{s}\sigma Gq\rangle^2}{9216\pi^2 T^4} \int dy
\frac{1}{y(1-y)}\,s\,\delta\left(s-\widetilde{m}_c^2\right) \nonumber\\
&&-\frac{41m_c \langle\bar{q}q\rangle \langle\bar{q}g_{s}\sigma Gq\rangle^2}{9216\pi^2T^2} \int dy\,
\frac{1}{y\left(1-y\right)}\delta\left(s-\widetilde{m}_c^2\right)\, .
\end{eqnarray}

{\bf  For the $\bar{D}\Sigma_c^*$ pentaquark molecular states},

\begin{eqnarray}
\rho_0^1(s)&=&\frac{11}{7864320\pi^8}\int dy dz \,yz\left(1-y-z\right)^4\left(s-\overline{m}_c^2\right)^4\left(7s-2\overline{m}_c^2\right)\, ,
\end{eqnarray}

\begin{eqnarray}
\rho_3^1(s)&=&- \frac{3m_c\langle\bar{q}q\rangle}{8192\pi^6} \int dy dz \,z\left(1-y-z\right)^2\,\left(s-\overline{m}_c^2\right)^3 \nonumber\\
&&+ \frac{5m_c\langle\bar{q}q\rangle}{36864\pi^6}\int dy dz \, z\left(1-y-z\right)^3 \left(s-\overline{m}_c^2\right)^3 \, ,
\end{eqnarray}

\begin{eqnarray}
\rho_4^1(s)&=&-\frac{11m_c^2}{1179648\pi^6}\langle\frac{\alpha_{s}GG}{\pi}\rangle \int dy dz\,
\left(\frac{z}{y^2}+\frac{y}{z^2}\right)\left(1-y-z\right)^4 \left(s-\overline{m}_c^2\right)\left(2s-\overline{m}_c^2\right) \nonumber\\
&&+\frac{1}{49152\pi^6}\langle\frac{\alpha_{s}GG}{\pi}\rangle \int dy dz\,z\left(1-y-z\right)^3\left(s-\overline{m}_c^2\right)^2\left(5s-2\overline{m}_c^2\right)
  \nonumber\\
&&-\frac{1}{98304\pi^6}\langle\frac{\alpha_{s}GG}{\pi}\rangle \int dy dz\,yz\left(1-y-z\right)^2\left(s-\overline{m}_c^2\right)^2\left(7s-4\overline{m}_c^2\right)\nonumber\\
&&-\frac{1}{3538944\pi^6}\langle\frac{\alpha_{s}GG}{\pi}\rangle \int dy dz\,z\left(1-y-z\right)^3\left(s-\overline{m}_c^2\right)^2\left(53s-20\overline{m}_c^2\right) \nonumber\\
&&-\frac{1}{294912\pi^6}\langle\frac{\alpha_{s}GG}{\pi}\rangle \int dy dz\,yz\left(1-y-z\right)^2 \left(s-\overline{m}_c^2\right)^2\left(4s-\overline{m}_c^2\right)
\, ,
\end{eqnarray}

\begin{eqnarray}
\rho_5^1(s)&=& \frac{31m_c\langle\bar{q}g_{s}\sigma Gq\rangle}{49152\pi^6} \int dy dz \, z\left(1-y-z\right)\left(s-\overline{m}_c^2\right)^2 \nonumber\\
&&-\frac{61m_c \langle\bar{q}g_{s}\sigma Gq\rangle }{98304\pi^6} \int dy dz \, \frac{z}{y} \left(1-y-z\right)^2
\left(s-\overline{m}_c^2\right)^2\nonumber\\
&&+\frac{31m_c \langle\bar{q}g_{s}\sigma Gq\rangle}{147456\pi^6}  \int dy dz \,\frac{z}{y}\left(1-y-z\right)^3\left(s-\overline{m}_c^2\right)^2\nonumber\\
&&-\frac{35m_c \langle\bar{q}g_{s}\sigma Gq\rangle}{98304\pi^6}  \int dy dz \,z\left(1-y-z\right)^2\left(s-\overline{m}_c^2\right)^2\, ,
\end{eqnarray}

\begin{eqnarray}
\rho_6^1(s)&=&\frac{\langle\bar{q}q\rangle^2}{1536\pi^4} \int dy dz \,yz\left(1-y-z\right)\left(s-\overline{m}_c^2\right)\left(23s-14\overline{m}_c^2\right) \, ,
\end{eqnarray}

\begin{eqnarray}
\rho_8^1(s)&=&-\frac{\langle\bar{q}q\rangle \langle\bar{q}g_{s}\sigma Gq\rangle}{3072\pi^4} \int dy dz \, yz\left(37s-28\overline{m}_c^2\right)\nonumber\\
&&-\frac{\langle\bar{q}q\rangle \langle\bar{q}g_{s}\sigma Gq\rangle}{18432\pi^4} \int dy dz \,z\left(1-y-z\right)\left(40s-29\overline{m}_c^2\right)\nonumber\\
&&-\frac{\langle\bar{q}q\rangle \langle\bar{q}g_{s}\sigma Gq\rangle}{18432\pi^4} \int dy dz \,y z \left(17s-13\overline{m}_c^2\right)\, ,
\end{eqnarray}

\begin{eqnarray}
\rho_9^1(s)&=&- \frac{11m_c\langle\bar{q}q\rangle^3}{2304\pi^2} \int dy   \, ,
\end{eqnarray}

\begin{eqnarray}
\rho_{10}^1(s)&=&\frac{\langle\bar{q}g_{s}\sigma Gq\rangle^2}{6144\pi^4} \int dy\, y\left(1-y\right)\left[7
+\frac{9s}{4}\delta\left(s-\widetilde{m}_c^2\right)\right] \nonumber\\
&&+ \frac{\langle\bar{q}g_{s}\sigma Gq\rangle^2}{73728\pi^4} \int dy dz \,z \Big[29+11s\,\delta\left(s-\overline{m}_c^2\right)\Big]\nonumber\\
&&+ \frac{\langle\bar{q}g_{s}\sigma Gq\rangle^2}{73728\pi^4} \int dy\, y\left(1-y\right)\Big[13 +4s\,\delta\left(s-\widetilde{m}_c^2\right)\Big]\nonumber\\
&&-\frac{\langle\bar{q}g_{s}\sigma Gq\rangle^2}{147456\pi^4} \int dy dz \,z \Big[1+s\,\delta\left(s-\overline{m}_c^2\right)\Big]\nonumber\\
&&+\frac{\langle\bar{q}g_{s}\sigma Gq\rangle^2}{1769472\pi^4} \int dy dz \,z \Big[121+49s\,\delta\left(s-\overline{m}_c^2\right)\Big]\, ,
\end{eqnarray}

\begin{eqnarray}
\rho_{11}^1&=&\frac{11m_c \langle\bar{q}q\rangle^2 \langle\bar{q}g_{s}\sigma Gq\rangle}{1536\pi^2} \int dy\,
 \left(1+\frac{s}{2T^2}\right)\delta\left(s-\widetilde{m}_c^2\right) \nonumber\\
 &&-\frac{5m_c \langle\bar{q}q\rangle^2 \langle\bar{q}g_{s}\sigma Gq\rangle}{1152\pi^2}\int dy\,\frac{1-y}{y}\delta\left(s-\widetilde{m}_c^2\right) \, ,
\end{eqnarray}

\begin{eqnarray}
\rho_{13}^1&=&-\frac{11m_c \langle\bar{q}q\rangle \langle\bar{q}g_{s}\sigma Gq\rangle^2}{6144\pi^2T^2}  \int dy\,
\left(1+\frac{s}{T^2}+\frac{s^2}{2T^4}\right) \delta\left(s-\widetilde{m}_c^2\right)\nonumber\\
&&+ \frac{5m_c \langle\bar{q}q\rangle \langle\bar{q}g_{s}\sigma Gq\rangle^2}{2304\pi^2T^2} \int dy\,
\frac{1-y}{y} \left(1+\frac{s}{T^2}\right)\delta\left(s-\widetilde{m}_c^2\right) \, ,
\end{eqnarray}

\begin{eqnarray}
\rho_0^0(s)&=&\frac{11}{7864320\pi^8}\int dy dz \, y\left(1-y-z\right)^4 \left(s-\overline{m}_c^2\right)^4\left(6s-\overline{m}_c^2\right) \, ,
\end{eqnarray}

\begin{eqnarray}
\rho_3^0(s)&=&- \frac{3m_c \langle\bar{q}q\rangle}{8192\pi^6} \int dy dz \,\left(1-y-z\right)^2\left(s-\overline{m}_c^2\right)^3
  \nonumber\\
&&+\frac{5m_c \langle\bar{q}q\rangle}{36864\pi^6} \int dy dz \,\left(1-y-z\right)^3\left(s-\overline{m}_c^2\right)^3 \, ,
\end{eqnarray}

\begin{eqnarray}
\rho_4^0(s)&=&-\frac{11m_c^2}{2359296\pi^6}\langle\frac{\alpha_{s}GG}{\pi}\rangle \int dy dz\,\left(\frac{1}{y^2}+\frac{y}{z^3}\right)\left(1-y-z\right)^4
\left(s-\overline{m}_c^2\right)\left(3s-\overline{m}_c^2\right) \nonumber\\
&&+\frac{11}{2359296\pi^6}\langle\frac{\alpha_{s}GG}{\pi}\rangle \int dy dz\,\frac{y}{z^2}\left(1-y-z\right)^4
\left(s-\overline{m}_c^2\right)^2\left(4s-\overline{m}_c^2\right)\nonumber\\
&&+\frac{1}{49152\pi^6}\langle\frac{\alpha_{s}GG}{\pi}\rangle \int dy dz\,\left(1-y-z\right)^3\left(s-\overline{m}_c^2\right)^2\left(4s-\overline{m}_c^2\right) \nonumber\\
&&-\frac{1}{32768\pi^6}\langle\frac{\alpha_{s}GG}{\pi}\rangle \int dy dz\,y\left(1-y-z\right)^2\left(s-\overline{m}_c^2\right)^2\left(2s-\overline{m}_c^2\right)
\nonumber\\
&&-\frac{1}{1179648\pi^6}\langle\frac{\alpha_{s}GG}{\pi}\rangle  \int dy dz\,\left(1-y-z\right)^3\left(s-\overline{m}_c^2\right)^2\left(14s-3\overline{m}_c^2\right)
\nonumber\\
&&-\frac{1}{98304\pi^6}\langle\frac{\alpha_{s}GG}{\pi}\rangle \int dy dz\,y\left(1-y-z\right)^2s\left(s-\overline{m}_c^2\right)^2\, ,
\end{eqnarray}

\begin{eqnarray}
\rho_5^0(s)&=&\frac{31m_c \langle\bar{q}g_{s}\sigma Gq\rangle}{49152\pi^6}  \int dy dz \,\left(1-y-z\right)\left(s-\overline{m}_c^2\right)^2 \nonumber\\
&&-\frac{61m_c \langle\bar{q}g_{s}\sigma Gq\rangle}{98304\pi^6}  \int dy dz \,\frac{\left(1-y-z\right)^2}{y}\left(s-\overline{m}_c^2\right)^2\nonumber\\
&&+\frac{31m_c \langle\bar{q}g_{s}\sigma Gq\rangle}{147456\pi^6}  \int dy dz \,\frac{\left(1-y-z\right)^3}{y}\left(s-\overline{m}_c^2\right)^2
\nonumber\\
&&-\frac{35m_c \langle\bar{q}g_{s}\sigma Gq\rangle}{98304\pi^6}  \int dy dz \,\left(1-y-z\right)^2\left(s-\overline{m}_c^2\right)^2\, ,
\end{eqnarray}

\begin{eqnarray}
\rho_6^0(s)&=& \frac{\langle\bar{q}q\rangle^2}{3072\pi^4} \int dy dz \,y\left(1-y-z\right) \left(s-\overline{m}_c^2\right)\left(37s-19\overline{m}_c^2\right) \, ,
\end{eqnarray}

\begin{eqnarray}
\rho_8^0(s)&=&- \frac{\langle\bar{q}q\rangle \langle\bar{q}g_{s}\sigma Gq\rangle}{3072\pi^4} \int dy dz \,y\left(28s-19\overline{m}_c^2\right)\nonumber\\
&&- \frac{\langle\bar{q}q\rangle \langle\bar{q}g_{s}\sigma Gq\rangle}{18432\pi^4} \int dy dz \,\left(1-y-z\right)\left(29s-18\overline{m}_c^2\right)\nonumber\\
&&- \frac{\langle\bar{q}q\rangle \langle\bar{q}g_{s}\sigma Gq\rangle}{18432\pi^4} \int dy dz \,y\left(13s-9\overline{m}_c^2\right)\, ,
\end{eqnarray}

\begin{eqnarray}
\rho_9^0(s)&=& -\frac{11m_c\langle\bar{q}q\rangle^3}{1152\pi^2} \int dy \, ,
\end{eqnarray}

\begin{eqnarray}
\rho_{10}^0(s)&=& \frac{\langle\bar{q}g_{s}\sigma Gq\rangle^2}{24576\pi^4}\int dy\,y
\Big[19-9s\delta\left(s-\widetilde{m}_c^2\right)\Big]\nonumber\\
&&+\frac{\langle\bar{q}g_{s}\sigma Gq\rangle^2}{73728\pi^4}\int dydz\,\Big[18+11s\,\delta\left(s-\overline{m}_c^2\right)\Big]\nonumber\\
&&+\frac{\langle\bar{q}g_{s}\sigma Gq\rangle^2}{18432\pi^4}\int dy\,y\left[\frac{9}{4}+s\,\delta\left(s-\widetilde{m}_c^2\right)\right]\nonumber\\
&&-\frac{\langle\bar{q}g_{s}\sigma Gq\rangle^2}{147456\pi^4}\int dydz\,s\,\delta\left(s-\overline{m}_c^2\right)\nonumber\\
&&+\frac{\langle\bar{q}g_{s}\sigma Gq\rangle^2}{1769472\pi^4}\int dydz\,\Big[72+49s\,\delta\left(s-\overline{m}_c^2\right)\Big]\, ,
\end{eqnarray}

\begin{eqnarray}
\rho_{11}^0&=& \frac{11m_c \langle\bar{q}q\rangle^2 \langle\bar{q}g_{s}\sigma Gq\rangle }{1536\pi^2} \int dy\,
\left(1 +\frac{s}{T^2}\right)\delta\left(s-\widetilde{m}_c^2\right) \nonumber\\
&&-\frac{5m_c \langle\bar{q}q\rangle^2 \langle\bar{q}g_{s}\sigma Gq\rangle}{1152\pi^2}\int dy\,\frac{1}{y}\,\delta\left(s-\widetilde{m}_c^2\right) \, ,
\end{eqnarray}

\begin{eqnarray}
\rho_{13}^0&=&-\frac{11m_c \langle\bar{q}q\rangle \langle\bar{q}g_{s}\sigma Gq\rangle^2}{6144\pi^2T^6} \int dy\,s^2\, \delta\left(s-\widetilde{m}_c^2\right)\nonumber\\
&&+\frac{5m_c \langle\bar{q}q\rangle \langle\bar{q}g_{s}\sigma Gq\rangle^2}{2304\pi^2T^4} \int dy\,\frac{1}{y}\,s\,\delta\left(s-\widetilde{m}_c^2\right)\, .
\end{eqnarray}

{ \bf For the $\bar{D}^*\Sigma_c$ pentaquark molecular states},
\begin{eqnarray}
\rho_0^1(s)&=&\frac{1}{786432\pi^8}\int dydz\, yz\left(1-y-z\right)^4\left(s-\overline{m}_c^2\right)^4\left(31s-9\overline{m}_c^2\right) \, ,
\end{eqnarray}

\begin{eqnarray}
\rho_3^1(s)&=&-\frac{19m_c\langle\bar{q}q\rangle }{12288\pi^6} \int dydz\, z\left(1-y-z\right)^2\left(s-\overline{m}_c^2\right)^3 \, ,
\end{eqnarray}

\begin{eqnarray}
\rho_4^1(s)&=&-\frac{m_c^2}{1179648\pi^6} \langle\frac{\alpha_{s}GG}{\pi}\rangle \int dydz\,
\left(\frac{z}{y^2}+\frac{y}{z^2}\right)\left(1-y-z\right)^4\left(s-\overline{m}_c^2\right)\left(89s-45\overline{m}_c^2\right)  \nonumber\\
&&+\frac{1}{73728\pi^6}\langle\frac{\alpha_{s}GG}{\pi}\rangle \int dydz\,z\left(1-y-z\right)^3\left(s-\overline{m}_c^2\right)^2\left(s+2\overline{m}_c^2\right)
\nonumber\\
&&-\frac{1}{49152\pi^6}\langle\frac{\alpha_{s}GG}{\pi}\rangle \int dydz\,yz\left(1-y-z\right)^2\left(s-\overline{m}_c^2\right)^2\left(5s-2\overline{m}_c^2\right)
 \nonumber\\
&&+\frac{1}{589824\pi^6}\langle\frac{\alpha_{s}GG}{\pi}\rangle \int dydz\,z\left(1-y-z\right)^4\left(s-\overline{m}_c^2\right)^2\left(7s-4\overline{m}_c^2\right)
 \nonumber\\
&&+\frac{1}{1769472\pi^6}\langle\frac{\alpha_{s}GG}{\pi}\rangle \int dydz\,z\left(1-y-z\right)^3\left(s-\overline{m}_c^2\right)^2\left(49s-25\overline{m}_c^2\right)
\nonumber\\
&&-\frac{1}{1769472\pi^6}\langle\frac{\alpha_{s}GG}{\pi}\rangle \int dydz\,y\left(1-y-z\right)^3\left(s-\overline{m}_c^2\right)^2\left(43s-16\overline{m}_c^2\right)
  \nonumber\\
&&+\frac{1}{7077888\pi^6}\langle\frac{\alpha_{s}GG}{\pi}\rangle \int dydz\,\left(1-y-z\right)^4\left(s-\overline{m}_c^2\right)^2\left(7s-4\overline{m}_c^2\right)\, ,
\end{eqnarray}

\begin{eqnarray}
\rho_5^1(s)&=&\frac{11m_c\langle\bar{q}g_{s}\sigma Gq\rangle}{4096\pi^6} \int dydz\, z\left(1-y-z\right)\left(s-\overline{m}_c^2\right)^2 \nonumber\\
&&+\frac{3m_c\langle\bar{q}g_{s}\sigma Gq\rangle}{32768\pi^6} \int dydz\,\left(1-y-z\right)^2\left(s-\overline{m}_c^2\right)^2 \, ,
\end{eqnarray}

\begin{eqnarray}
\rho_6^1(s)&=&\frac{\langle\bar{q}q\rangle^2}{1536\pi^4}  \int dydz\,  yz\left(1-y-z\right)\left(s-\overline{m}_c^2\right)\left(31s-15\overline{m}_c^2\right)\, ,
\end{eqnarray}

\begin{eqnarray}
\rho_8^1(s)&=&-\frac{\langle\bar{q}q\rangle \langle\bar{q}g_{s}\sigma Gq\rangle}{1536\pi^4} \int dydz\, y z\left(23s-15\overline{m}_c^2\right) \nonumber\\
&&+\frac{\langle\bar{q}q\rangle \langle\bar{q}g_{s}\sigma Gq\rangle}{256\pi^4} \int dydz\,y\left(1-y-z\right)\left(3s-2\overline{m}_c^2\right)\nonumber\\
&&-\frac{\langle\bar{q}q\rangle \langle\bar{q}g_{s}\sigma Gq\rangle}{9216\pi^4} \int dydz\,y z \left(8s-5\overline{m}_c^2\right)\nonumber\\
&&+\frac{\langle\bar{q}q\rangle \langle\bar{q}g_{s}\sigma Gq\rangle}{9216\pi^4} \int dydz\,z\left(1-y-z\right)\left(23s-18\overline{m}_c^2\right) \nonumber\\
&&-\frac{\langle\bar{q}q\rangle \langle\bar{q}g_{s}\sigma Gq\rangle}{9216\pi^4} \int dydz\,y\left(1-y-z\right)\left(25s-16\overline{m}_c^2\right) \, ,
\end{eqnarray}

\begin{eqnarray}
\rho_9^1(s)&=&-\frac{11m_c\langle\bar{q}q\rangle^3}{1152\pi^2}  \int dy\, ,
\end{eqnarray}

\begin{eqnarray}
\rho_{10}^1(s)&=& \frac{\langle\bar{q}g_{s}\sigma Gq\rangle^2}{4096\pi^4}\int dy \,y\left(1-y\right)
\left[5+\frac{8s}{3}\delta\left(s-\widetilde{m}_c^2\right)\right]\nonumber\\
&&-\frac{\langle\bar{q}g_{s}\sigma Gq\rangle^2}{512\pi^4}\int dydz\,y\left[1+\frac{s}{2}\,\delta\left(s-\overline{m}_c^2\right)\right]\nonumber\\
&&+ \frac{\langle\bar{q}g_{s}\sigma Gq\rangle^2}{36864\pi^4}\int dy \,y\left(1-y\right)
\Big[5+3s\,\delta\left(s-\widetilde{m}_c^2\right)\Big]\nonumber\\
&&-\frac{\langle\bar{q}g_{s}\sigma Gq\rangle^2}{36864\pi^4} \int dydz\,z \Big[18+5s\,\delta\left(s-\overline{m}_c^2\right)\Big]\nonumber\\
&&+\frac{\langle\bar{q}g_{s}\sigma Gq\rangle^2}{2304\pi^4} \int dydz\,y\left[1+\frac{9s}{16}\delta\left(s-\overline{m}_c^2\right)\right]\nonumber\\
&&+\frac{\langle\bar{q}g_{s}\sigma Gq\rangle^2}{73728\pi^4} \int dydz\,z\Big[4+s\,\delta\left(s-\overline{m}_c^2\right)\Big]\nonumber\\
&&-\frac{\langle\bar{q}g_{s}\sigma Gq\rangle^2}{884736\pi^4} \int dydz\,z\Big[196+73s\,\delta\left(s-\overline{m}_c^2\right)\Big] \nonumber\\
&&+\frac{11\langle\bar{q}g_{s}\sigma Gq\rangle^2}{221184\pi^4} \int dydz\,\left(1-y-z\right)\Big[4+s\,\delta\left(s-\overline{m}_c^2\right)\Big]\, ,
\end{eqnarray}

\begin{eqnarray}
\rho_{11}^1(s)&=& \frac{11m_c \langle\bar{q}q\rangle^2 \langle\bar{q}g_{s}\sigma Gq\rangle}{768\pi^2} \int dy \,  \left(1+\frac{s}{2T^2}\right)\delta\left(s-\widetilde{m}_c^2\right) \nonumber\\
&&+\frac{m_c \langle\bar{q}q\rangle^2 \langle\bar{q}g_{s}\sigma Gq\rangle}{6912\pi^2} \int dy \,\frac{1-y}{y}\delta\left(s-\widetilde{m}_c^2\right) \nonumber\\
&&-\frac{7m_c \langle\bar{q}q\rangle^2 \langle\bar{q}g_{s}\sigma Gq\rangle}{1536\pi^2} \int dy \, \delta\left(s-\widetilde{m}_c^2\right) \, ,
\end{eqnarray}

\begin{eqnarray}
\rho_{13}^1(s)&=&-\frac{11m_c \langle\bar{q}q\rangle \langle\bar{q}g_{s}\sigma Gq\rangle^2}{3072\pi^2T^2} \int dy \,
 \left(1+\frac{s}{T^2}+\frac{s^2}{2T^4}\right) \delta\left(s-\widetilde{m}_c^2\right)\nonumber\\
&&-\frac{m_c \langle\bar{q}q\rangle \langle\bar{q}g_{s}\sigma Gq\rangle^2}{13824\pi^2T^2} \int dy \, \frac{1-y}{y}
\left(1+\frac{s}{T^2}\right)\delta\left(s-\widetilde{m}_c^2\right)\nonumber\\
&&+\frac{7m_c \langle\bar{q}q\rangle \langle\bar{q}g_{s}\sigma Gq\rangle^2}{3072\pi^2T^2} \int dy \,
\left(1+\frac{s}{T^2}\right)\delta\left(s-\widetilde{m}_c^2\right)\nonumber\\
&&+\frac{5m_c \langle\bar{q}q\rangle \langle\bar{q}g_{s}\sigma Gq\rangle^2}{165888\pi^2T^2} \int dy \,
\frac{1}{y} \delta\left(s-\widetilde{m}_c^2\right)\, ,
\end{eqnarray}

\begin{eqnarray}
\rho_0^0(s)&=&\frac{11}{3932160\pi^8}\int dy dz \,y\left(1-y-z\right)^4 \left(s-\overline{m}_c^2\right)^4\left(6s-\overline{m}_c^2\right)\, ,
\end{eqnarray}

\begin{eqnarray}
\rho_{3}^0(s)&=& -\frac{m_c \langle\bar{q}q\rangle}{1536\pi^6}\int dy dz \, \left(1-y-z\right)^2\left(s-\overline{m}_c^2\right)^3\, ,
\end{eqnarray}

\begin{eqnarray}
\rho_4^0(s)&=&-\frac{11m_c^2}{1179648\pi^6}\langle\frac{\alpha_{s}GG}{\pi}\rangle \int dy dz \,
\left(\frac{1}{y^2}+\frac{y}{z^3}\right)\left(1-y-z\right)^4\left(s-\overline{m}_c^2\right)\left(3s-\overline{m}_c^2\right)
 \nonumber\\
&&+\frac{11}{1179648\pi^6}\langle\frac{\alpha_{s}GG}{\pi}\rangle \int dy dz \,\frac{y}{z^2}\left(1-y-z\right)^4\left(s-\overline{m}_c^2\right)^2 \left(4s-\overline{m}_c^2\right) \nonumber\\
&&-\frac{1}{16384\pi^6}\langle\frac{\alpha_{s}GG}{\pi}\rangle \int dy dz \,y\left(1-y-z\right)^2\left(s-\overline{m}_c^2\right)^2 \left(4s-\overline{m}_c^2\right) \nonumber\\
&&-\frac{7}{196608\pi^6}\langle\frac{\alpha_{s}GG}{\pi}\rangle \int dy dz \, \left(1-y-z\right)^3 \left(s-\overline{m}_c^2\right)^2 \left(2s-\overline{m}_c^2\right) \nonumber\\
&&+\frac{7}{196608\pi^6}\langle\frac{\alpha_{s}GG}{\pi}\rangle \int dy dz \, \frac{y}{z}\left(1-y-z\right)^3\left(s-\overline{m}_c^2\right)^2 \left(4s-\overline{m}_c^2\right) \nonumber\\
&&+\frac{1}{2359296\pi^6}\langle\frac{\alpha_{s}GG}{\pi}\rangle \int dy dz \,\frac{1}{z}\left(1-y-z\right)^4\left(s-\overline{m}_c^2\right)^2 \left(2s-\overline{m}_c^2\right) \, ,
\end{eqnarray}

\begin{eqnarray}
\rho_{5}^0(s)&=& \frac{9m_c \langle\bar{q}g_{s}\sigma Gq\rangle}{8192\pi^6}\int dy dz \,\left(1-y-z\right)\left(s-\overline{m}_c^2\right)^2 \nonumber\\
&&+\frac{m_c \langle\bar{q}g_{s}\sigma Gq\rangle}{8192\pi^6}\int dy dz \,\frac{1}{z}\left(1-y-z\right)^2\left(s-\overline{m}_c^2\right)^2 \, ,
\end{eqnarray}

\begin{eqnarray}
\rho_{6}^0(s)&=&\frac{19\langle\bar{q}q\rangle^2}{1536\pi^4} \int dy dz \, y\left(1-y-z\right)\left(s-\overline{m}_c^2\right)\left(3s-\overline{m}_c^2\right)\, ,
\end{eqnarray}

\begin{eqnarray}
\rho_{8}^0(s)&=&-\frac{41\langle\bar{q}q\rangle \langle\bar{q}g_{s}\sigma Gq\rangle}{3072\pi^4}  \int dy dz \, y \left(2s-\overline{m}_c^2\right)\nonumber\\
&&+\frac{\langle\bar{q}q\rangle \langle\bar{q}g_{s}\sigma Gq\rangle}{1536\pi^4}  \int dy dz \,\left(1-y-z\right)\left(4s-3\overline{m}_c^2\right) \nonumber\\
&&+\frac{19\langle\bar{q}q\rangle \langle\bar{q}g_{s}\sigma Gq\rangle}{3072\pi^4} \int dy dz \,\frac{y}{z}\left(1-y-z\right)\left(2s-\overline{m}_c^2\right) \, ,
\end{eqnarray}

\begin{eqnarray}
\rho_{9}^0(s)&=&-\frac{11 m_c\langle\bar{q}q\rangle^3}{288\pi^2} \int dy\, ,
\end{eqnarray}

\begin{eqnarray}
\rho_{10}^0(s)&=& \frac{11\langle\bar{q}g_{s}\sigma Gq\rangle^2}{6144\pi^4} \int dy \, y \Big[1+s\,\delta\left(s-\widetilde{m}_c^2\right)\Big]\nonumber\\
&&- \frac{7\langle\bar{q}g_{s}\sigma Gq\rangle^2}{36864\pi^4}\int dy dz\,\Big[3+s\,\delta\left(s-\overline{m}_c^2\right)\Big]\nonumber\\
&&-\frac{41\langle\bar{q}g_{s}\sigma Gq\rangle^2}{24576\pi^4}\int dy dz\,\frac{y}{z}\Big[1+s\,\delta\left(s-\overline{m}_c^2\right)\Big]\nonumber\\
&&+\frac{\langle\bar{q}g_{s}\sigma Gq\rangle^2}{12288\pi^4}\int dy dz\,\frac{1}{z}\left(1-y-z\right)\Big[3+s\,\delta\left(s-\overline{m}_c^2\right)\Big]\, ,
\end{eqnarray}

\begin{eqnarray}
\rho_{11}^0&=& \frac{11m_c \langle\bar{q}q\rangle^2 \langle\bar{q}g_{s}\sigma Gq\rangle}{384\pi^2} \int dy\, \left(1+\frac{s}{T^2}\right)\delta\left(s-\widetilde{m}_c^2\right) \nonumber\\
&&-\frac{7m_c \langle\bar{q}q\rangle^2 \langle\bar{q}g_{s}\sigma Gq\rangle}{768\pi^2} \int dy\,
\frac{1}{1-y}\delta\left(s-\widetilde{m}_c^2\right) \, ,
\end{eqnarray}

\begin{eqnarray}
\rho_{13}^0&=& -\frac{11m_c \langle\bar{q}q\rangle \langle\bar{q}g_{s}\sigma Gq\rangle^2}{1536\pi^2T^6}\int dy\,
s^2\, \delta\left(s-\widetilde{m}_c^2\right)\nonumber\\
&&+ \frac{7m_c \langle\bar{q}q\rangle \langle\bar{q}g_{s}\sigma Gq\rangle^2}{1536\pi^2T^4}\int dy\,
\frac{1}{1-y}s\, \delta\left(s-\widetilde{m}_c^2\right)\, .
\end{eqnarray}

{ \bf  For the $\bar{D}^*\Sigma_c^*$ pentaquark molecular states},
\begin{eqnarray}
\rho_0^1(s)&=&\frac{1}{1966080\pi^8} \int dy dz \, yz\left(1-y-z\right)^4\left(4+y+z\right)\left(s-\overline{m}_c^2\right)^4\left(7s-2\overline{m}_c^2\right) \, ,
\end{eqnarray}

\begin{eqnarray}
\rho_3^1(s)&=&-\frac{m_c\langle\bar{q}q\rangle}{6144\pi^6} \int dy dz \,z\left(1-y-z\right)^2\left(2+y+z\right)\left(s-\overline{m}_c^2\right)^3 \, ,
\end{eqnarray}

\begin{eqnarray}
\rho_4^1(s)&=&-\frac{m_c^2}{294912\pi^6}\langle\frac{\alpha_{s}GG}{\pi}\rangle \int dy dz \,
\left(\frac{z}{y^2}+\frac{y}{z^2}\right)\left(1-y-z\right)^4\left(4+y+z\right)\left(s-\overline{m}_c^2\right)\left(2s-\overline{m}_c^2\right) \nonumber\\
&&-\frac{1}{73728\pi^6}\langle\frac{\alpha_{s}GG}{\pi}\rangle \int dy dz \,
z\left(1-y-z\right)^3\left(s-\overline{m}_c^2\right)^2\left(7s-4\overline{m}_c^2\right) \nonumber\\
&&+\frac{1}{98304\pi^6}\langle\frac{\alpha_{s}GG}{\pi}\rangle \int dy dz \,
z\left(1-y-z\right)^4\left(s-\overline{m}_c^2\right)^2\left(3s-2\overline{m}_c^2\right) \nonumber\\
&&-\frac{1}{49152\pi^6}\langle\frac{\alpha_{s}GG}{\pi}\rangle \int dy dz \,
yz\left(1-y-z\right)^2\left(s-\overline{m}_c^2\right)^2\left(5s-2\overline{m}_c^2\right) \nonumber\\
&&+\frac{1}{884736\pi^6}\langle\frac{\alpha_{s}GG}{\pi}\rangle \int dy dz \,
z\left(1-y-z\right)^3\left(s-\overline{m}_c^2\right)^2\left(23s-14\overline{m}_c^2\right) \nonumber\\
&&-\frac{1}{3538944\pi^6}\langle\frac{\alpha_{s}GG}{\pi}\rangle \int dy dz \,
z\left(1-y-z\right)^4\left(s-\overline{m}_c^2\right)^2\left(31s-22\overline{m}_c^2\right) \nonumber\\
&&-\frac{1}{442368\pi^6}\langle\frac{\alpha_{s}GG}{\pi}\rangle \int dy dz \,
yz\left(1-y-z\right)^3\left(s-\overline{m}_c^2\right)^2\left(5s-2\overline{m}_c^2\right) \, ,
\end{eqnarray}

\begin{eqnarray}
\rho_5^1(s)&=& \frac{3m_c\langle\bar{q}g_{s}\sigma Gq\rangle}{4096\pi^6} \int dy dz \,z\left(1-y-z\right)\left(s-\overline{m}_c^2\right)^2 \nonumber\\
&&-\frac{3m_c\langle\bar{q}g_{s}\sigma Gq\rangle}{8192\pi^6} \int dy dz \,z\left(1-y-z\right)^2\left(s-\overline{m}_c^2\right)^2\, ,
\end{eqnarray}

\begin{eqnarray}
\rho_6^1(s)&=&\frac{\langle\bar{q}q\rangle^2}{128\pi^4} \int dy dz \, yz\left(1-y-z\right) \left(s-\overline{m}_c^2\right)\left(2s-\overline{m}_c^2\right)\, ,
\end{eqnarray}

\begin{eqnarray}
\rho_8^1(s)&=&-\frac{\langle\bar{q}q\rangle\langle\bar{q}g_{s}\sigma Gq\rangle}{256\pi^4}  \int dy dz \,y z\left(3s-2\overline{m}_c^2\right) \nonumber\\
&&+\frac{\langle\bar{q}q\rangle\langle\bar{q}g_{s}\sigma Gq\rangle}{1536\pi^4}  \int dy dz \,z\left(1-y-z\right)\left(3s-2\overline{m}_c^2\right)\, ,
\end{eqnarray}

\begin{eqnarray}
\rho_9^1(s)&=&-\frac{5m_c\langle\bar{q}q\rangle^3}{576\pi^2}  \int dy\, ,
\end{eqnarray}

\begin{eqnarray}
\rho_{10}^1(s)&=&\frac{\langle\bar{q}g_{s}\sigma Gq\rangle^2}{2048\pi^4}  \int dy\, y\left(1-y\right)\Big[2+s\,\delta\left(s-\widetilde{m}_c^2\right)\Big]\nonumber\\
&&-\frac{3\langle\bar{q}g_{s}\sigma Gq\rangle^2}{16384\pi^4} \int dy dz \,z \Big[2+s\,\delta\left(s-\overline{m}_c^2\right)\Big]\, ,
\end{eqnarray}

\begin{eqnarray}
\rho_{11}^1(s)&=&\frac{5m_c\langle\bar{q}q\rangle^2 \langle\bar{q}g_{s}\sigma Gq\rangle}{384\pi^2} \int dy\, \left(1+\frac{s}{2T^2}\right)\delta\left(s-\widetilde{m}_c^2\right) \nonumber\\
&&+\frac{m_c\langle\bar{q}q\rangle^2 \langle\bar{q}g_{s}\sigma Gq\rangle}{3456\pi^2} \int dy\,
\frac{1-y}{y} \delta\left(s-\widetilde{m}_c^2\right)\, ,
\end{eqnarray}

\begin{eqnarray}
\rho_{13}^1(s)&=&-\frac{5m_c\langle\bar{q}q\rangle \langle\bar{q}g_{s}\sigma Gq\rangle^2}{1536\pi^2T^2} \int dy\,
 \left(1+\frac{s}{T^2}+\frac{s^2}{2T^4}\right) \delta\left(s-\widetilde{m}_c^2\right)\nonumber\\
&&-\frac{m_c\langle\bar{q}q\rangle \langle\bar{q}g_{s}\sigma Gq\rangle^2}{6912\pi^2T^2} \int dy\,
 \frac{1-y}{y} \left(1+\frac{s}{T^2}\right)\delta\left(s-\widetilde{m}_c^2\right)\, ,
\end{eqnarray}

\begin{eqnarray}
\rho_0^0(s)&=&\frac{1}{1966080\pi^8}\int dy dz \,y\left(1-y-z\right)^4\left(4+y+z\right)\left(s-\overline{m}_c^2\right)^4\left(6s-\overline{m}_c^2\right)\, ,
\end{eqnarray}

\begin{eqnarray}
\rho_{3}^0(s)&=&-\frac{m_c\langle\bar{q}q\rangle}{6144\pi^6} \int dy dz \,\left(1-y-z\right)^2\left(2+y+z\right) \left(s-\overline{m}_c^2\right)^3 \, ,
\end{eqnarray}

\begin{eqnarray}
\rho_4^0(s)&=&-\frac{m_c^2}{589824\pi^6}\langle\frac{\alpha_{s}GG}{\pi}\rangle \int dy dz \,
\left(\frac{1}{y^2}+\frac{y}{z^3}\right)\left(1-y-z\right)^4\left(4+y+z\right)\left(s-\overline{m}_c^2\right)\left(3s-\overline{m}_c^2\right) \nonumber\\
&&+\frac{1}{589824\pi^6}\langle\frac{\alpha_{s}GG}{\pi}\rangle \int dy dz \,
\frac{y}{z^2}\left(1-y-z\right)^4\left(4+y+z\right)\left(s-\overline{m}_c^2\right)^2\left(4s-\overline{m}_c^2\right) \nonumber\\
&&-\frac{1}{24576\pi^6}\langle\frac{\alpha_{s}GG}{\pi}\rangle \int dy dz \,
\left(1-y-z\right)^3\left(s-\overline{m}_c^2\right)^2\left(2s-\overline{m}_c^2\right)\nonumber\\
&&+\frac{1}{294912\pi^6}\langle\frac{\alpha_{s}GG}{\pi}\rangle \int dy dz \,
\left(1-y-z\right)^4\left(s-\overline{m}_c^2\right)^2\left(8s-5\overline{m}_c^2\right)\nonumber\\
&&-\frac{1}{49152\pi^6}\langle\frac{\alpha_{s}GG}{\pi}\rangle \int dy dz \,
y\left(1-y-z\right)^2\left(s-\overline{m}_c^2\right)^2\left(4s-\overline{m}_c^2\right)\nonumber\\
&&+\frac{1}{884736\pi^6}\langle\frac{\alpha_{s}GG}{\pi}\rangle \int dy dz \,
\left(1-y-z\right)^3\left(s-\overline{m}_c^2\right)^2\left(20s-11\overline{m}_c^2\right)\nonumber\\
&&-\frac{1}{3538944\pi^6}\langle\frac{\alpha_{s}GG}{\pi}\rangle \int dy dz \,
\left(1-y-z\right)^4\left(s-\overline{m}_c^2\right)^2\left(28s-19\overline{m}_c^2\right)\nonumber\\
&&-\frac{1}{442368\pi^6}\langle\frac{\alpha_{s}GG}{\pi}\rangle \int dy dz \,
y\left(1-y-z\right)^3\left(s-\overline{m}_c^2\right)^2\left(4s-\overline{m}_c^2\right)\, ,
\end{eqnarray}

\begin{eqnarray}
\rho_{5}^0(s)&=&\frac{3m_c\langle\bar{q}g_{s}\sigma Gq\rangle}{8192\pi^6} \int dy dz \,\left(1-y-z\right)\left(1+y+z\right)\left(s-\overline{m}_c^2\right)^2 \, ,
\end{eqnarray}

\begin{eqnarray}
\rho_{6}^0(s)&=&\frac{\langle\bar{q}q\rangle^2}{256\pi^4} \int dy dz \,y\left(1-y-z\right)\left(s-\overline{m}_c^2\right)\left(3s-\overline{m}_c^2\right)\, ,
\end{eqnarray}

\begin{eqnarray}
\rho_{8}^0(s)&=&- \frac{\langle\bar{q}q\rangle\langle\bar{q}g_{s}\sigma Gq\rangle}{256\pi^4} \int dy dz \, y\left(2s-\overline{m}_c^2\right) \nonumber\\
&&+\frac{\langle\bar{q}q\rangle\langle\bar{q}g_{s}\sigma Gq\rangle}{1536\pi^4} \int dy dz \,
\left(1-y-z\right)\left(2s-\overline{m}_c^2\right)\, ,
\end{eqnarray}

\begin{eqnarray}
\rho_{9}^0(s)&=& -\frac{5m_c\langle\bar{q}q\rangle^3}{288\pi^2} \int dy\, ,
\end{eqnarray}

\begin{eqnarray}
\rho_{10}^0(s)&=& \frac{\langle\bar{q}g_{s}\sigma Gq\rangle^2}{2048\pi^4}  \int dy\,y\Big[1+s\,\delta\left(s-\widetilde{m}_c^2\right)\Big]\nonumber\\
&&-\frac{3\langle\bar{q}g_{s}\sigma Gq\rangle^2}{16384\pi^4} \int dy dz \, \Big[1+s\,\delta\left(s-\overline{m}_c^2\right)\Big]\, ,
\end{eqnarray}

\begin{eqnarray}
\rho_{11}^0(s)&=& \frac{5m_c\langle\bar{q}q\rangle^2 \langle\bar{q}g_{s}\sigma Gq\rangle}{384\pi^2} \int dy\,
\left(1+\frac{s}{T^2}\right)\delta\left(s-\widetilde{m}_c^2\right) \nonumber\\
&&+ \frac{m_c\langle\bar{q}q\rangle^2 \langle\bar{q}g_{s}\sigma Gq\rangle}{3456\pi^2} \int dy\,
 \frac{1}{y}   \delta\left(s-\widetilde{m}_c^2\right)\, ,
\end{eqnarray}

\begin{eqnarray}
\rho_{13}^0(s)&=& -\frac{5m_c\langle\bar{q}q\rangle \langle\bar{q}g_{s}\sigma Gq\rangle^2}{1536\pi^2T^6}  \int dy\,
s^2\, \delta\left(s-\widetilde{m}_c^2\right)\nonumber\\
&&-\frac{m_c\langle\bar{q}q\rangle \langle\bar{q}g_{s}\sigma Gq\rangle^2}{6912\pi^2T^4}  \int dy\,
\frac{1}{y}\, s\,  \delta\left(s-\widetilde{m}_c^2\right)\, ,
\end{eqnarray}
where $\int dydz =\int_{y_i}^{y_f}dy \int_{z_i}^{1-y}dz$, $\int dy=\int_{y_i}^{y_f}dy$, $y_{f}=\frac{1+\sqrt{1-4m_{c}^{2}/s}}{2}$, $y_{i}=\frac{1-\sqrt{1-4m_{c}^{2}/s}}{2}$, $z_{i}=\frac{ym_{c}^{2}}{ys-m_{c}^{2}}$, $\overline{m}_{c}^{2}=\frac{(y+z)m_{c}^{2}}{yz}$, $\widetilde{m}_{c}^{2}=\frac{m_{c}^{2}}{y(1-y)}$, $\int_{y_{i}}^{y_{f}}dy\rightarrow\int_{0}^{1}$, $\int_{z_{i}}^{1-y}dz\rightarrow\int_{0}^{1-y}dz$, when the $\delta$ functions $\delta(s-\overline{m}_{c}^{2})$ and $\delta(s-\widetilde{m}_{c}^{2})$ appear.

\section*{Acknowledgements}
This  work is supported by National Natural Science Foundation,
Grant Number 11775079, and the Fundamental Research Funds for the
Central Universities, Grant Number 2016MS155.

\end{document}